\documentclass[12pt]{article}

\usepackage{epsf,amsfonts,hyperref}
\usepackage{graphicx}

\renewcommand{\theequation}{\thesection.\arabic{equation}}

\renewcommand{\appendix}[1]{
    \setcounter{section}{0}
    \setcounter{equation}{0}
    \renewcommand{\thesection}{\Alph{section}}
		\renewcommand{\theequation}{\Alph{section}-\arabic{equation}}
}
\newcommand\encadremath[1]{\vbox{\hrule\hbox{\vrule\kern8pt
\vbox{\kern8pt \hbox{$\displaystyle #1$}\kern8pt}
\kern8pt\vrule}\hrule}}
\def\enca#1{\vbox{\hrule\hbox{
\vrule\kern8pt\vbox{\kern8pt \hbox{$\displaystyle #1$}
\kern8pt} \kern8pt\vrule}\hrule}}

\newcommand\figureframex[3]{
\begin{figure}[bth]
\hrule\hbox{\vrule\kern8pt
\vbox{\kern8pt \vbox{
\begin{center}
{\mbox{\epsfxsize=#1.truecm\epsfbox{#2}}}
\end{center}
\caption{#3}
}\kern8pt}
\kern8pt\vrule}\hrule
\end{figure}
}
\newcommand\figureframey[3]{
\begin{figure}[bth]
\hrule\hbox{\vrule\kern8pt
\vbox{\kern8pt \vbox{
\begin{center}
{\mbox{\epsfysize=#1.truecm\epsfbox{#2}}}
\end{center}
\caption{#3}
}\kern8pt}
\kern8pt\vrule}\hrule
\end{figure}
}

\newcommand{\genus}{{\mathbf g \!\!\! /\;\;}}
\renewcommand{\thesection}{\arabic{section}}
\renewcommand{\theequation}{\arabic{section}-\arabic{equation}}
\makeatletter
\@addtoreset{equation}{section}
\makeatother
\newtheorem{theorem}{Theorem}[section]

\newtheorem{remark}{Remark}[section]
\newtheorem{proposition}{Proposition}[section]
\newtheorem{lemma}{Lemma}[section]
\newtheorem{corollary}{Corollary}[section]
\newtheorem{definition}{Definition}[section]
\def\br{\begin{remark}\rm\small}
\def\er{\end{remark}}
\def\bt{\begin{theorem}}
\def\et{\end{theorem}}
\def\bd{\begin{definition}}
\def\ed{\end{definition}}
\def\bp{\begin{proposition}}
\def\ep{\end{proposition}}
\def\bl{\begin{lemma}}
\def\el{\end{lemma}}
\def\bc{\begin{corollary}}
\def\ec{\end{corollary}}
\def\beaq{\begin{eqnarray}}
\def\eeaq{\end{eqnarray}}
\newcommand{\proof}[1]{{\noindent \bf proof:}\par
{#1} $\square$}

\newcommand{\eq}[1]{Eq.~(\ref{#1})}

\newcommand{\beq}{\begin{equation}}
\newcommand{\eeq}{\end{equation}}
\newcommand{\bea}{\begin{eqnarray}}
\newcommand{\eea}{\end{eqnarray}}

\newcommand{\vs}{\vspace{0.7cm}}
\renewcommand{\and}{{\qquad {\rm and} \qquad}}

\newcommand{\virg}{{\qquad , \qquad}}

 \newcommand{\Tr}{{\,\rm Tr}\:}
\newcommand{\tr}{{\,\rm tr}\:}

\newcommand{\Res}{\mathop{\,\rm Res\,}}

\newcommand{\td}[1]{{\tilde{#1}}}

\renewcommand{\l}{\lambda}
\newcommand{\om}{\omega}

\newcommand{\ee}[1]{{{\rm e}^{#1}}}

\renewcommand{\d}{{{\partial}}}

\newcommand{\C}{{\mathbf C}}

\newcommand{\Pint}{{\int\kern -1.em -\kern-.25em}}

\renewcommand{\Im}{{\mathrm{Im}}}

\renewcommand{\l}{\lambda}
\renewcommand{\L}{\Lambda}

\newcommand{\ovl}{\overline}
\newcommand{\bfx}{{\mathbf x}}

\newcommand{\acycle}{{\cal A}}
\newcommand{\bcycle}{{\cal B}}

\newcommand{\curve}{{\cal E}}
\newcommand{\Surf}{\Sigma}

\newcommand{\pbar}{\ovl{p}}
\newcommand{\qbar}{\ovl{q}}

\newcommand{\bfa}{{\mathbf{a}}}

\newcommand{\bfalpha}{{\mathbf{\alpha}}}

\newcommand{\primef}{{E}}

\newcommand{\Bergmann}{{\underline{B}}}

\begin{document}
\sloppy


\pagestyle{empty}
\hfill SPhT-T07/021
\addtolength{\baselineskip}{0.20\baselineskip}
\begin{center}
\vspace{26pt}
{\large \bf {Invariants of algebraic curves and topological expansion}}
\newline
\vspace{26pt}

{\sl B.\ Eynard}\hspace*{0.05cm}\footnote{ E-mail: eynard@spht.saclay.cea.fr }, {\sl N.\ Orantin}\hspace*{0.05cm}\footnote{ E-mail: orantin@spht.saclay.cea.fr }\\
\vspace{6pt}
Service de Physique Th\'{e}orique de Saclay,\\
F-91191 Gif-sur-Yvette Cedex, France.\\
\end{center}

\vspace{20pt}
\begin{center}
{\bf Abstract}:
\end{center}

%


For any arbitrary algebraic curve, we define an infinite sequence of invariants.
We study their properties, in particular their variation under a variation of the curve, and their modular properties. We also study their limits when the curve becomes singular.
In addition we find that they can be used to define a formal series, which satisfies formally an Hirota equation, and we thus obtain a new way of constructing a tau function attached to an algebraic curve.

These invariants are constructed in order to coincide with the topological expansion of a matrix formal integral, when the algebraic curve is chosen as the large $N$ limit of the matrix model's spectral curve.
Surprisingly, we find that the same invariants also give the topological expansion of other models, in particular the matrix model with an external field, and the so-called double scaling limit of matrix models, i.e. the $(p,q)$ minimal models of conformal field theory.

As an example to illustrate the efficiency of our method, we apply it to the Kontsevitch integral, and we give a new and extremely easy proof that Kontsevitch integral depends only on odd times, and that it is a KdV $\tau$-function.




\vspace{26pt}
\pagestyle{plain}
\setcounter{page}{1}


\newpage

\tableofcontents


\section{Introduction}

Computing the topological expansion of various matrix integrals has been an interesting problem for more than 30 years. The reason for it, is that formal matrix integrals (cf \cite{eynform} for a definition of formal integrals), are known to be combinatorics generating functionals.
Some formal matrix integrals enumerate maps, or colored maps of given topology \cite{BIPZ, ZJDFG, PDFgraph}, 
some count intersection numbers (the Kontsevitch integral and its generalizations \cite{kontsevitch}), 
and physicists have also tried to reach the limit of continuous maps through critical limits, and thus tried to recover Liouville's field theory \cite{KPZ, BookPDF}.

\medskip

Many methods have been invented to compute those formal matrix integrals, and the most successful
is undoubtedly the "loop equations" method \cite{Kazakovloop, Virasoro}, which is in fact nothing but integration by parts, or Tutte's equations \cite{tutte,tutte2}, or Schwinger--Dyson equations, or Ward identities, or Virasoro constraints, or W-algebra \cite{ZJDFG}.
Until recently, those loop equations were solved only for the first few orders (mostly planar or torus), and case by case (for each matrix model).
One of the most remarkable methods was obtained in \cite{ACKM}.
Let us also mention that other methods were invented using orthogonal polynomials \cite{Mehta} (only in the case were the formal integral comes from an actual convergent integral),
or topological string theory methods \cite{MMtopo,DV,BCOV} using the so-called holomorphic anomaly equations.

\medskip

In 2004, a new method for computing the large $N$ expansion of matrix integrals was introduced in \cite{eynloop1mat}, and further developped in \cite{eyno, ec1loopF}.
The starting point of that method was not new, it was the same as in \cite{ACKM}, it consists in solving the loop equations recursively in powers of the expansion parameter $1/N^2$, where $N$ is the size of the matrix.
To leading order, loop equations become algebraic equations, and give rise to an algebraic curve $\curve(x,y)=0$ where $\curve$ is some polynomial in 2 variables, which we call the ``classical  spectral curve''.

The new feature which was introduced in \cite{eynloop1mat} was to use contour integrals and functions on the curve rather than on the $x$-plane as in \cite{ACKM}.
When written on the curve, the loop equations, together with the Cauchy residue formula and the Riemann bilinear identity, simplify enormously, and take a very universal structure which can be written {\bf entirely in terms of geometric properties of the curve}.
In other words, the solution of loop equations of many different matrix models, depends only on the properties of the spectral curve, and not on the matrix model which gives that curve.
In particular, they can be written for any arbitrary algebraic curve, even for curves which don't come from matrix models.
It is thus tempting to define "free energies" for any algebraic curve.
This is what we do in this article.

\bigskip

Therefore, in this article, for any arbitrary algebraic curve $\curve(x,y)=0$, we define an infinite sequence of complex numbers $F^{(g)}(\curve)$, computed as residues of meromorphic forms on the curve.
Out of these $F^{(g)}(\curve)$'s, we build a formal power series:
\beq
\ln{Z_N(\curve)}=-\sum_{g=0}^\infty N^{2-2g} F^{(g)}(\curve)
\eeq
and we study its properties.

\medskip

We compute the variations of $F^{(g)}$ under variations of the curve (variations of its complex structure, its moduli, and modular transformations).
We show that the $F^{(g)}$'s are invariant under some transformations of the curve,
namely under transformations of the curve which preserve the symplectic form up to a sign $\pm dx\wedge dy$.

\smallskip

We also show that $Z_N(\curve)$ satisfies bilinear Hirota equations, and thus $Z_N(\curve)$ is a formal $\tau$-function, and we construct the associated formal Baker-Hakiezer function \cite{BBT}.

We thus have a notion of a $\tau$ function associated to an algebraic curve.
Such notion has already been encountered in the litterature \cite{BBT}, and it is not clear whether our definition coincides with other existing definitions.
What can be understood so far, is that we are defining a sort of quantum deformation of a classical $\tau$-function whose spectral curve is $\curve$.
The classical $\tau$ function being only the dispersionless limit $ \ln{Z_{\infty}(\curve)} = - F^{(0)}(\curve)$, while our $Z_N(\curve)$ concerns the full system.

\medskip

Almost by definition, if $\curve$ is the algebraic curve coming from the large $N$ limit of the loop equations of a matrix model, then $Z_N(\curve)$ is the matrix integral.

What is more interesting is to see what is $Z_N(\curve)$ for curves not coming from the large $N$ limit of the loop equations of a matrix model.

We study in details a few examples.

\smallskip

- The double scaling limit of a matrix model. It has been well known since \cite{KazakovRMTcrit, ZJDFG}, that if we fine tune the parameters of a matrix model so that the algebraic curve $\curve$ develops a singularity,
the free energies become singular and the most singular part of the free energies form the KP-hierarchy $\tau$-function (KdV hierarchy for the 1-matrix model).
We show, by looking at the double scaling limit of matrix models, that the KP $\tau$-function (resp. KdV $\tau$-function), coincides with our definition for the classical limit of the $(p,q)$ systems (resp. $(p,2)$).

- It has been well known since the works of Kontsevitch \cite{kontsevitch}, that the KdV $\tau$-function can be represented by another matrix integral called Kontsevitch integral.
Kontsevitch introduced that integral as a counting function for intersection numbers, and proved that it is a KdV $\tau$-function.
One of the key features is that it depends on the eigenvalues of a diagonal matrix $\L$, only through the quantities $t_k=\tr \L^{-k}$ for odd $k$ (cf \cite{IZK,ZJDFG}).
Another important known property is that, if $t_k=0$ for $k> p$, it coincides with the $(p,2)$ $\tau$-function found from the double scaling limit of the 1-matrix model, i.e. the $(p,2)$ conformal minimal model.

Here, we prove that the Kontsevitch matrix integral coincides with our $Z_N(\curve)$ when $\curve$ is the large $N$ limit of the Schwinger--Dyson equation of the Kontsevitch integral.
The remarkable fact, is that for our $Z_N(\curve)$, the above properties (i.e. the fact that it depends only on odd $t_k$'s and the fact that it gives the $(p,2)$ $\tau$-function if $t_k=0$ for $k>p$) are trivial.
We thus provide a new proof of those properties, and maybe a new interpretation.

\section{Main results of this article}

In this section we just sketch briefly the contents of the main body of the article.

\subsection{Definitions}

Given a polynomial of two variables $\curve(x,y)$, we construct an infinite sequence of 
multilinear meromorphic forms over the curve of equation $\curve(x,y)=0$, which we call:
\beq
W_k^{(g)}(p_1,p_2,\dots,p_k) \virg k, g \in {\mathbf N}.
\eeq
In particular $W_0^{(g)}=-F^{(g)}$ are complex numbers $F^{(g)}(\curve)$.

The $F^{(g)}$'s and the $W_k^{(g)}$'s are defined in terms of residues near the branchpoints of the curve only, i.e. they depend only on the {\bf local behavior of the curve near its branch points}.

Then we show some properties:
\begin{itemize}
\item the $W_k^{(g)}$'s are {\bf symmetric} in their $k$ variables;
\item there is a "loop insertion operator" which increases $k\to k+1$:
\beq
D_{B(p_{k+1},.)} W_k^{(g)}(p_1,p_2,\dots,p_k) = W_{k+1}^{(g)}(p_1,p_2,\dots,p_k,p_{k+1});
\eeq
\item there is an inverse operator which contracts $k\to k-1$:
\beq
\Res_{p_k\to {\rm branch\,points}} \Phi(p_k)\,W_k^{(g)}(p_1,p_2,\dots,p_k) = (2g+k-3)\, W_{k-1}^{(g)}(p_1,p_2,\dots,p_{k-1}).
\eeq
\end{itemize}

The $F^{(g)}$'s and the $W_k^{(g)}$'s are defined in a way which mimics the solution of matrix models loop equations, and almost by definition, they coincide with matrix model's free energy and correlation functions when the polynomial $\curve$ is chosen as the classical large $N$ limit of the matrix model's spectral curve:
\beq
\ln{\left( \int dM \exp{-N\Tr V(M) }\right)} = - \sum_{g=0}^\infty N^{2-2g}\,F^{(g)}(\curve_{\rm 1MM})
\eeq
and
\beq
\left< \Tr {dx_1\over x_1-M}\,\Tr {dx_2\over x_2-M}\,\dots\,\Tr {dx_k\over x_k-M}\right> =  \sum_{g=0}^\infty N^{2-2g-k}\,W^{(g)}_k(p(x_1),p(x_2),\dots,p(x_k))
\eeq

The same construction works also for the 2-matrix model and the matrix model in an external field:
\bea
F^{(g)}_{\rm 1MM} =  F^{(g)}(\curve_{\rm 1MM}) \cr
F^{(g)}_{\rm 2MM} =  F^{(g)}(\curve_{\rm 2MM}) \cr
F^{(g)}_{\rm ext.field} =  F^{(g)}(\curve_{\rm ext.field}) .
\eea
in particular it works for the Kontsevitch integral
\beq
F^{(g)}_{\rm Kontsevitch} =  F^{(g)}(\curve_{\rm Kontsevitch}) 
\eeq
where the LHS is the topological expansion of the corresponding matrix integral, and the RHS is the functional defined in this article, applied to the curve $\curve(x,y)=0$ coming from the large $N$ limit of the Schwinger--Dyson equations of the corresponding matrix model.

Let us emphasize that not every curve $\curve$ is the large $N$ limit of a matrix model's spectral curve, and thus our functional $F^{(g)}(\curve)$ is defined beyond matrix models, and is really an algebro-geometric object.
It has many remarkable properties, and we list below some of the most important ones:

\subsection{Remarkable properties}

{\bf Theorem \ref{thdiagrepr} Diagrammatic representation:}
\beq
W_{k+1}^{(g)}(p,p_1,\dots,p_k) = \sum_{G\in {\cal G}_{k+1}^{(g)}(p,p_1,\dots,p_k)}\,\, w(G) = w\left(\sum_{G\in {\cal G}_{k+1}^{(g)}(p,p_1,\dots,p_k)}\,\, G\right)
\eeq
where  ${\cal G}_{k+1}^{(g)}(p,p_1,\dots,p_k)$ is a set of trivalent graphs (built on trees), and $w$ is a Feynman-like weight function associating values to edges and integrals (residues in fact) to vertices of the graph.

This theorem is important because it makes all formulae particularly easy to remember, and many theorems below can be proved in a diagrammatic way.

\medskip
{\bf Theorem \ref{thsinglimit} Singular limits:} if the curve becomes singular, the functional $F^{(g)}$ commutes with the singular limit, i.e.:
\beq
{\rm lim}\,\, F^{(g)}(\curve) = F^{(g)}({\rm lim}\,\, \curve).
\eeq

\medskip
\noindent {\bf Theorem \ref{thHirota} Integrability:} the formal series:
\beq
\ln{Z_N(\curve)} = - \sum_{g=0}^\infty N^{2-2g}\,F^{(g)}(\curve)
\eeq
obeys Hirota's bilinear equations, and thus is a $\tau$ function.

\medskip
\noindent {\bf Theorem \ref{homogeneity} Homogeneity :} $F^{(g)}(\curve)$ is homogeneous of degree $2-2g$ in the moduli of the curve:
\beq
(2-2g)F^{(g)} = \sum_k t_k {\partial F^{(g)}\over \partial t_k}.
\eeq

\medskip
\noindent {\bf Theorem \ref{variat} Deformations :} if the curve $\curve$ is deformed into $\curve+\delta\curve$,
the differential $ydx$ is deformed into $ydx\to ydx+\delta(ydx)$ where $\delta(ydx)$ is a meromorphic one-form which we denote $\delta(ydx)= - \Omega$, and which can be written as:
$\Omega  = \int_{\partial\Omega} W_2^{(0)} \L$ for some appropriate contour $\partial\Omega$ and some appropriate function $\L$. Then we have:
\beq
\delta W_{k}^{(g)}(p_1,p_2,\dots,p_k) = \int_{\partial\Omega} \L(p_{k+1})\,W_{k+1}^{(g)}(p_1,p_2,\dots,p_k,p_{k+1}) .
\eeq

\medskip
\noindent {\bf Theorem \ref{symplinv} Symplectic invariance :} $F^{(g)}(\curve)$ is unchanged under the following changes of curve $\curve(x,y)$:
\beq\label{symplecticinvintro}
\begin{array}{l}
y\to y+ R(x) \virg \hbox{$R(x)$=rational fraction of $x$,} \cr
y\to c y \,\, , \,\,\, x\to c^{-1}x\,  \virg \hbox{$c$=complex number,} \cr
y\to - y \,\, , \,\,\, x\to x\, , \cr
y\to x \,\, , \,\,\, x\to y\, .
\end{array}
\eeq
all those transformations conserve the symplectic form $dx\wedge dy$ up to the sign.

\medskip
\noindent {\bf Theorem \ref{thmodular} Modular transformations :} 
the modular dependence of $F^{(g)}(\curve)$ is only in the Bergmann kernel (defined in section \ref{secdefBergmann}), and thus the modular transformations of $F^{(g)}(\curve)$ are derived from those of the Bergmann kernel.
Under a modular transformation, the Bergmann kernel is changed into $B(p,q) \to B(p,q) + 2i\pi\,du(p) \kappa du(q)$, and thus we introduce a new kernel for any arbitrary symmetric matrix $\kappa$:
\beq
B_\kappa(p,q) \to B(p,q) + 2i\pi\,du(p) \kappa du(q)
\eeq
We thus define some $F^{(g)}_\kappa(\curve)$, and we compute:
\beq
{\partial F^{(g)}\over \partial \kappa}
\eeq
We also remark that when $\kappa = (\overline\tau-\tau)^{-1}$, $F^{(g)}_\kappa(\curve)$ is modular invariant.

\subsection{Some applications, Kontsevitch's integral}

{\bf $(p,q)$ minimal models, KP and KdV Hierarchies:}
it is well known \cite{ZJDFG, DKK} that some rational singular limits of matrix models correspond to $(p,q)$ minimal models, and theorem \ref{thsinglimit} implies that:
\beq
F^{(g)}_{(p,q)} =  F^{(g)}(\curve_{(p,q)}) 
\eeq
and it is well known that $(p,q)$ minimal models are some reductions of KdV hierarchy for $q=2$ and KP hierarchy for general $(p,q)$.

Notice that the $x\leftrightarrow y$ symmetry of theorem \ref{symplinv} (i.e. \eq{symplecticinvintro}) implies the famous $(p,q)\leftrightarrow (q,p)$ duality \cite{ZJDFG, BookPDF, KharMar}.

\bigskip
{\bf Kontsevitch integral's properties:}
Kontsevitch's integral is defined as:
\beq
Z_{\rm Kontsevitch}(\L) = \int dM\, \ee{-N\Tr {M^3\over 3} - M\L^2}
\virg
\ln{Z_{\rm Kontsevitch}} = -\sum_{g=0}^\infty N^{2-2g}\, F^{(g)}_{\rm Kontsevitch}
\eeq
and we define the Kontsevitch's times:
\beq
t_k={1\over N}\Tr \L^{-k}.
\eeq
It is straightforward to write the Schwinger-Dyson equations and find the classical spectral curve:
\beq
\curve_{\rm Kontsevitch} = \left\{
\begin{array}{l}
x(z) = z + {1\over 2N}\Tr {1\over \L} {1\over z-\L} \cr
y(z) = z^2 + t_1
\end{array}\right. .
\eeq
According to theorem \ref{thMext}, we have:
\beq
F^{(g)}_{\rm Kontsevitch} = F^{(g)}(\curve_{\rm Kontsevitch}).
\eeq
Using the $x\leftrightarrow y$ invariance of eq.\ref{symplecticinvintro}, we see that the only branch point in $y$ is located at $z=0$, and since the $F^{(g)}$'s depend only on the local behavior near the branchpoint, we may perform a Taylor expansion of $x(z)$ near $z=0$:
\beq
\curve_{\rm Kontsevitch}(t_1,t_2,\dots) = \left\{
\begin{array}{l}
x(z) = z - {1\over 2} \sum_{k=0}^\infty t_{k+2} z^k \cr
y(z) = z^2 + t_1
\end{array}\right. .
\eeq
From the symplectic invariance theorem \ref{symplinv}, (i.e. eq.\ref{symplecticinvintro}), we may add to $x$ any rational function of $y$ i.e. of $z^2$, thus we may substract to $x$ its even part, and thus the following curves are related by symplectic invariance:
\beq
\curve_{\rm Kontsevitch}(t_1,t_2,t_3,\dots) \sim \curve_{\rm Kontsevitch}(t_1,0,t_3,0,t_5,\dots) .
\eeq
We thus have a very easy proof that {\bf $F^{(g)}_{\rm Kontsevitch}$ depends only on odd times}.

Moreover, if $t_k=0$ for $k>p+2$, we have:
\beq
\curve_{\rm Kontsevitch}(t_1,t_2,\dots,t_{p+2},0,\dots) = \left\{
\begin{array}{l}
x(z) = z - {1\over 2} \sum_{k=0}^{p} t_{k+2} z^k \cr
y(z) = z^2 + t_1
\end{array}\right.
\eeq
which is exactly the curve of the $(p,2)$ minimal model, i.e. it satisfies KdV hierarchy.
We thus have a very easy proof that {\bf $Z_{\rm Kontsevitch}$ is a KdV tau-function}.

\medskip

Those are old and classical results about the Kontsevitch integral, and we just propose a new proof, in order to illustrate the power of the tools we introduce.

\section{Algebraic curves, reminder and notations}

We begin by recalling some elements of algebraic geometry, which are used to fix the notations.
We refer the reader to
\cite{Farkas} or \cite{Fay} for further details about algebro-geometric concepts.

\bigskip

\noindent
\begin{tabular}{|l@{ $\,\,\longrightarrow\,\,$ }p{260pt}|}
\hline
&\underline{\bf Summary of notations} \\
\hline
$\curve(x,y)=0$ & classical spectral curve. \\
$d_1+1= \deg_{x} \curve$  & $x$-degree of the polynomial $\curve$. \\
$d_2+1 = \deg_{y} \curve$  & $y$-degree of the polynomial $\curve$ (number of sheets). \\
$\bfa=\{a_i\} $ & set of branch points $dx(a_i)=0$. \\
$\bfalpha=\{\alpha_i\} $ & poles of $ydx$. \\
$\genus$ & genus  of the curve. \\
$\underline\acycle_i \cap \underline\bcycle_j =\delta_{ij}$ & cannonical basis of non-contractible cycles. \\
$du_i$  & cannonical holomorphic forms $\oint_{\underline\acycle_j} du_i = \delta_{ij}$. \\
$\tau_{ij} = \oint_{\underline\bcycle_j} du_i$ & Riemann's matrix of periods. \\
$u_i(p) = \int_{p_0}^p du_i$ & Abel map. \\
$\acycle=\underline\acycle - \kappa (\underline\bcycle-\tau\underline\acycle)$ & $\kappa$-modified $\acycle$-cycles \\
$\bcycle=\underline\bcycle-\tau\underline\acycle$ & $\kappa$-modified $\bcycle$-cycles, $\acycle_i \cap \bcycle_j =\delta_{ij}$ \\
$dS_{q_1,q_2}(p)$ & 3rd kind differential with simple poles $q_1$ and $q_2$, such that $\Res_{q_1} dS_{q_1,q_2} = 1=- \Res_{q_2} dS_{q_1,q_2}$ and $\oint_{\acycle_i} dS_{q_1,q_2} =0$. \\
$B(p,q)$ & Bergmann kernel, i.e. 2nd kind differential with double pole at $p=q$, no residue and vanishing $\acycle$-cycle integrals. \\
$z={\vec{n}+\tau\vec{m}\over 2}$ & regular odd characterisic, i.e. $\sum_{i=1}^g n_i m_i =$odd. \\
$dh_z = \sum_i \left. {\partial \theta_z(\vec{v})\over \partial v_i}\right|_{v=0}\,du_i$ & Holomorphic form with only double zeroes. \\
$\underline\primef(p,q)={\theta_z(u(p)-u(q),\tau)\over \sqrt{dh_z(p)dh_z(q)}}$ & Prime form independent of $z$, with a simple zero at $p=q$.\\
$\Phi(p)=\int_o^p y dx$ & some antiderivative of $ydx$ defined on the universal covering.\\
$\pbar$, $x(\pbar)=x(p)$ & if $p$ is near a branchpoints $a$, then $\pbar\neq p$ is the unique other point near $a$ such that $x(\pbar)=x(p)$.\\
$p^i(p)$, $x(p^i)=x(p)$ &  The $p^i$'s, $i=0,\dots,d_2$,  are the pre-images of $x(p)$ on the curve. By convention $p^0(p)=p$.\\
$D_\Omega = \delta_\Omega + \tr (\kappa \,\delta_\Omega\tau \,\kappa {\partial \over \partial \kappa})$ & Covariant variation wrt $\Omega = \delta(y dx)$.\\
\hline
\end{tabular}

\bigskip

Consider an (embedded) algebraic curve given by its equation:
\beq
\curve(x,y)=0
\eeq
where $\curve$ is an almost arbitrary polynomial of two variables. This is equivalent to considering a compact
Riemann surface $\overline{\Sigma}$ and 2 meromorphic functions $x$ and $y$, such that 
\beq
\forall p \in \overline{\Sigma} \virg \; \curve(x(p),y(p))=0.
\eeq

We only require that $\curve(x,y)$ is  not factorizable, and that all branchpoints (zeroes of $dx$) are simple, i.e. near a branchpoint $a_i$,
$y$ behaves like a square root $\sqrt{x-x(a_i)}$.

\subsection{Some properties of algebraic curves} \label{sectgeomalg}

\subsubsection{Sheets}

For each complex $x$, there exist $d_2+1=\deg_y\curve$ solutions for $y$ of $\curve(x,y)=0$.
This means that there are exactly $d_2+1$ points on the Riemann surface $\overline{\Sigma}$ for which $x(p)=x$:
$\overline{\Sigma}$ has a sheet structure with $d_2+1$ $x$-sheets.
We call them:
\beq
x(p)=x \quad \leftrightarrow\quad p=p^i(x)\,\, ,\,\,\, i=0,\dots, d_2.
\eeq

\begin{figure}
\beq
\begin{array}{r}
{\epsfxsize 7cm\epsffile{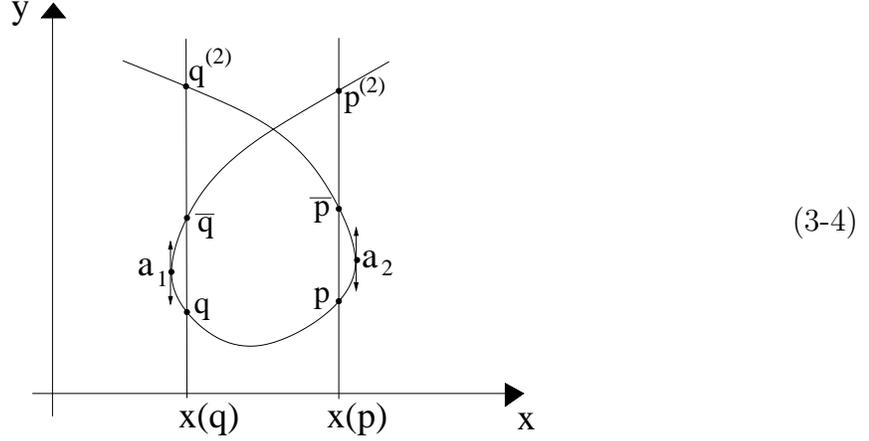}}
\end{array} \eeq
\caption{Example of an algebraic curve with two $x$-branch points $a_1$ and $a_2$ and a three sheeted structure
($x$ has three preimages). One can see that the map
$p \to \overline{p}$ is not globally defined, for instance when $q \to p$, we have $\qbar \to p^{(2)}$. The notion of conjugated point depends on the branch point.}
\label{branch}
\end{figure}

\subsubsection{Branch points and conjugated points}

Let $a_i$, $i=1,\dots, n$, $\bfa=\{a_1,\dots,a_n\}$ be the set of branch-points, solutions of $dx=0$:
\beq
\forall a\in \bfa \virg dx(a)=0.
\eeq
Since we assume that the branch-points are simple zeros of $dx$, we have the following property:
if $p$ is in the vicinity of a branch-point $a_i$, there is a unique point $\pbar\neq p$, such that $x(\pbar)=x(p)$, which is also in the vicinity of $a_i$.
$\pbar$ depends on $i$, and in general, $\pbar$ is not globaly defined (see fig. \ref{branch} for an example).

Notice that $\pbar$ is one of the $p^k$ defined in the previous section.


\subsubsection{Genus, cycles, Abel map}

\begin{figure}
\beq
\begin{array}{r}
{\epsfxsize 5cm\epsffile{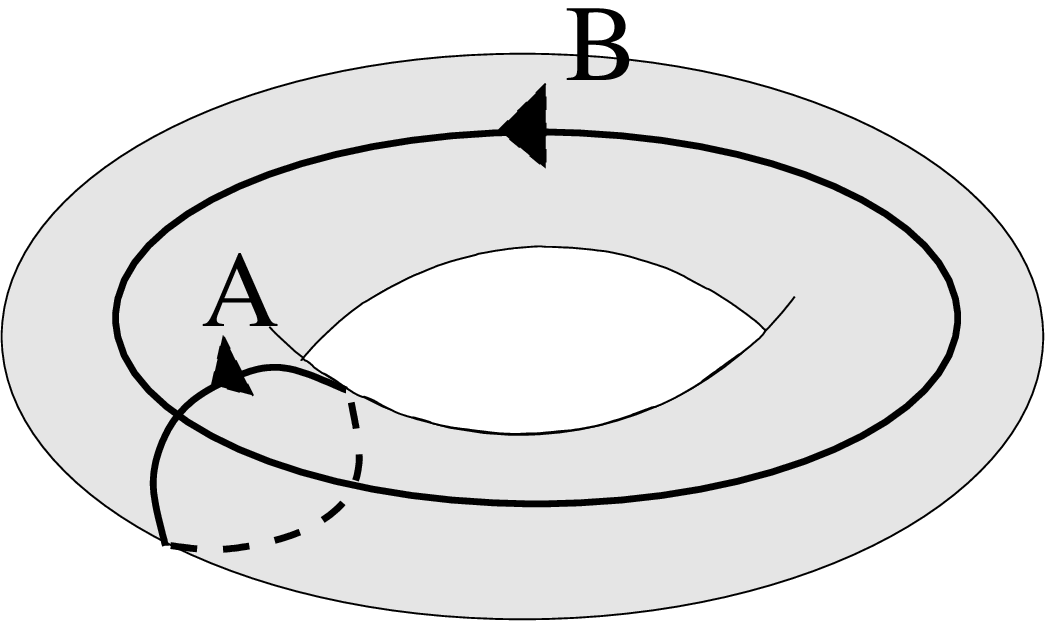}}
\end{array}
\Leftrightarrow
\begin{array}{r}
{\epsfxsize 4cm\epsffile{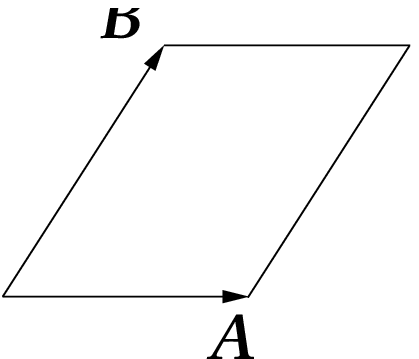}}
\end{array} \eeq
\caption{Example of canonical cycles and the corresponding fundamental domain in the case of the torus (genus $\genus=1$).}
\label{tor1}
\end{figure}

If the curve has genus $\genus$, there are $2\genus$ homologicaly independent non-trivial cycles, and we may choose a (not unique) cannonical basis:
\beq
\underline\acycle_i\cap \underline\bcycle_j=\delta_{ij}
\virg
\underline\acycle_i\cap \underline\acycle_j=0
\virg
\underline\bcycle_i\cap \underline\bcycle_j=0.
\eeq
The simply connected domain obtained by removing all $\underline\acycle$ and $\underline\bcycle$-cycles from the curve is called the ``{\bf fundamental domain}''
(see fig.(\ref{tor1}) for the example of the torus).

On a genus $\genus$ curve, there are $\genus$ linearly independent holomorphic forms $du_1,\dots, du_\genus$, which we choose normalized on the $\underline\acycle$-cycles:
\beq
\oint_{\underline\acycle_j} du_i = \delta_{ij}.
\eeq

The {\bf Riemann matrix of period} is defined by the $\underline\bcycle$-cycles
\beq
\tau_{ij} = \oint_{\underline\bcycle_j} du_i.
\eeq
They have the property that
\beq
\tau_{ij}=\tau_{ji} \virg \Im\,\tau>0.
\eeq

Given a  base point $p_0$ on the curve (we assume it is not on any $\underline\acycle$ or $\underline\bcycle$-cycle), we define the {\bf Abel map}
\beq
u_i(p) = \int_{p_0}^{p} du_i
\eeq
where the integration path is in the fundamental domain.
The $\genus$-dimensional vector $u(p)=(u_1(p),\dots,u_\genus(p))$ maps the curve into its Jacobian.

\subsubsection{Theta-functions and prime forms}

We say that $z\in \C^\genus$ is a {\bf characteristic} if there exist two vectors with integer coefficients $\vec{a}\in Z^\genus$ and $\vec{b}\in Z^\genus$ such that:
\beq
z={\vec{a}+\tau.\vec{b}\over 2}.
\eeq
$z$ is called an odd characteristic if
\beq
\sum_{i=1}^\genus a_i b_i = {\rm odd}.
\eeq
Given a characteristic $z={\vec{a}+\tau \vec{b}\over 2}$, and given a symmetric matrix $\tau_{ij}=\tau_{ji}$ such that $\Im\tau$ is positive definite, and a vector $v\in \C^\genus$, we define the $\theta_z$ function:
\beq
\theta_z(v,\tau) = \sum_{\vec{n}\in Z^\genus} \ee{i\pi (\vec{n}-\vec{b}/2)^t \tau (\vec{n}-\vec{b}/2)}\,\ee{2i\pi (v+\vec{a}/2)^t.(\vec{n}+\vec{b}/2)}.
\eeq
If $z$ is an odd characteristic, $\theta_z$ is an odd function of $v$, and in particular $\theta_z(0,\tau)=0$
and we define the following holomorphic form:
\beq
dh_z(p) = \sum_{i=1}^\genus du_i(p)\,\,. \,\left.{\partial \theta_z(v)\over \partial v_i}\right|_{v=0}.
\eeq
All its $\genus-1$ zeroes are double zeroes, so that it makes sense to consider its square root defined on the fundamental domain.
The {\bf prime form} is:
\beq
\underline{\primef}(p,q) = {\theta_z(u(p)-u(q))\over \sqrt{dh_z(p)\,dh_z(q)}}.
\eeq
It is independent of $z$, and it vanishes only if $p=q$ (and with a simple zero), and has no pole.

\subsubsection{Bergmann kernel}\label{secdefBergmann}

We define a bilinear meromorphic form called the ``{\bf Bergmann kernel}'' \cite{BergSchif, Fay2}:
\beq
\Bergmann(p,q) = \hbox{Bergmann kernel}
\eeq
as the unique $1$-form in $p$, which has a double pole with no residue at $p=q$ and no other pole, and which is normalized such that
\beq
\Bergmann(p,q) \sim_{p\to q} {dz(p)dz(q)\over (z(p)-z(q))^2} + {\rm finite}
\virg
\oint_{\underline\acycle_I} \Bergmann =0
\eeq
where $z$ is any local coordinate on the curve in the vicinity of $q$.
The Bergmann kernel depends only on the complex structure of the curve, and not on the details of $\curve$.
For instance if $\curve$ has genus zero, $\Bergmann$ is the Bergmann kernel of the projective complex plane (the Riemann sphere),
and if the curve has genus $1$, $\Bergmann$ is related to the Weierstrass function.


\medskip

{\bf Properties:}

\beq
\Bergmann(p,q)=\Bergmann(q,p)
\virg
\oint_{q\in\underline\bcycle_i} \Bergmann(p,q) = 2i\pi\, du_i(p).
\eeq
For any odd characteristic $z$, we have:
\beq
\Bergmann(p,q) = d_p d_q \ln{(\theta_z(u(p)-u(q)))}
\eeq
If $f(p)$ is any meromorphic function, its differential is given by
\beq\label{Bergmanndiff}
df(p) = \Res_{q\to p} \Bergmann(p,q) f(q).
\eeq

\subsubsection{3rd type differentials}

Given two points $q_1$ and $q_2$ on the curve , we define the 1-form $\underline{dS}_{q_1,q_2}$ by:
\beq
\underline{dS}_{q_1,q_2}(p) = \int_{q_2}^{q_1} \Bergmann(p,q)
\eeq
where the integration path is in the fundamental domain.

$\underline{dS}_{q_1,q_2}$ is the unique meromorphic form with only simple poles at $q_1$ and $q_2$, such that:
\beq
\Res_{q_1} \underline{dS}_{q_1,q_2} = 1 = - \Res_{q_2} \underline{dS}_{q_1,q_2}
\virg
\oint_{\underline\acycle_i} \underline{dS}_{q_1,q_2} = 0
.
\eeq

\subsubsection{Modified set of cycles:}\label{secgeneralizedcycles}

In order to easily deal with modular properties of the objects we are going to introduce, it is convenient to  define some modified cycles and kernels with an arbitrary symmetric matrix $\kappa$. When $\kappa=0$, all those quantities reduce to the unmodified ones. The modular transformations of the modified objects, merely amount to a change of $\kappa$.

\medskip

We thus choose an arbitrary {\bf $\genus\times \genus$ symmetric matrix $\kappa$} with complex coefficients, and we define another set of cycles:
\bea
\acycle_i &=& \underline\acycle_i - \sum_j \kappa_{ij} (\underline\bcycle_j - \sum_l \tau_{jl} \underline\acycle_l ) \cr
\bcycle_i &=& \underline\bcycle_i - \sum_j \tau_{ij} \underline\acycle_j
\eea
They satisfy:
\beq
\acycle_i\cap \bcycle_j=\delta_{ij}
\virg
\acycle_i\cap \acycle_j=0
\virg
\bcycle_i\cap \bcycle_j=0
\eeq
and we straightforwardly have
\beq
\oint_{\acycle_i} du_j = \delta_{ij}
\virg
\oint_{\bcycle_i} du_j = 0.
\eeq

\subsubsection{Modified Bergmann kernel}\label{secdefgeneBergmann}

We also define the modified Bergmann kernel, normalized on $\acycle$ instead of $\underline\acycle$:
\beq
B(p,q) = \Bergmann(p,q) + 2i\pi\, \sum_{i,j}\, du_i(p)\, \kappa_{ij}\, du_j(q).
\eeq
It is such that
\beq
B(p,q)=B(q,p)
\virg
\oint_{\acycle_I} B =0
\virg
\oint_{q\in\bcycle_i} B(p,q) = 2i\pi\, du_i(p)
\eeq
and if $f(p)$ is any meromorphic function, its differential is given by:
\beq\label{geneBergmanndiff}
df(p) = \Res_{q\to p} B(p,q) f(q)
.
\eeq

$\bullet$ For $\kappa=0$ we have $\Bergmann=B$.

$\bullet$ For $\kappa=(\overline\tau-\tau)^{-1}$, $B$ is the Schiffer kernel \cite{BergSchif, Fay2},
and it is modular invariant.

\subsubsection{Modified prime form}

Similarly we define a modified prime form:
\beq
\primef(p,q) 
=
\underline{\primef}(p,q)  \, \ee{2i\pi u^t(p)\kappa u(q)}
\eeq
It vanishes only if $p=q$ (with a simple zero), and has no pole.

\subsubsection{Modified 3rd type differentials}

In the same fashion, we define the modified 3rd type differentials ${dS}_{q_1,q_2}$ by:
\beq
{dS}_{q_1,q_2}(p) = \int_{q_2}^{q_1} B(p,q)
\eeq
where the integration path is in the fundamental domain.

$dS_{q_1,q_2}$ is the unique meromorphic form with only simple poles at $q_1$ and $q_2$, such that:
\beq
\Res_{q_1} dS_{q_1,q_2} = 1 = - \Res_{q_2} dS_{q_1,q_2}
\virg
\oint_{\acycle_i} dS_{q_1,q_2} = 0
.
\eeq

\medskip

{\bf Properties:}

\beq
dS_{q_1,q_2} = - dS_{q_2,q_1}
\eeq

\beq
dS_{q_1,q_2}(p) = d_p \ln{\left(\theta_z(u(p)-u(q_1))\over \theta_z(u(p)-u(q_2))\right)} + 2i\pi\, \sum_{i,j}\, du_i(p) \kappa_{ij} (u_j(q_1)-u_j(q_2))
\eeq

\beq
\oint_{\bcycle_i} dS_{q_1,q_2} = 2i\pi (u_i(q_1)-u_i(q_2))
\eeq

\beq
d_{q_1} \left(dS_{q_1,q_2}(p)\right) =  B(q_1,p)
\eeq

\beq
\int_{p_1}^{p_2} dS_{q_1,q_2} = \int_{q_1}^{q_2} dS_{p_1,p_2}
\eeq

\medskip

{\bf Cauchy residue formula:} for any meromorphic function $f(p)$ we have:
\beq
f(p) = - \Res_{q_1\to p} dS_{q_1,q_2}(p) f(q_1).
\eeq

\subsubsection{Bergmann tau function}

The {\bf Bergmann $\tau$-function} $\tau_{Bx}$ was introduced and studied in \cite{KoKo,KoKo2,EKK}, it is such that
\beq\label{deftauBx}
{\d \ln{(\tau_{Bx})}\over \d x(a_i)} =  \Res_{p\to a_i} {B(p,\pbar)\over dx(p)}
\eeq
It is well defined because the Rauch variational formula \cite{Rauch} implies that the RHS is a closed form.
Notice that $\tau_{Bx}$ is  defined only up to a multiplicative constant which will play no role in all the sequel.

\subsection{Examples: genus 0 and 1}

\noindent $\bullet$ {\bf Genus 0}

If the curve $\curve$ has a genus $\genus = 0$, it is conformaly equivalent to the Riemann sphere, i.e. the complex plane with a point at $\infty$, and there exists a rationnal parametrization of the curve. 
It means that there exists two rational functions $X(p)$ and $Y(p)$ such that:
\beq
\curve(x,y) = 0 \quad \leftrightarrow \quad \exists p\in {\bf C}\, , \,\, x=X(p)\, , \, y=Y(p)
\eeq

In this case, the Bergmann kernel is the Bergmann kernel of the Riemann sphere:
\beq
B(p,q)=\Bergmann(p,q) = {dp dq \over (p-q)^2} = d_p\,d_q\,\ln{(p-q)}.
\eeq
The prime form is:
\beq
\primef(p,q) =\underline\primef(p,q)= {p-q\over \sqrt{dp\,dq}} .
\eeq

\noindent $\bullet$ {\bf Genus 1}

If the curve has genus $\genus=1$, then it can be parametrized on a rhombus corresonding to the fundamental domain of a torus (see fig. \ref{tor1}). 
It means that there exists two elliptical functions $X(p)$ and $Y(p)$ such that (see \cite{ellipticalf} for elliptical functions):
\bea
\curve(x,y) = 0 \quad \leftrightarrow \quad \exists p\in {\bf C}\, , \,\, x=X(p)\, , \, y=Y(p) \cr
X(p+1)=X(p+\tau)=X(p) \virg
Y(p+1)=Y(p+\tau)=Y(p) 
\eea

Then, the Bergmann kernel is the corresponding Weierstrass function \cite{ellipticalf}:
\beq
\Bergmann(p,q) = (\wp(p-q,\tau)+{\pi\over \Im\tau})\,\,dp dq .
\eeq
The prime form is:
\beq
\underline\primef(p,q) = {\theta_1(p-q,\tau)\over \theta'_1(0,\tau) \sqrt{dp\,dq}} .
\eeq

When $\kappa={-1\over 2i \Im\tau}$, the modified Bergmann kernel is the Schiffer kernel, and if $\genus=1$ it is the Weierstrass function:
\beq
B(p,q) = \wp(p-q,\tau)\,\,dp dq .
\eeq

\subsection{Riemann bilinear identity}

If $\om_1$ and $\om_2$ are two meromorphic forms on the curve.
Let $p_0$ be an arbitrary base point, we consider the function $\Phi_1$ defined on the fundamental domain by
\beq
\Phi_1(p) = \int_{p_0}^p \om_1
.
\eeq
We have
\beq\label{Riemannbilinear}
\Res_{p\to {\rm all\, poles}} \Phi_1(p)\om_2(p) = {1\over 2i\pi}\, \sum_{i=1}^\genus \oint_{\underline\acycle_i} \om_1 \oint_{\underline\bcycle_i} \om_2 - \oint_{\underline\bcycle_i} \om_1 \oint_{\underline\acycle_i} \om_2 .
\eeq
Note that this identity holds also for the modified cycles with any $\kappa$:
\beq\label{Riemannbilinear2}
\Res_{p\to {\rm all\, poles}} \Phi_1(p)\om_2(p) = {1\over 2i\pi}\, \sum_{i=1}^\genus \oint_{\acycle_i} \om_1 \oint_{\bcycle_i} \om_2 - \oint_{\bcycle_i} \om_1 \oint_{\acycle_i} \om_2 .
\eeq
In particular with $\om_1(p)=B(p,q)$, we have:
\beq
\Res_{p\to {\rm all\, poles}} dS_{p,p_0}(q)\, \om (p) = - \sum_{i=1}^\genus du_i(q)\,\oint_{\acycle_i} \om
\eeq
and:
\beq\label{blineardSCauchy}
\om(q) = \Res_{p\to {\rm poles\, of}\,\om} dS_{p,p_0}(q)\, \om (p) + \sum_{i=1}^\genus du_i(q)\,\oint_{\acycle_i} \om
.\eeq

\subsection{Moduli of the curve} \label{sectmoduliofcurve}

The curve $\curve(x,y)=0$ is  parameterized by:
\begin{itemize}
\item a genus $\genus$ compact Riemann surface $\bar\Sigma$,  with periods $\tau_{ij}$.
\item punctures $\alpha_i$ at the poles of $x$ and $y$, whose moduli are given by the negative coefficients of the Laurent series of $ydx$ near the poles.
\item the $\acycle$-cycle integrals of $ydx$, called filling fractions.
\end{itemize}



\subsubsection{Filling fractions}

We define:
\beq
\epsilon_i = {1\over 2i\pi} \oint_{\acycle_i} ydx
\eeq
which are called ``filling fractions'' by analogy with matrix models (\cite{eynm2mg1}).

\subsubsection{Moduli of the poles}

Consider a pole $\alpha$ of $ydx$, define the ``{\bf temperatures}'':
\beq
t_\alpha=\Res_\alpha ydx
.
\eeq
Notice that:
\beq\label{sumresydxzero}
\sum_\alpha t_\alpha=0.
\eeq

Then consider the 3 cases:

$\bullet$ Either $\alpha$ is a pole of $x$ of degree $d_\alpha$, then we define the local parameter near $\alpha$ as:
\beq
z_\alpha(p) = x(p)^{1\over d_\alpha};
\eeq

$\bullet$ or $\alpha$ is not a pole of $x$ neither a branchpoint (thus it is a pole of $y$), then we define the local parameter near $\alpha$ as:
\beq
z_\alpha(p) = {1\over x(p)-x(\alpha)}.
\eeq

$\bullet$ or $\alpha$ is not a pole of $x$, and it is a branchpoint (thus it is a pole of $y$), then we define the local parameter near $\alpha$ as:
\beq
z_\alpha(p) = {1\over \sqrt{x(p)-x(\alpha)}}.
\eeq

\bigskip

In all cases, in the vicinity of $\alpha$, we define the ``{\bf potential}''
\beq
V_\alpha(p) = \Res_{q\to \alpha} y(q)dx(q)\, \ln{\left(1-{ z_\alpha(p)\over z_\alpha(q)}\right)}
\eeq
which is a polynomial in $z_\alpha(p)$:
\beq
V_\alpha(p) = \sum_{k=1}^{\deg V_\alpha} t_{\alpha,k}\, z_\alpha^k(p).
\eeq
The {\bf $t_{\alpha,k}$ are the moduli of the pole $\alpha$}.

\bigskip

We have the following properties:
\beq
dV_\alpha(p) = \Res_{q\to \alpha} y(q)dx(q)\,  {dz_\alpha(p)\over z_\alpha(p)-z_\alpha(q)} ,
\eeq
\beq
\Res_\alpha dV_\alpha=0
\eeq
and
\beq
y(p)dx(p) \mathop{{\sim}}_{p\to\alpha} dV_\alpha(p) - t_\alpha\, {dz_\alpha(p)\over z_\alpha(p)} +O({dz_\alpha(p)\over z_\alpha(p)^2}).
\eeq
We have from \eq{blineardSCauchy}:
\beq\label{ydxmoduli}
y(p)dx(p) = -\sum_\alpha  \Res_{q\to\alpha} B(p,q) V_{\alpha}(q) + \sum_\alpha t_\alpha dS_{\alpha,o}(p) + 2i\pi \sum_i \epsilon_i du_i(p) .
\eeq

If we introduce
\beq
B_{\alpha,k}(p) = - \Res_{q\to\alpha} B(p,q) \,z_\alpha(q)^k,
\eeq
we can turn this expression to
\beq\label{ydxmodulitka}
ydx = \sum_{\alpha,k} t_{\alpha,k} B_{\alpha,k} + \sum_\alpha t_\alpha dS_{\alpha,o} + 2i\pi \sum_i \epsilon_i du_i(p)
\eeq
in order to exhibit the moduli of the curve.

\section{Definition of Correlation functions and free energies}

In all this section, the curve $\curve(x,y)=0$ and a symmetric matrix $\kappa$ are given and fixed.
The unfamiliar reader may choose $\kappa=0$ since most usual applications (matrix models) correspond to that case.

\subsection{Notations}

Consider an arbitrary point $p\in\bar\Sigma$, and a point $q$ of $\bar\Sigma$ which is in the vicinity of a branch point $a_i$ (so that $\qbar$ is well defined).
We define:
\bd\label{defdiagrule}
 Diagrammatic rules:
\beq
{\mathbf{\rm vertex:}}\qquad \om(q) = (y(q)-y(\qbar))dx(q)
\eeq
\beq
{\mathbf{\rm line-propagator:}}\qquad B(p,q)
\eeq
\beq
{\mathbf{\rm arrow-propagator:}}\qquad dE_{q}(p) = {1\over 2} \int_{q}^{\qbar} B(\xi,p)
\eeq
where the integration path is a  path which lies entirely in a vincinity of $a_i$ (thus it is uniquely defined)\footnote{This definition is the opposite
of the notation used in \cite{eyno,CEO} since the integral goes from $q$ to $\qbar$ instead of going from
$\qbar$ to $q$.}.
\beq
{\mathbf{\rm root:}}\qquad \Phi(q) = \int_{o}^q y dx
\eeq
where $o$ is an arbitrary base point on the curve, i.e. $\Phi$ is an arbitrary antiderivative of $y dx$, i.e.  $d\Phi=ydx$.
\ed
The reason why we call these objects diagramatic rules and vertices, propagator or root, is explained in section \ref{sectiondiagrepresent} below.

Notice that $dE$ depends on $i$, i.e. on which branchpoint we are considering, but we omit to mention the index $i$ in order to make the notations easier to read.
In all what follows it is always clear which $i$ is being considered.

\bigskip
{\bf Notation for subset of indices:}

Given a set of points of the curve $\{ p_{1}, p_{2},\dots, p_{n}\}$, 
if $K=\{ i_1,i_2,\dots,i_k\}$ is any subset of $\{ 1,2,\dots, n\}$, we denote:
\beq\label{notationsubsetint}
{\mathbf p_K} = \{ p_{i_1}, p_{i_2},\dots, p_{i_k}\}
\eeq

\subsection{Correlation functions and free energies}

\subsubsection{Correlation functions}

The {\bf $k$-point correlation functions to order $g$, $W_{k}^{(g)}$}, 
are meromorphic multilinear forms, defined
by the following recursive triangular system:
\bd\label{defloopfctions}
Correlation functions
\beq
W_k^{(g)}=0 \quad {\rm if}\,\, g<0
\eeq
\beq
W_1^{(0)}(p) = 0
\eeq
\beq\encadremath{
W_2^{(0)}(p_1,p_2) = B(p_1,p_2)
}\eeq
and define recursively (remember that ${\bf p_K}$ is a $k$-uplet of points cf eq.\ref{notationsubsetint}):
\beq\label{defWkgrecursive}\encadremath{
\begin{array}{l} 
W_{k+1}^{(g)}(p,{\bf p_K}) 
= \Res_{q\to {\bf a}} {dE_{q}(p)\over \om(q)}\,\Big(  \cr
\qquad \sum_{m=0}^g \sum_{J\subset K} W_{|J|+1}^{(m)}(q,{\bf p_J})W_{k-|J|+1}^{(g-m)}(\qbar,{\bf p_{K/J}})
+ W_{k+2}^{(g-1)}(q,\qbar,{\bf p_K}) \Big) \cr
\end{array}
}\eeq
\ed
This system is triangular because all terms in the RHS have lower $2g+k$ than in the LHS and
given $W_1^{(0)}$ and $W_2^{(0)}$, it has a unique solution.

Notice that $W_{k+1}^{(g)}(p,p_1,\dots,p_k)$ is a multilinear meromorphic form in each of its arguments, it is clearly symmetric in the last $k$-ones, and we prove below (theorem \ref{thsymWk}) that it is in fact symmetric in all its arguments.

More properties of $W_{k+1}^{(g)}$ are studied below in section \ref{propcorrel}.

\subsubsection{Free energies}

We define the {\bf free energies} which are complex numbers:

\bd\label{defFg}
Free energies.

For $g>1$
\beq\encadremath{
F^{(g)} = {1\over 2-2g}\,\sum_i \Res_{q\to a_i} \Phi(q) W_{1}^{(g)}(q)
}\eeq
and
\beq\encadremath{
F^{(1)}=-{1\over 2}\ln{(\tau_{Bx})}\,\,-{1\over 24}\ln{\left(\prod_i y'(a_i) \right)}- \ln\left(\det \kappa\right)
}\eeq
where
\beq
y'(a_i) = {dy(a_i)\over dz_i(a_i)}
\virg
z_i(p) = \sqrt{x(p)-x(a_i)}
\eeq
and $\tau_{Bx}$ is the Bergmann $\tau$-function defined in \eq{deftauBx}.

$F^{(0)}$ is defined in the next section.

\ed

\subsubsection{Leading order free energy  \texorpdfstring{$F^{(0)}$}{F0}.}

Let us define $F^{(0)}$ as follows
:

\beq\label{defF0}\encadremath{
F^{(0)}
= {1\over 2}\sum_{\alpha} \Res_\alpha V_\alpha ydx + {1\over 2} \sum_{\alpha} t_\alpha \mu_\alpha
- {1\over 4i\pi}\sum_i \oint_{\acycle_i} ydx \oint_{\bcycle_i} ydx 
}\eeq
where $\mu_\alpha$ is given by 
\beq
\mu_\alpha
= \int_\alpha^{o} (ydx - dV_\alpha + t_\alpha {dz_\alpha\over z_\alpha}) + V_\alpha(o) - t_\alpha \ln{(z_\alpha(o))} 
\eeq
Notice that $\mu_\alpha$ depends on some base point $o$, but the sum $\sum_\alpha t_\alpha \mu_\alpha$ does not.

\subsubsection{Special free energies and correlation functions}
\label{sectdefspfree}

All the quantities defined so far, were defined with the $\kappa$-modified cycles and modified Bergmann kernel.
Let us also define them for $\kappa=0$ (for instance $F^{(1)}$ obviously needs another definition).

Therefore we also define the unmodified quantities corresponding to $\kappa=0$, as:
\beq
\label{defspcorr}
\forall k,g, \qquad 
\underline{W}_{k}^{(g)}(p_1,\dots,p_k)
:= \left. W_{k}^{(g)}(p_1,\dots,p_k)
\right|_{\kappa = 0},
\eeq
\beq \label{defspfree}
{\rm for}\,\, g>1, \qquad
\underline{F}^{(g)} := {1\over 2-2g}\,\sum_i \Res_{q\to a_i} \Phi(q) \underline{W}_{1}^{(g)}(q)
= \left.
F^{(g)} \right|_{\kappa = 0} \,\, ,
\eeq
\beq
\underline{F}^{(1)}=-{1\over 2}\ln{(\tau_{\underline{B}x})}\,\,-{1\over 24}\ln{\left(\prod_i y'(a_i) \right)} \,\, ,
\eeq
\beq
{\rm and}\qquad
\underline{F}^{(0)}
= {1\over 2}\sum_{\alpha} \Res_\alpha V_\alpha ydx + {1\over 2} \sum_{\alpha} t_\alpha \mu_\alpha
- {1\over 4i\pi}\sum_i \oint_{\underline\acycle_i} ydx \oint_{\underline\bcycle_i} ydx \, .
\eeq

\br
The special functions, except $\underline{F}^{(1)}$ and $\underline{F}^{(0)}$, are obtained by changing $B$ and $dS$ by
$\Bergmann$ and $\underline{dS}$ in the diagrammatic rules defined in section \ref{sectiondiagrepresent}.
\er

\subsection{Tau function}

\bd
We define the {\bf tau-function} as the formal power series in $N^{-2}$:
\beq\encadremath{
\ln{(Z_N(\curve))} = -\sum_{g=0}^\infty N^{2-2g} F^{(g)}.
}\eeq
\ed
We show in section \ref{sectintegrability} that $Z_N(\curve)$ is indeed a tau-function because it obeys Hirota equations, order by order in $N^{-2}$.

\subsection{Properties of correlation functions}\label{propcorrel}

The loop functions defined in definition \ref{defloopfctions} satisfy the following theorems, whose {\bf proofs can be found
in Appendix \ref{appprop}}:

\bt\label{thW30}
The correlation function $W_3^{(0)}$ is worth:
\beq
W_3^{(0)}(p,p_1,p_2) = \Res_{q\to \bfa} {B(q,p)B(q,p_1)B(q,p_2)\over dx(q) dy(q)}
\eeq
\et
In particular, $W_3^{(0)}$ is symmetric in its 3  variables.

\bt\label{thpolesWkgbp}
For $(k,g)\neq (1,0)$, the loop function $W_{k+1}^{(g)}(p,p_1,\dots,p_k)$ has poles (in any of its variables $p,p_1,\dots,p_k$) only at the branch points.
\et

\bt\label{thWkcycle}
For every ${\cal A}$ cycle we have:
\beq
\forall i=1,\dots,\genus \qquad \oint_{p\in{\cal A}_i} W_{k+1}^{(g)}(p,p_1,\dots,p_k) = 0
,
\eeq
\beq
\forall i=1,\dots,\genus, \forall m=1,\dots,k \qquad \oint_{p_m\in{\cal A}_i} W_{k+1}^{(g)}(p,p_1,\dots,p_k) = 0
.
\eeq
\et

\bt\label{thsumWk}
For every $k$ and $g$, we have:
\beq\label{thsumWk1}
\sum_i {W_{k+1}^{(g)}(p^i,p_1,\dots,p_k) \over dx(p^i)}= \delta_{k,1}\delta_{g,0}\,{dx(p_1)\over (x(p)-x(p_1))^2}
\eeq
and if $k\geq 1$:
\beq\label{thsumWk2}
\sum_i {W_{k+1}^{(g)}(p_1,p^i,p_2,\dots,p_k) \over dx(p^i)} = \delta_{k,1}\delta_{g,0}\,{dx(p_1)\over (x(p)-x(p_1))^2}
\eeq
where we recall that $p^i$ are all the points such that $x(p^i)=x(p)$ (see section \ref{sectgeomalg}).
\et

\bt\label{thPkpol}
For $(k,g)\neq (0,1)$,
\bea
P_k^{(g)}(x(p),{\bf p_K})
&=& {1\over dx(p)^2}\,\sum_i \Big[-2y(p^i)dx(p) W_{k+1}^{(g)}(p^i,{\bf p_K}) + W_{k+2}^{(g-1)}(p^i,p^i,{\bf p_K}) \cr
&& \qquad \qquad + \sum_{m=0}^g \sum_{J\subset K} W_{j+1}^{(m)}(p^i,{\bf p_J})W_{k-j+1}^{(g-m)}(p^i,{\bf p_{K/J}}) \Big] \cr
\eea
 is a rational function of $x(p)$, with no poles at branch-points.
\et

\bt\label{thsymWk}
$W_k^{(g)}$ is a symmetric function of its $k$ variables.
\et

\bc\label{corResxyWk}
\beq
\forall i,\qquad \Res_{p\to a_i} W^{(g)}_{k+1}(p,p_1,\dots,p_k) =0
,
\eeq
\beq
\forall i,\qquad \Res_{p\to a_i} x(p) W^{(g)}_{k+1}(p,p_1,\dots,p_k) =0
,
\eeq
\beq
\sum_i \Res_{p\to a_i} y(p) W^{(g)}_{k+1}(p,p_1,\dots,p_k) =0
,
\eeq
\beq
\sum_i \Res_{p\to a_i} x(p)y(p) W^{(g)}_{k+1}(p,p_1,\dots,p_k) =0
.
\eeq
\ec

\bt\label{thintPhi}
For $k\geq 1$ we have:
\beq
\Res_{p_{k+1}\to \bfa,p_1,\dots,p_k} \Phi(p_{k+1}) W^{(g)}_{k+1}({\bf p_K},p_{k+1}) = (2g+k-2) W^{(g)}_k({\bf p_K}) + \delta_{g,0}\delta_{k,1} y(p_1)dx(p_1)
.
\eeq
\et
Notice that for $k=0$ and $g\geq 2$, it holds by definition if we define $W^{(g)}_0= - F^{(g)}$.

\subsection{Diagrammatic representation \label{sectiondiagrepresent}}

The recursive  definitions of $W_k^{(g)}$ and $F^{(g)}$ can be represented {\bf graphically}.

We represent the multilinear form $W_k^{(g)}(p_1,\dots,p_k)$ as a blob-like ``surface'' with $g$ holes and
$k$ legs (or punctures) labeled with the variables $p_1,\dots, p_k$, and $F^{(g)}$ with $0$ legs and
$g$ holes.
\beq
W_{k+1}^{(g)}(p,p_1,\dots,p_k):=\begin{array}{r}
{\epsfxsize 4.5cm\epsffile{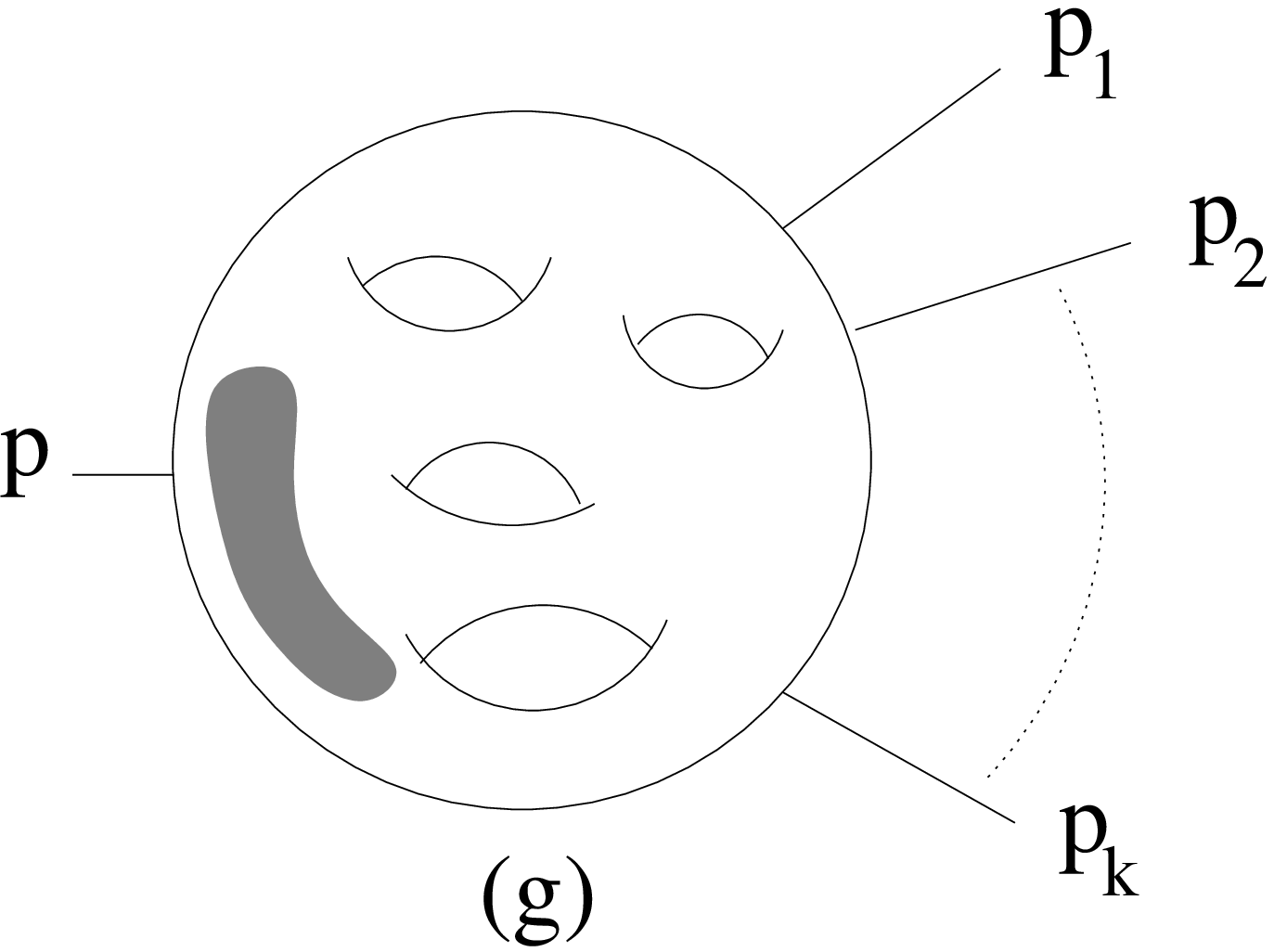}}
\end{array}
\virg
F^{(g)}:= \begin{array}{r}
{\epsfxsize 2.5cm\epsffile{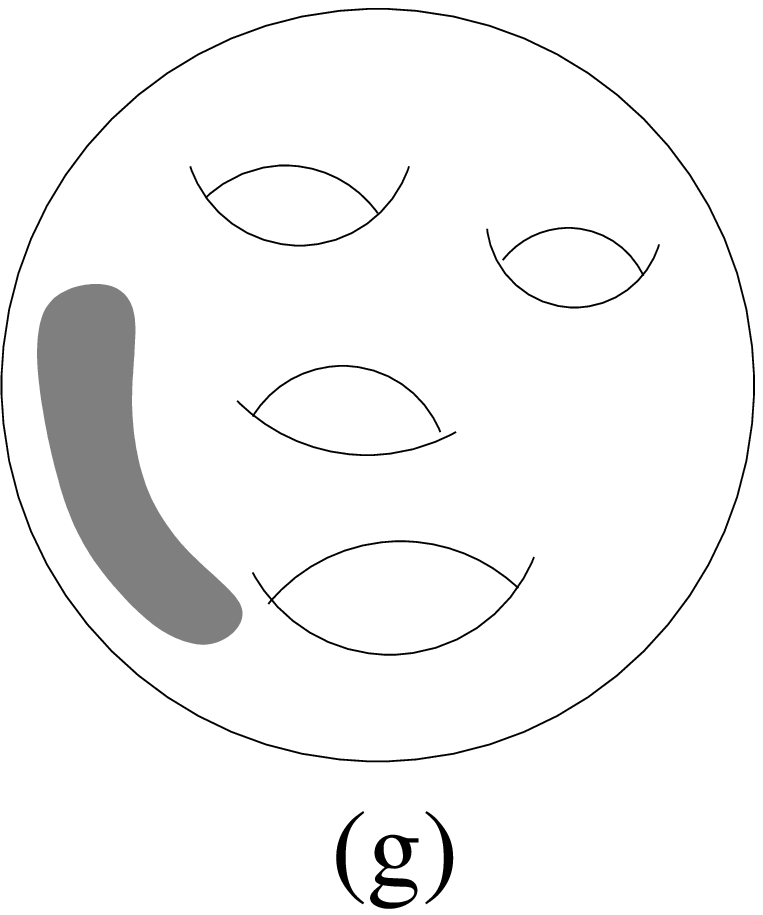}}
\end{array}
\eeq

We represent the Bergmann kernel $B(p,q)$ (which is also $W_2^{(0)}$, i.e. a blob with 2 legs and no hole) as a straight non-oriented line between $p$ and $q$
\beq
B(p,q):= \begin{array}{r}
{\epsfxsize 2cm\epsffile{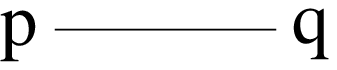}}
\end{array}.
\eeq

We represent ${dE_{q}(p)\over \om(q)}$ as a straight arrowed line with the arrow from $p$ towards $q$, and with a tri-valent vertex whose legs are $q$ and $\qbar$
\beq
{dE_{q}(p)\over \om(q)}:= \begin{array}{r}
{\epsfxsize 3cm\epsffile{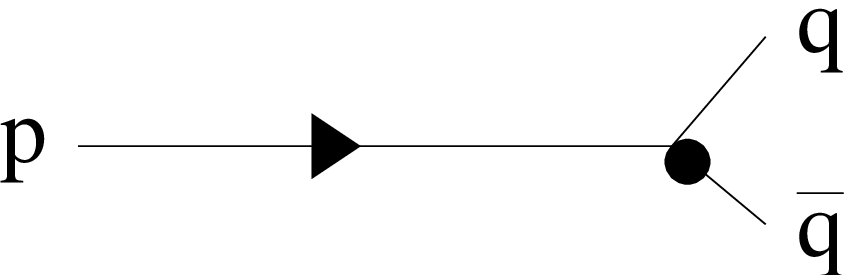}}
\end{array}.
\eeq

\medskip

\subsubsection*{Graphs}

\bd\label{defgraphs}
For any $k\geq 0$ and $g\geq 0$ such that $k+2g\geq 3$, we define:

Let ${\cal G}_{k+1}^{(g)}(p,p_1,\dots,p_k)$ be the set of connected trivalent graphs defined as follows:
\begin{enumerate}

\item there are $2g+k-1$ tri-valent vertices called vertices.
\item there is one 1-valent vertex labelled by $p$, called the root.
\item there are $k$ 1-valent vertices labelled with $p_1,\dots,p_k$ called the leaves.
\item There are $3g+2k-1$ edges.
\item Edges can be arrowed or non-arrowed. There are $k+g$ non-arrowed edges and $2g+k-1$ arrowed edges.
\item The edge starting at $p$ has an arrow leaving from the root $p$.
\item The $k$ edges ending at the leaves $p_1,\dots, p_k$ are non-arrowed.
\item The arrowed edges form a "spanning\footnote{It goes through all vertices.} planar\footnote{planar tree means that the left child and right child are not equivalent. The right child is marked by a black disk on the outgoing edge.} binary skeleton\footnote{a binary skeleton tree is a binary tree from which we have removed the leaves, i.e. a tree with vertices of valence 1, 2 or 3.} tree" with root $p$. The arrows are oriented from root towards leaves. In particular, this induces a partial ordering of all vertices.
\item There are $k$ non-arrowed edges going from a vertex to a leaf, and $g$ non arrowed edges joining two inner vertices. Two inner vertices can be connected by a non arrowed edge only if one is the parent of the other along the tree.
\item If an arrowed edge and a non-arrowed inner edge come out of a vertex, then the arrowed edge is the left child. This rule
only applies when the non-arrowed edge links this vertex to one of its descendants (not one of its parents).

\end{enumerate}

\ed

We have the following useful lemma:
\bl\label{lemgraphaddleg}
There is a $1$ to $3g+2k-1$ map from ${\cal G}_{k+1}^{(g)}(p,{\bf p_K})$ to ${\cal G}_{k+2}^{(g)}(p,{\bf p_K},p_{k+1})$.
\el
\proof{If $G$ is  a graph in ${\cal G}_{k+2}^{(g)}(p,p_1,\dots,p_k,p_{k+1})$, remove the non-arrowed edge attached to the leaf $p_{k+1}$ and remove the corresponding vertex, and merge the incoming and the other outgoing edges of that vertex. You clearly get a graph $G'\in{\cal G}_{k+1}^{(g)}(p,p_1,\dots,p_k)$. It is clear that the same graph is obtained $3g+2k-1$ times (the number of edges of $G'$.
And it is clear that from any $G'\in{\cal G}_{k+1}^{(g)}(p,p_1,\dots,p_k)$, you can obtain $3g+2k-1$ graphs $G\in{\cal G}_{k+2}^{(g)}(p,p_1,\dots,p_k,p_{k+1})$ by adding a new vertex on any edge, and linking this new vertex to the leaf $p_{k+1}$.}

\vs

\noindent {\bf Example of ${\cal G}_{1}^{(2)}(p)$}

As an example, let us build step by step all the graphs of ${\cal G}_{1}^{(2)}(p)$, i.e. $g=2$ and $k=0$.

We first draw all planar binary skeleton trees with one root $p$ and $2g+k-1=3$ arrowed edges:
\beq
\begin{array}{r}
{\epsfxsize 2cm\epsffile{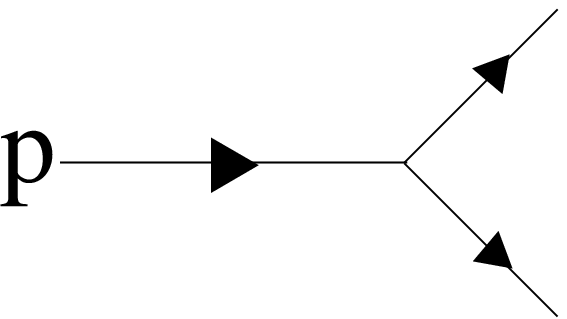}}
\end{array}
\virg
\begin{array}{r}
{\epsfxsize 2.5cm\epsffile{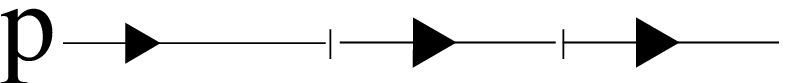}}
\end{array}.
\eeq
Then, we draw $g+k=2$ non-arrowed edges in all possible ways such that every vertex is trivalent, also satisfying rule 9) of definition.\ref{defgraphs}. There is only
one possibility for the first graph and two for the second one:
\beq
\begin{array}{r}
\begin{array}{r}
{\epsfxsize 2.5cm\epsffile{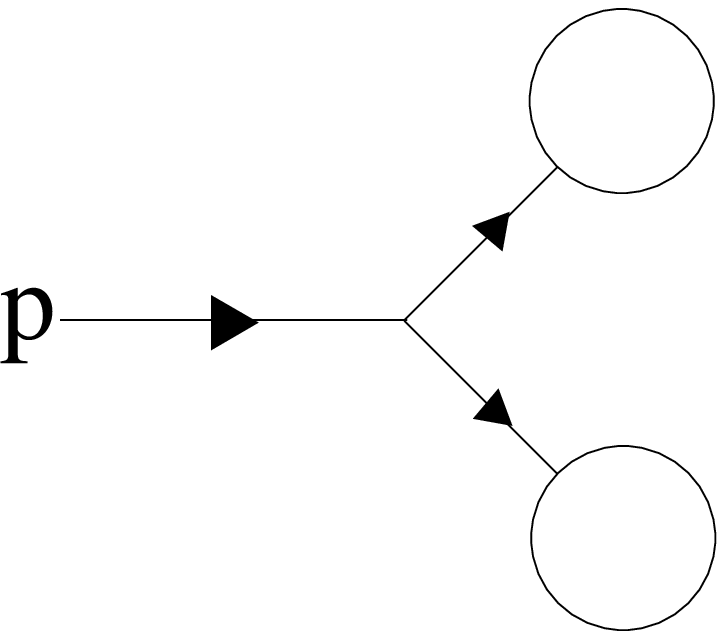}}
\end{array}
\virg
{\epsfxsize 2.2cm\epsffile{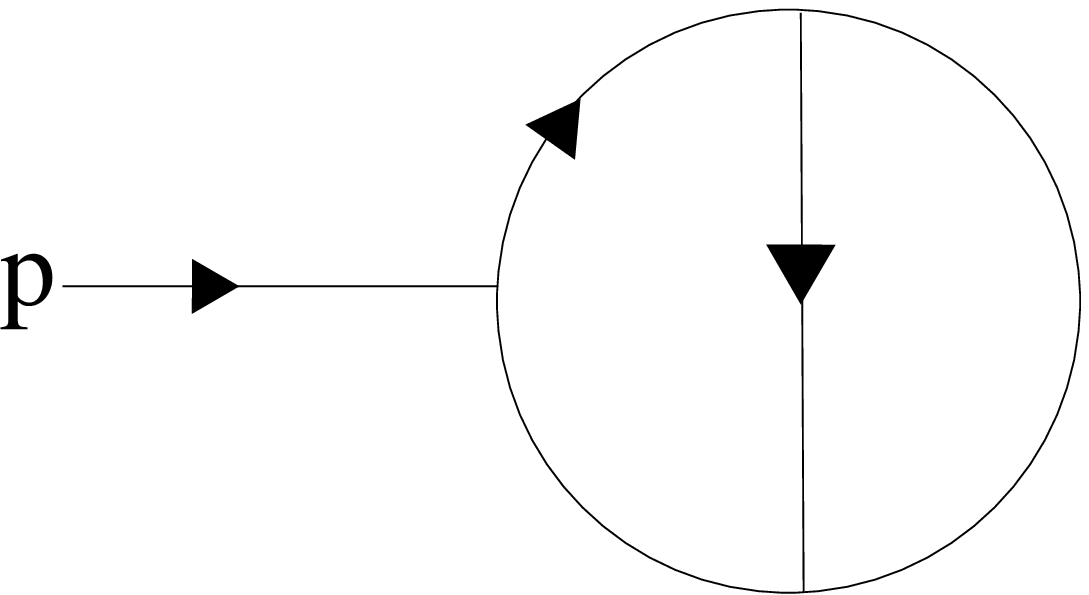}}
\end{array}
\virg
\begin{array}{r}
{\epsfxsize 3cm\epsffile{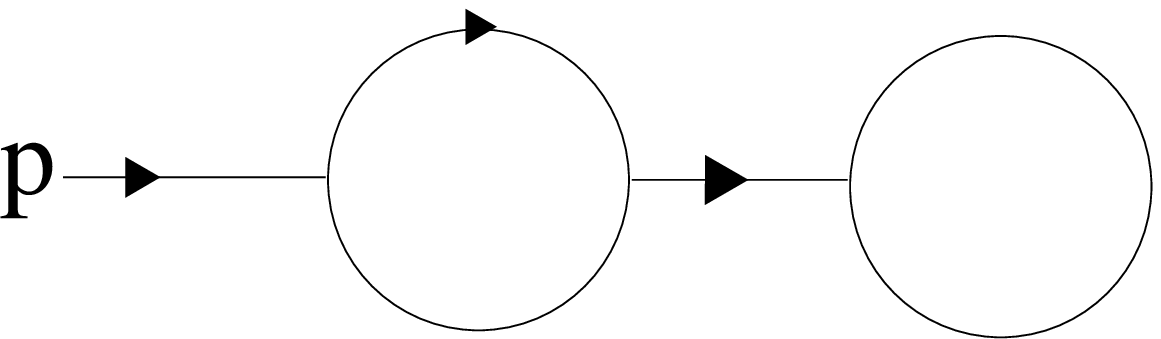}}
\end{array}.
\eeq

It just remains to specify the left and right children for each vertex. The only possibilities in accordance with rule 10) of def.\ref{defgraphs} are\footnote{ Note that the graphs are not
necessarily planar.}:
\bea
\begin{array}{r}
{\epsfxsize 2.2cm\epsffile{y123.eps}}
\end{array}
\virg
\begin{array}{r}
{\epsfxsize 2.5cm\epsffile{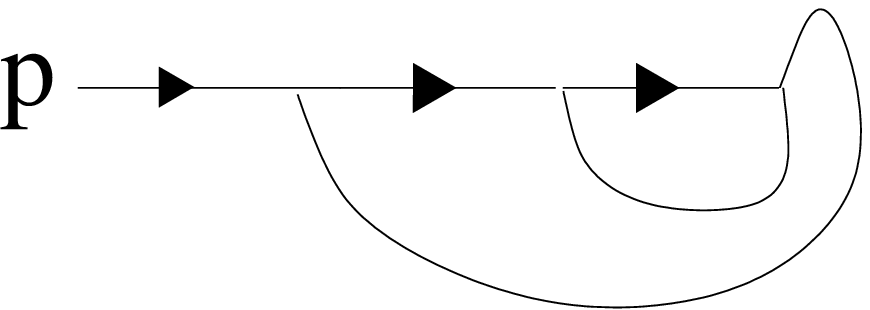}}
\end{array}
\virg
\begin{array}{r}
{\epsfxsize 3cm\epsffile{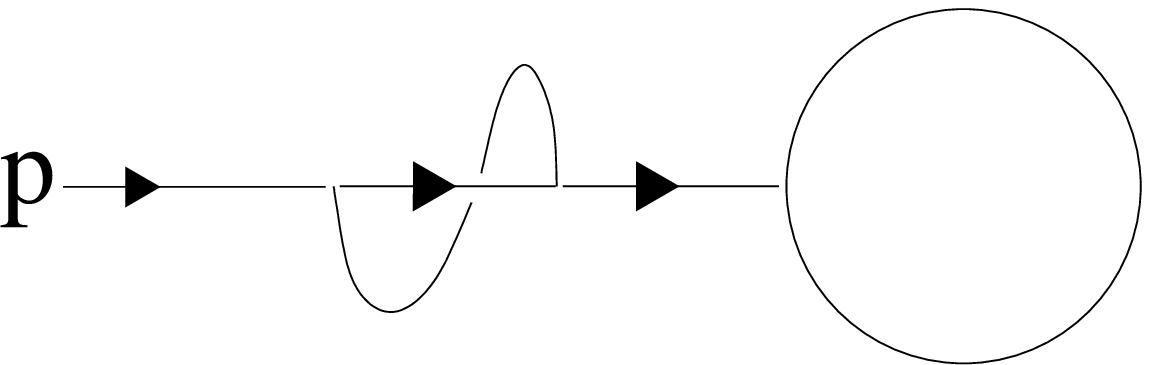}}
\end{array} \cr
\begin{array}{r}
{\epsfxsize 2.5cm\epsffile{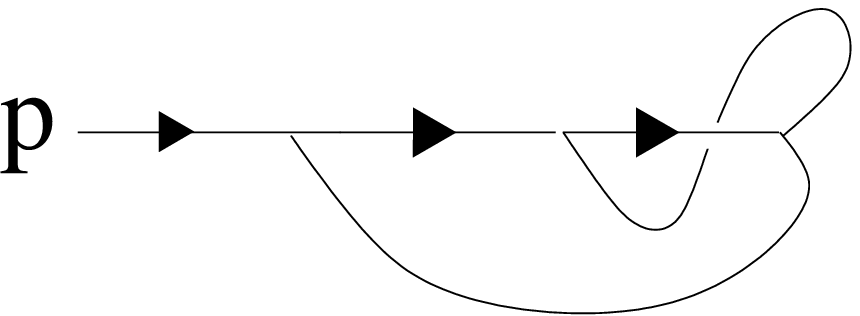}}
\end{array}
\virg
\begin{array}{r}
{\epsfxsize 3cm\epsffile{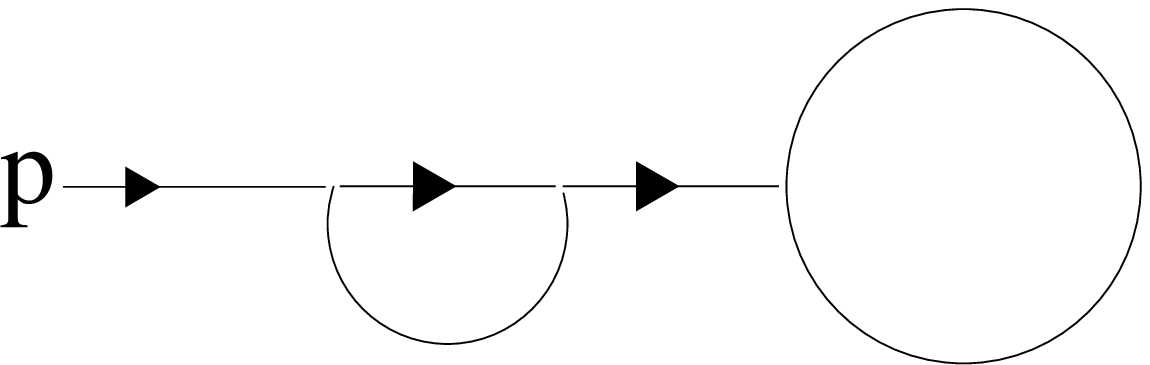}}
\end{array} .\cr
\eea

In order to simplify the drawing, we can draw a black dot to specify the right child. This way one gets only planar graphs.
\bea
\begin{array}{r}
{\epsfxsize 2.2cm\epsffile{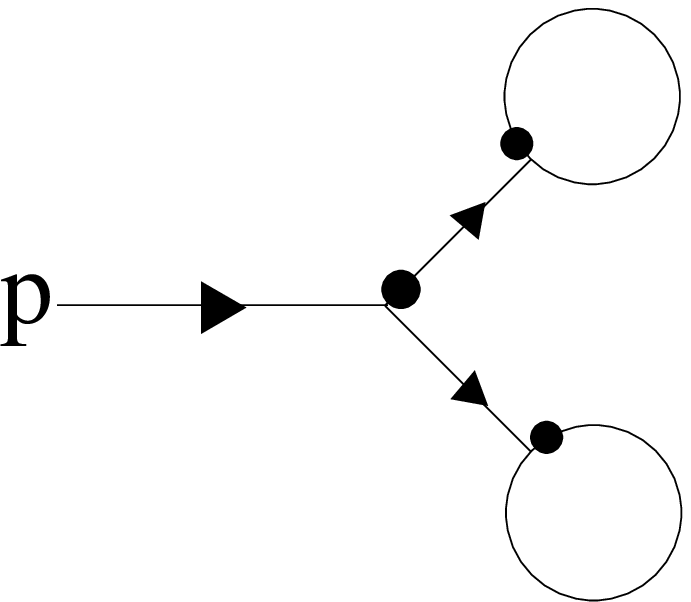}}
\end{array}
\virg
\begin{array}{r}
{\epsfxsize 2.5cm\epsffile{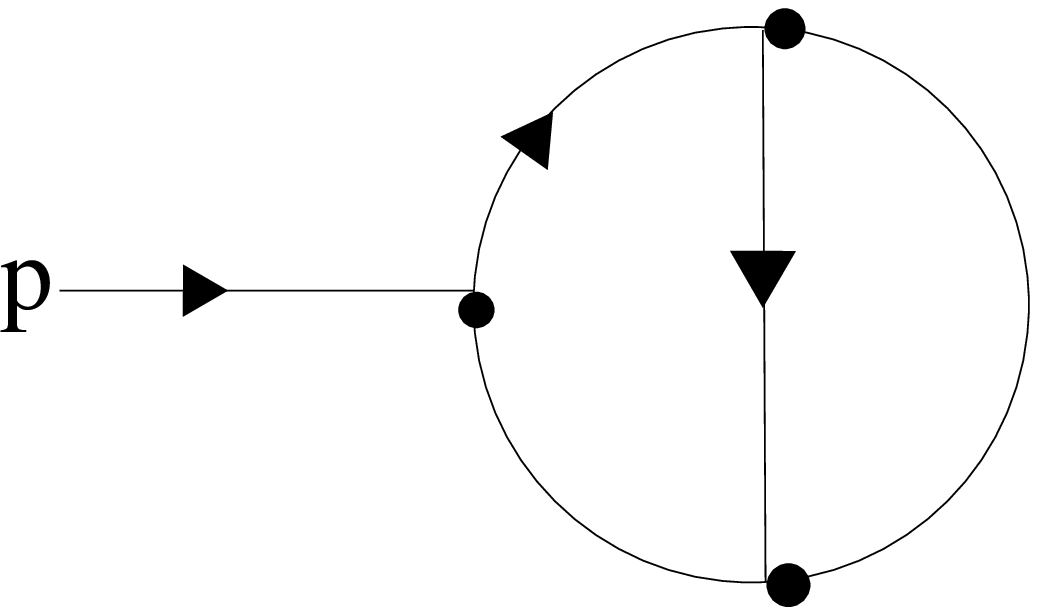}}
\end{array}
\virg
\begin{array}{r}
{\epsfxsize 3cm\epsffile{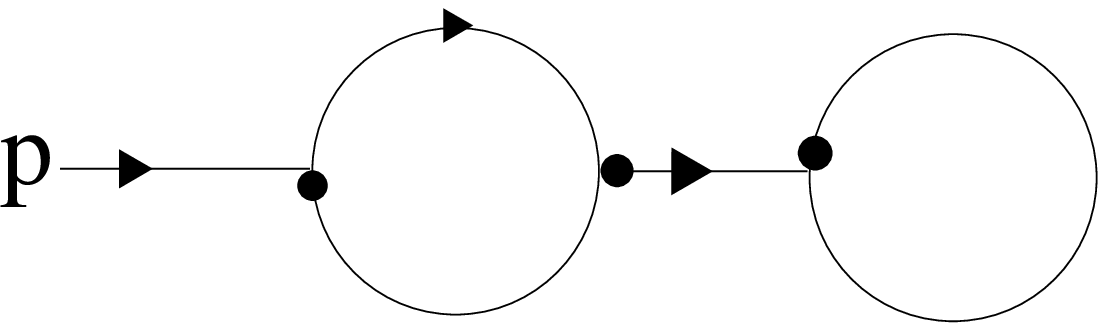}}
\end{array} \cr
\begin{array}{r}
{\epsfxsize 2.5cm\epsffile{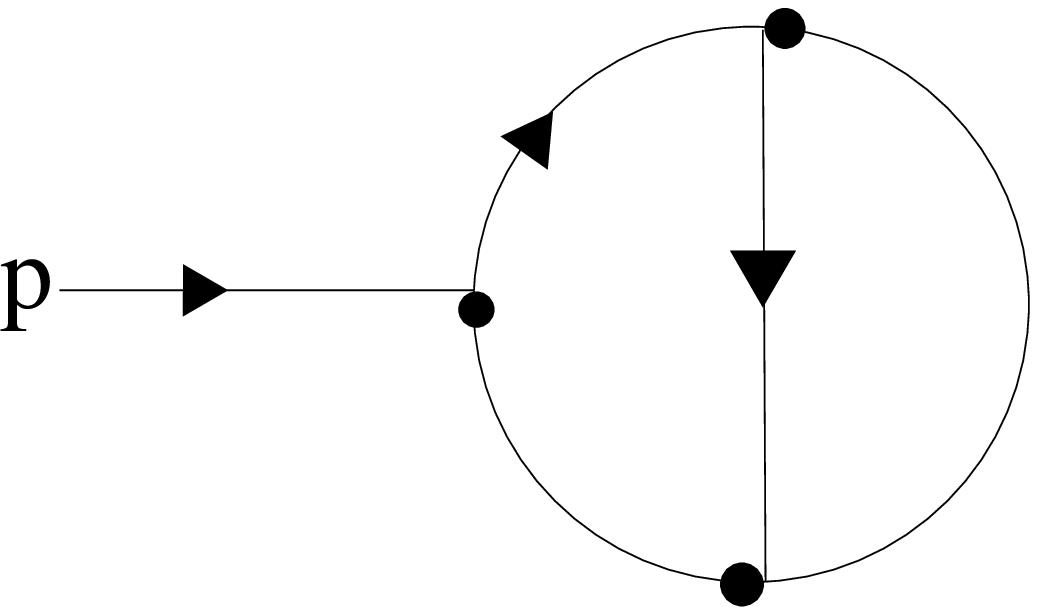}}
\end{array}
\virg
\begin{array}{r}
{\epsfxsize 3cm\epsffile{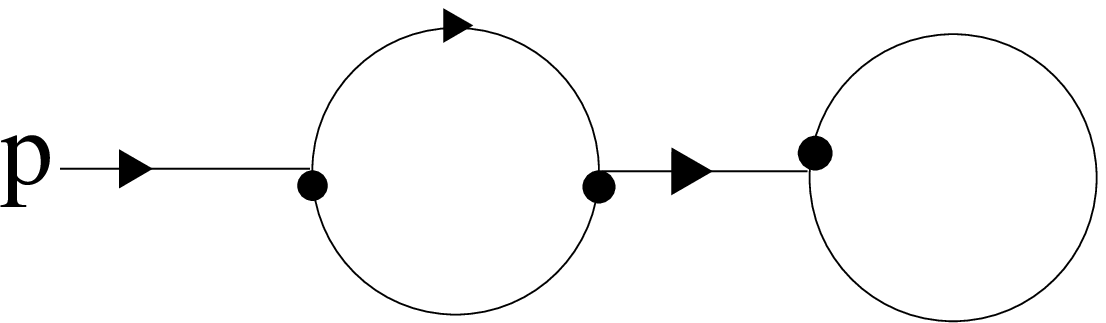}}
\end{array} \cr
\eea
Remark that without the prescriptions 9) and 10), one would get 13 different graphs whereas we only have 5.

\subsubsection*{Weight of a graph}

Consider a graph $G\in {\cal G}_{k+1}^{(g)}(p,p_1,\dots,p_k)$.
Then, to each vertex $i=1,\dots,2g+k-1$ of $G$, we associate a label $q_i$, and we associate $q_i$ to the beginning of the left child edge, and $\qbar_i$ to the right child edge.
Thus, each edge (arrowed or not), links two labels which are points on the Riemann surface $\bar\Surf$.
\begin{itemize}
\item To an arrowed edge going from $q'$ towards $q$, we associate a factor ${dE_q(q')\over (y(q)-y(\qbar))dx(q)}$.
\item To a non arrowed edge going between $q'$ and $q$ we associate a factor $B(q',q)$.
\item Following the arrows backwards (i.e.  from leaves to root), for each vertex $q$, we take a residue at $q\to \bfa$, i.e. we sum over all branchpoints.
\end{itemize}
After taking all the residues, we get the weight of the graph:
\beq
w(G)
\eeq
which is a multilinear form in $p,p_1,\dots,p_k$.

Similarly, we define weights of linear combinations of graphs by:
\beq
w(\alpha G_1 + \beta G_2) = \alpha w(G_1) + \beta w(G_2)
\eeq
and for a disconnected graph, i.e. a product of two graphs:
\beq
w( G_1  G_2) = w(G_1)  w(G_2)
.
\eeq

\bt\label{thdiagrepr}
We have:
\beq
W_{k+1}^{(g)}(p,p_1,\dots,p_k) = \sum_{G\in {\cal G}_{k+1}^{(g)}(p,p_1,\dots,p_k)}\,\, w(G) = w\left(\sum_{G\in {\cal G}_{k+1}^{(g)}(p,p_1,\dots,p_k)}\,\, G\right)
\eeq
\et
\proof{This is precisely what the recursion equations \ref{defWkgrecursive} of def.\ref{defloopfctions} are doing.
Indeed, one can represent them diagrammatically by
\beq
\begin{array}{r}
{\epsfxsize 10cm\epsffile{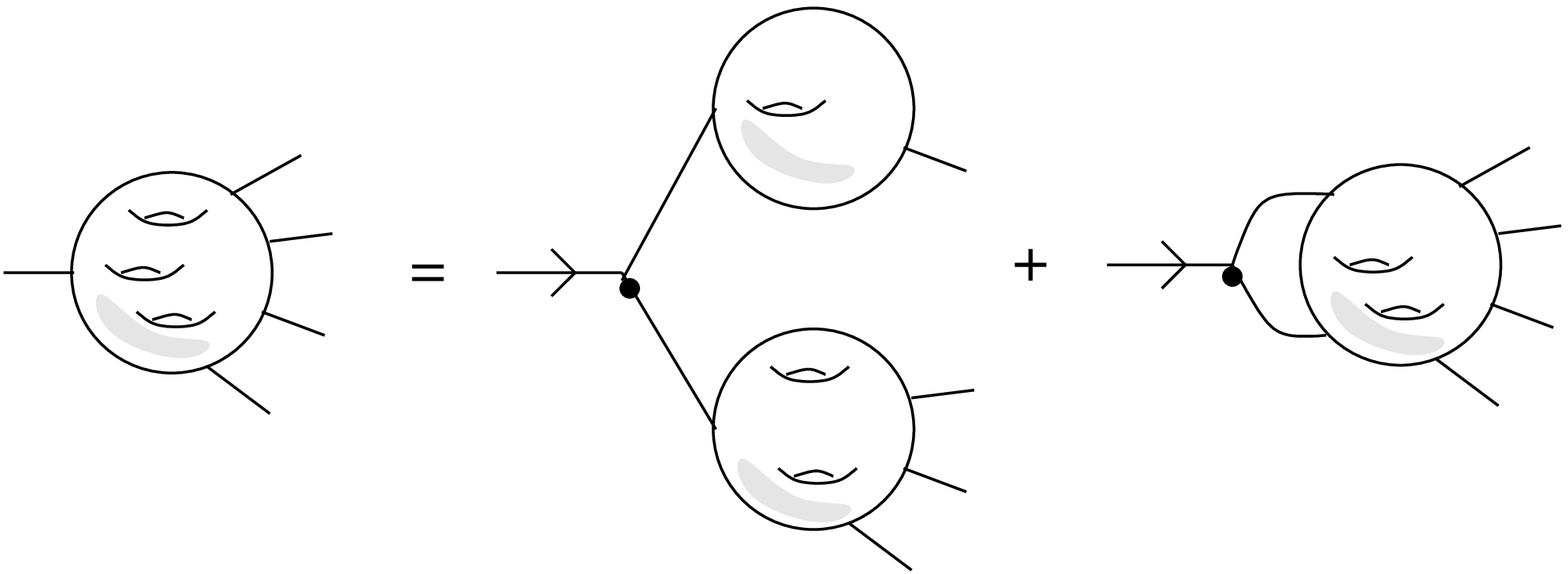}}
\end{array}.
\eeq
}

\bigskip

Such graphical notations are very convenient, and are a good support for intuition and even help proving some relationships.
It was immediately noticed after \cite{eynloop1mat} that those diagrams look very much like Feynman graphs, and there was a hope that they could be the Feynman's graphs for the Kodaira--Spencer theory.
But they ARE NOT Feynman graphs, because Feynman graphs can't have non-local restrictions like the fact that non oriented lines can join only a vertex and one of its descendent.

Those graphs are merely a notation for the recursive definition \ref{defloopfctions}.

\bl\label{lemmasymfactor} {\bf Symmetry factor:}
The weight of two graphs differing by the exchange of the right and left children of a vertex are the same. Indeed, the distinction between right and left child is just a way of encoding symmetry factors.
\el

\proof{
This property follows directly from theorem \ref{thsumWk} and the definition \eq{defWkgrecursive}.
Consider one term contributing to the first part of RHS of \eq{defWkgrecursive}:
\beq
\begin{array}{l}
\displaystyle \Res_{q \to {\bf a}} {dE_{q}(p) \over \omega(q)} W_{|J|+1}^{(m)}(q,{\bf P_J}) W_{k-|J|+1}^{(g-m)}(\qbar,{\bf P_{K/J}}) \cr
\displaystyle \qquad \quad =  - \Res_{q \to {\bf a}} {dE_{q}(p) \over \omega(q)} W_{|J|+1}^{(m)}(q,{\bf P_J}) W_{k-|J|+1}^{(g-m)}(q,{\bf P_{K/J}}) \cr
\displaystyle \qquad \quad =  \Res_{q \to {\bf a}} {dE_{q}(p) \over \omega(q)} W_{|J|+1}^{(m)}(\qbar,{\bf P_J}) W_{k-|J|+1}^{(g-m)}(q,{\bf P_{K/J}}) .\cr
\end{array}
\eeq
where the equalities are obtained by adding terms without residues at the branch points to the integrand and
using theorem \ref{thsumWk}. 
One can perform the same trick for the second term in \eq{defWkgrecursive} and this proves the lemma.}

\subsection{Examples.}

Let us present some examples of correlation functions and free energy for low order.

\subsubsection*{Correlation functions.}

To leading order, one has the first correlation functions given by:
\beq
W_{2}^{(0)}(p,q) = B(p,q).
\eeq

\bea
W_{3}^{(0)}(p,p_1,p_2)&=& \begin{array}{r}
{\epsfxsize 3.5cm\epsffile{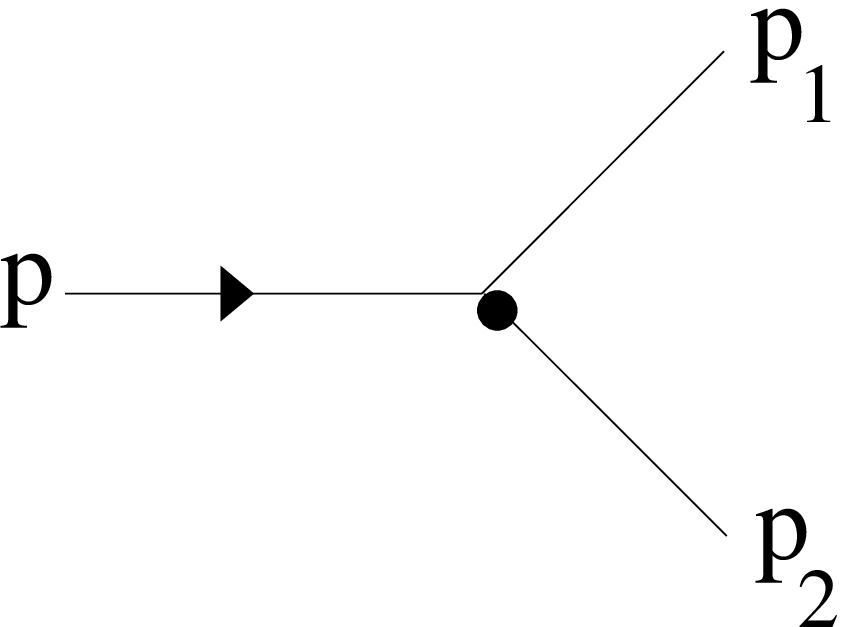}}
\end{array}
+
\begin{array}{r}
{\epsfxsize 3.5cm\epsffile{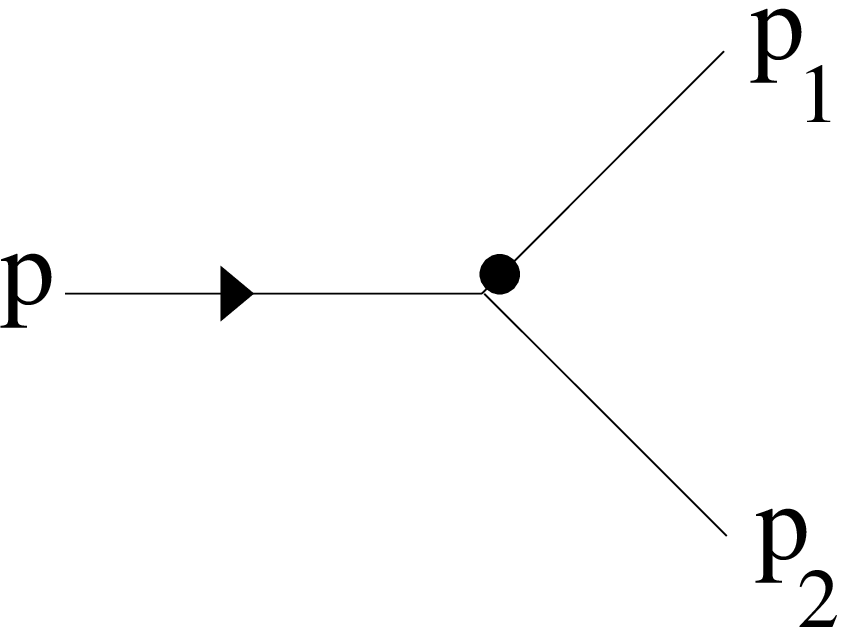}}
\end{array} \cr
&=& \Res_{q \to {\bf a}} {dE_q(p) \over \omega(q)} \left[ B(q,p_1) B(\overline{q},p_2) + B(\overline{q},p_1) B(q,p_2) \right] \cr
\eea

\bea
W_4^{(0)}(p,p_1,p_2,p_3)
&=& 
\begin{array}{r}
{\epsfxsize 4cm\epsffile{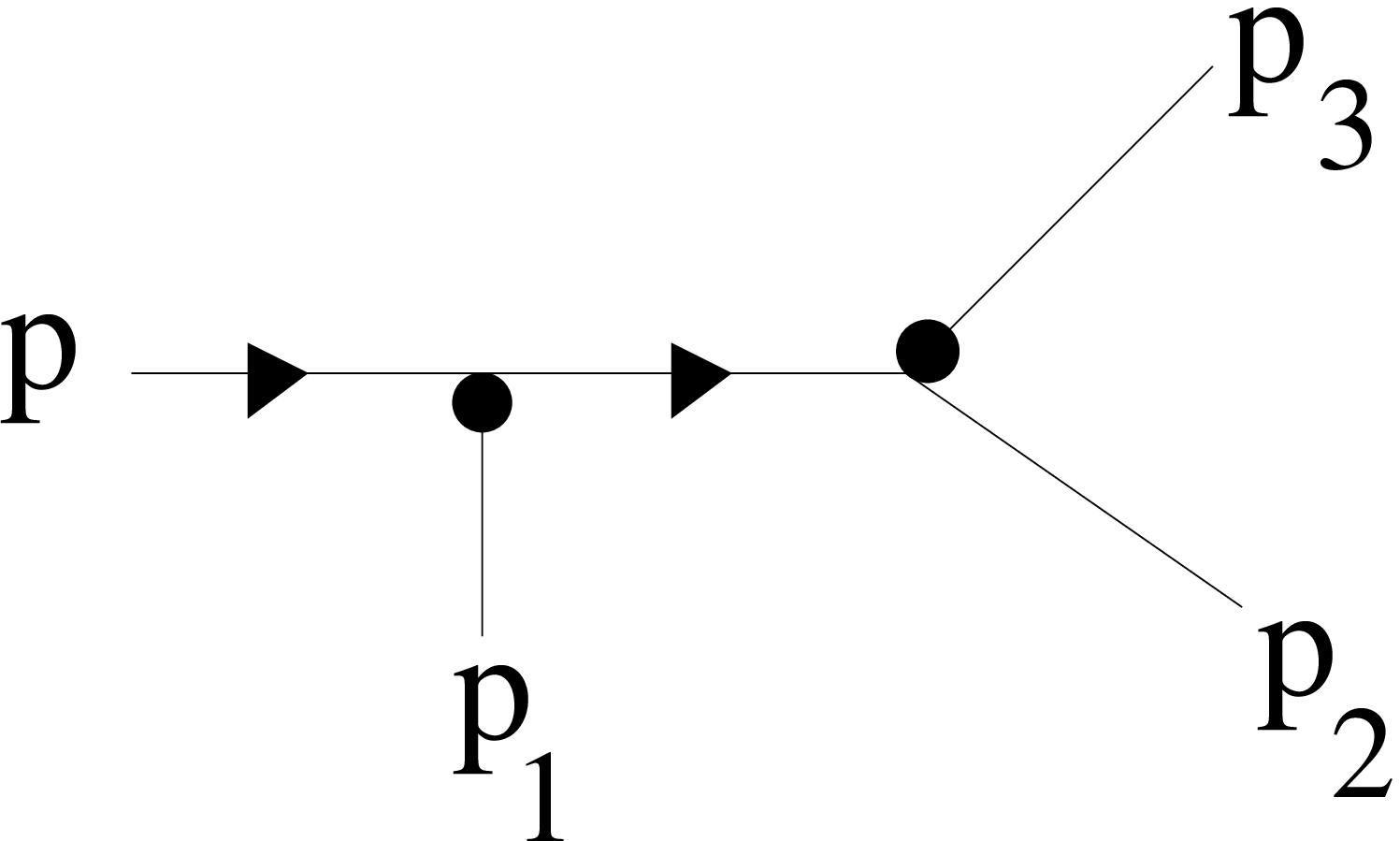}}
\end{array}
+ \;\;\;\; \hbox{perm. (1,2,3)}\cr
&& + \begin{array}{r}
{\epsfxsize 4cm\epsffile{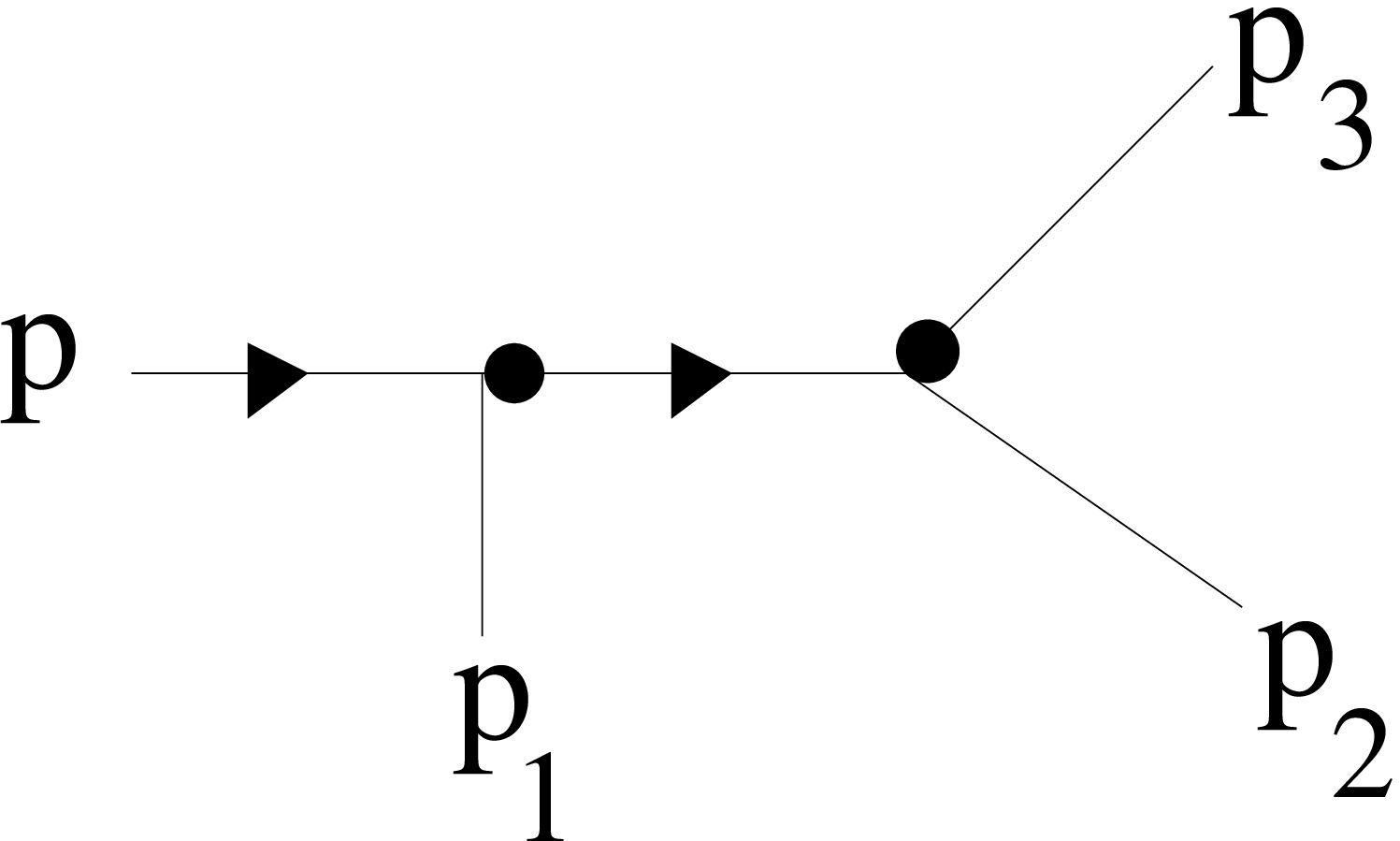}}
\end{array}
+ \;\;\;\; \hbox{perm. (1,2,3)}\cr
&=& \Res_{q \to {\bf a}} \Res_{r \to {\bf a}} {dE_q(p) \over \omega(q)} {dE_r(q) \over \omega(r)}
\left[ B(\qbar,p_1)B(r,p_2)B(\overline{r},p_3) \right. \cr
&& + B(\qbar,p_1)B(\overline{r},p_2)B(r,p_3) + B(\qbar,p_2)B(r,p_1)B(\overline{r},p_3) \cr
&& + B(\qbar,p_2)B(\overline{r},p_1)B(r,p_3) + B(\qbar,p_3)B(r,p_2)B(\overline{r},p_1) \cr
&& \left. + B(\qbar,p_3)B(\overline{r},p_2)B(r,p_1)\right] \cr
&& + \Res_{q \to {\bf a}} \Res_{r \to {\bf a}} {dE_{q}(p) \over \omega(q)} {dE_r(\qbar) \over \omega(r)}
\left[ B(q,p_1)B(r,p_2)B(\overline{r},p_3) \right. \cr
&& + B(q,p_1)B(\overline{r},p_2)B(r,p_3) + B(q,p_2)B(r,p_1)B(\overline{r},p_3)\cr
&& + B(q,p_2)B(\overline{r},p_1)B(r,p_3) + B(q,p_3)B(r,p_2)B(\overline{r},p_1) \cr
&& \left. + B(q,p_3)B(\overline{r},p_2)B(r,p_1)\right] \cr
\eea

First orders for the one point correlation function read:
\bea
W_{1}^{(1)}(p) &=& \begin{array}{r}
{\epsfxsize 4cm\epsffile{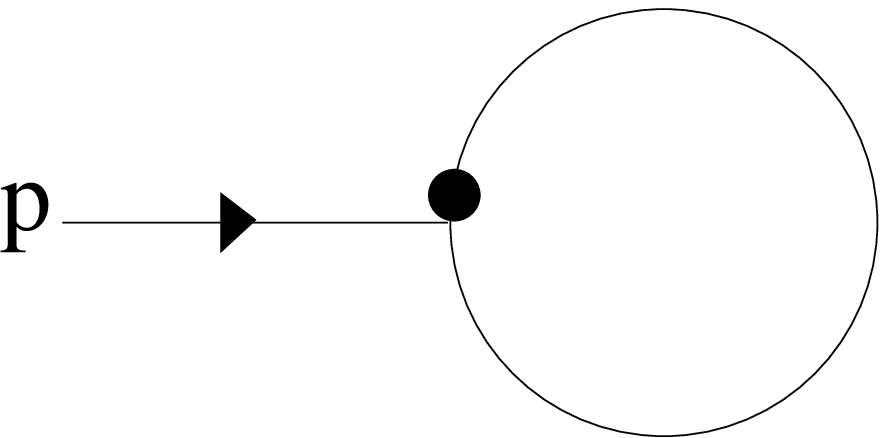}}
\end{array} \cr
&=& \Res_{q \to {\bf a}} {dE_q(p) \over \omega(q)} B(q, \qbar)\cr
\eea

\bea
W_1^{(2)}(p)&=&
\begin{array}{r}
{\epsfxsize 4cm\epsffile{w121.eps}}
\end{array}
+
\begin{array}{r}
{\epsfxsize 4cm\epsffile{w123.eps}}
\end{array} \cr
&& +
\begin{array}{r}
{\epsfxsize 4cm\epsffile{w124.eps}}
\end{array}
+
\begin{array}{r}
{\epsfxsize 4cm\epsffile{w125.eps}}
\end{array} \cr
&& + \begin{array}{r}
{\epsfxsize 3.1cm\epsffile{w122.eps}}
\end{array}
\cr
&=&  \Res_{q \to {\bf a}} \Res_{r \to {\bf a}} \Res_{s \to {\bf a}}  {dE_q(p) \over \omega(q)}
{dE_r(q) \over \omega(r)} {dE_s(\qbar) \over \omega(s)} B(r,\overline{r}) B(s, \overline{s})\cr
&&  +\Res_{q \to {\bf a}} \Res_{r \to {\bf a}} \Res_{s \to {\bf a}}  {dE_q(p) \over \omega(q)}
{dE_r(q) \over \omega(r)} {dE_s(\overline{r}) \over \omega(s)} B(r,\qbar) B(s,\overline{s})\cr
&&  +\Res_{q \to {\bf a}} \Res_{r \to {\bf a}} \Res_{s \to {\bf a}}  {dE_q(p) \over \omega(q)}
{dE_r(q) \over \omega(r)} {dE_s(r) \over \omega(s)} \left[ B(\qbar,\overline{r}) B(s,\overline{s})
 \right. \cr
&& \left. + B(\overline{s},\qbar) B(s,\overline{r}) + B(s,\qbar) B(\overline{s},\overline{r}) \right]\cr
&=&
2 \begin{array}{r}
{\epsfxsize 4cm\epsffile{w121.eps}}
\end{array}
 +2 
 \begin{array}{r}
{\epsfxsize 3cm\epsffile{w124.eps}}
\end{array}  
+ \begin{array}{r}
{\epsfxsize 2.8cm\epsffile{w122.eps}}
\end{array}
\eea
where the last expression is obtained using lemma \ref{lemmasymfactor}.

\subsubsection*{Free energy.}

The second order free energy reads
\bea
-2 F^{(2)}
&=&  \Res_{p \to {\bf a}} \Res_{q \to {\bf a}} \Res_{r \to {\bf a}} \Res_{s \to {\bf a}}  {\Phi(p) dE_q(p) \over \omega(q)}
{dE_r(q) \over \omega(r)} {dE_s(\qbar) \over \omega(s)} B(r,\overline{r}) B(s, \overline{s})\cr
&&  +\Res_{p \to {\bf a}} \Res_{q \to {\bf a}} \Res_{r \to {\bf a}} \Res_{s \to {\bf a}}  { \Phi(p) dE_q(p) \over \omega(q)}
{dE_r(q) \over \omega(r)} {\Phi(p) dE_s(\overline{r}) \over \omega(s)} B(r,\qbar) B(s,\overline{s})\cr
&&  +\Res_{p \to {\bf a}} \Res_{q \to {\bf a}} \Res_{r \to {\bf a}} \Res_{s \to {\bf a}}  { \Phi(p) dE_q(p) \over \omega(q)}
{dE_r(q) \over \omega(r)} {dE_s(r) \over \omega(s)} \left[ B(\qbar,\overline{r}) B(s,\overline{s})
 \right. \cr
&& \left. + B(\overline{s},\qbar) B(s,\overline{r}) + B(s,\qbar) B(\overline{s},\overline{r}) \right]\cr
\eea

\subsection{Remark: Teichmuller pant gluings}

Every Riemann surface of genus $g$ with $k$ punctures can be decomposed into 
$2g+k$ pants whose boundaries are $3g+k$ closed geodesics (in the metric with constant negative curvature).
The number of ways (in the combinatorial sense) of gluing $2g+k$ pants by their boundaries is clearly the same as the number of diagrams of ${\cal G}_{k}^{(g)}$, and each diagram corresponds to one pant decomposition.

Example with $k=1$ and $g=2$:
$$
W_1^{(2)}={\epsfxsize 2.2cm\epsffile{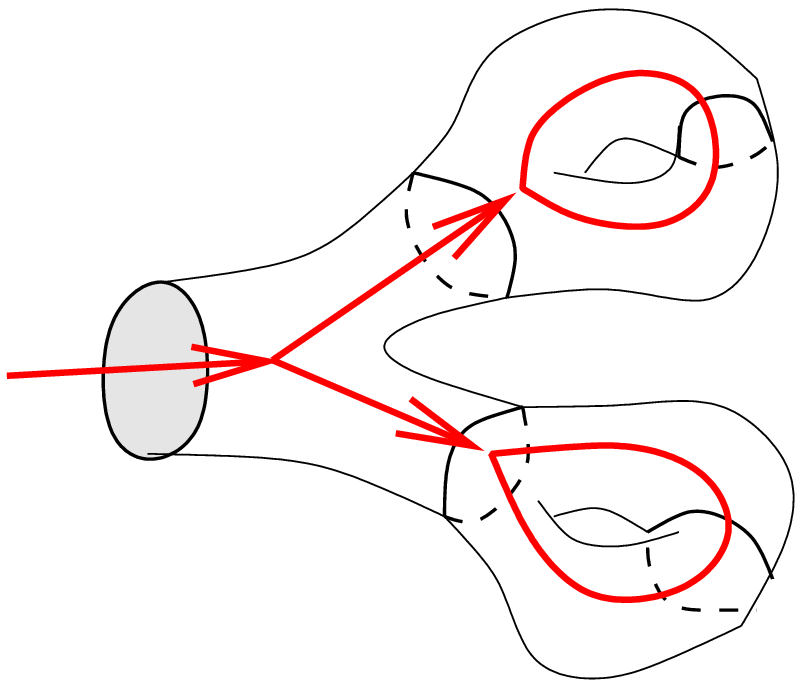}}+2 {\epsfxsize 3cm\epsffile{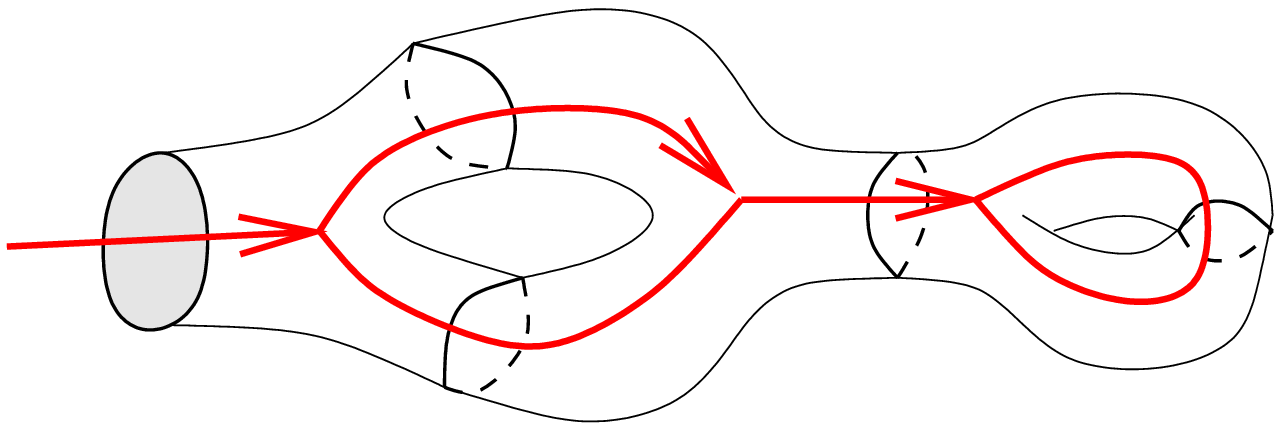}}+2 {\epsfxsize 3cm\epsffile{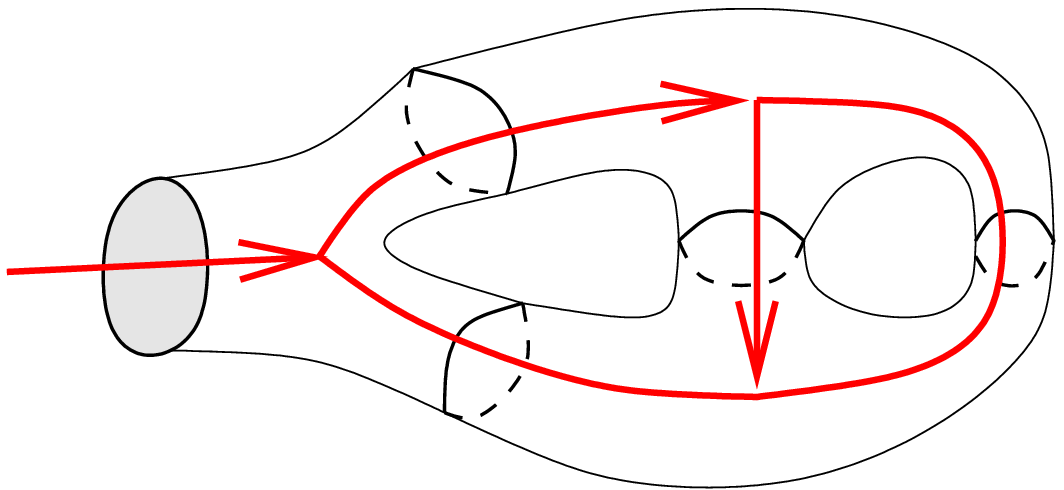}}
$$


\section{Variations of the curve}
\label{sectcompstruct}

The goal of this section is to study how the various $F^{(g)}$ and correlation functions change under the variations of moduli of the curve.


Consider an infinitesimal variation of the curve $\curve\to \curve+\delta \curve$.
It induces a variation of the function $y(x)$ at fixed $x$:
\beq
\delta_\Omega\,\, y|_x \,dx = - \Omega.
\eeq
If we use a local coordinate $z$, we may prefer to work at fixed $z$ instead of fixed $x$, we have a Poisson structure:
\beq\label{deltaydxxdy}
\delta_\Omega\, y|_z\,dx - \delta_\Omega\, x|_z\, dy = - \Omega
\eeq

The possible $\Omega$'s can be classified as first type (holomorphic), second type (residueless, and vanishing $\acycle$-cycles) and third type (only simple poles and vanishing $\acycle$-cycles), 
see \cite{Marco2} for this classification.


\subsection{Rauch variational formula}

Equation \ref{deltaydxxdy} implies that the variation of position of a branch point $a_i$ is given by:
\beq
\delta_\Omega\, x(a_i) =  {\Omega(a_i)\over dy(a_i)}.
\eeq
We assume here that ${\Omega\over dy}$ has no pole at branchpoints.
Then, Rauch variational formula \cite{Rauch, Fay} implies that the change of the Bergmann kernel is
\bea
\left. \delta_\Omega \underline{B}(p,q)\right|_{x(p),x(q)}
&=&  \sum_i {\Omega(a_i)\over dy(a_i)} \Res_{r\to a_i} {\underline{B}(r,p)\underline{B}(r,q)\over dx(r)} \cr
&=&  \sum_i \Res_{r\to a_i} {\Omega(r)\underline{B}(r,p)\underline{B}(r,q)\over dx(r)dy(r)} .
\eea
In particular after integrating over a $\underline\bcycle$-cycle we have:
\bea
\left. \delta_\Omega du(p)\right|_{x(p)}
&=&  \sum_i \Res_{r\to a_i} {\Omega(r)\underline{B}(r,p)du(r)\over dx(r)dy(r)}, \cr
\eea
and integrating again over a $\underline\bcycle$-cycle:
\bea
 \delta_\Omega \tau
&=&  2i\pi \sum_i \Res_{r\to a_i} {\Omega(r)du(r) du^t(r)\over dx(r)dy(r)} .\cr
\eea

Let us  compute the variations of the $\kappa$-modified Bergmann kernel:
\bea
\left. \delta_\Omega B(p,q)\right|_{x(p),x(q)}
&=& \left. \delta_\Omega \underline{B}(p,q)\right|_{x(p),x(q)} + 2i\pi \left. \delta_\Omega du^t(p)\right|_{x(p)} \, \kappa du(q) \cr
&& + 2i\pi du^t(p) \,\kappa\, \left. \delta_\Omega du(q)\right|_{x(q)}  \cr
&=&  \Res_{r\to \bfa} {\Omega(r)\underline{B}(r,p)\underline{B}(r,q)\over dx(r)dy(r)} \cr
&& + 2i\pi  \Res_{r\to \bfa} {\Omega(r)\underline{B}(r,p)du^t(r)\kappa du(q)\over dx(r)dy(r)} \cr
&& + 2i\pi \Res_{r\to \bfa} {\Omega(r)\underline{B}(r,q)du^t(p) \,\kappa\, du(r)\over dx(r)dy(r)}  \cr
&=&  \Res_{r\to \bfa} {\Omega(r)\underline{B}(r,p)\underline{B}(r,q)\over dx(r)dy(r)}
\cr
&& +  \Res_{r\to \bfa} {\Omega(r)\underline{B}(r,p) (B(r,q) - \underline{B}(r,q))\over dx(r)dy(r)} \cr
&& +  \Res_{r\to \bfa} {\Omega(r) (B(r,p) - \underline{B}(r,p))\underline{B}(r,q)\over dx(r)dy(r)}  \cr
&=&  \Res_{r\to \bfa} {\Omega(r) B(r,p)B(r,q)\over dx(r)dy(r)} \cr
&& +  4\pi^2 \, \Res_{r\to \bfa} {\Omega(r) du^t(p) \kappa du(r) du^t(r) \kappa du(q)\over dx(r)dy(r)}  \cr
&=&  \Res_{r\to \bfa} {\Omega(r) B(r,p)B(r,q)\over dx(r)dy(r)}  -  2i\pi \, du^t(p) \kappa \, \delta_\Omega\tau\, \kappa du(q) 
\eea
i.e.
\bea\label{eqDOmBResB}
 \left( \delta_\Omega + \tr (\kappa \,\delta_\Omega\tau \,\kappa {\partial \over \partial \kappa})\right)_{x(p),x(q)}\,\,  B(p,q)
&=&  \Res_{r\to \bfa} {\Omega(r) B(r,p)B(r,q)\over dx(r)dy(r)}  \cr
&=& - 2 \Res_{r\to \bfa} {\Omega(r) dE_r(p)B(r,q)\over \om(r)}   .\cr
\eea
We thus define the {\bf covariant variation}:
\beq
\encadremath{
D_\Omega = \delta_\Omega + \tr (\kappa \,\delta_\Omega\tau \,\kappa {\partial \over \partial \kappa})
.}
\eeq
It is more convenient to rewrite \eq{eqDOmBResB} as follows:
\bea
D_\Omega B(p,q) &=& - 2 \Res_{r\to \bfa} {\Omega(r) dE_r(p)B(r,q)\over \om(r)}   \cr
&=& - 2 \Res_{r\to \bfa} \Res_{s\to r} {\Omega(r) dE_r(p) B(s,q)\over (y(r)- y(\overline{r})) (x(s) - x(r))}   \cr
&=& 2 \Res_{r\to \bfa} \Res_{s\to \overline{r}} {\Omega(r) dE_r(p) B(s,q)\over (y(r)- y(\overline{r})) (x(s) - x(r))}   \cr
&=& 2 \Res_{r\to \bfa} {\Omega(r) dE_r(p)B(\overline{r},q)\over \om(r)}   \cr
&=& \Res_{r\to \bfa} {dE_r(p)\over \om(r)} \left[ \Omega(r) B(\overline{r},q) + \Omega(\overline{r}) B(r,q) \right]  \, . 
\eea
because now we recognize the propagator and vertex of def.\ref{defdiagrule}.
Similarly, by integrating once with respect to $q$, near a branch point $a_j$ we get:
\bea
\left. D_\Omega\, dE_{q}(p)\right|_{x(p),x(q)}  &=&  - 2 \Res_{r\to \bfa} { dE_{r}(p) \over \om(r)}\, \Omega(r) dE_{q}(r) \cr
&=& \Res_{r\to \bfa} { dE_{r}(p) \over \om(r)}\, \left[ \Omega(r) dE_{q}(\overline{r}) + \Omega(\overline{r}) dE_{q}(r) \right] \cr
\eea

Those two relations can be  depicted:
$$ \mbox{\epsfxsize=12cm{\epsfbox{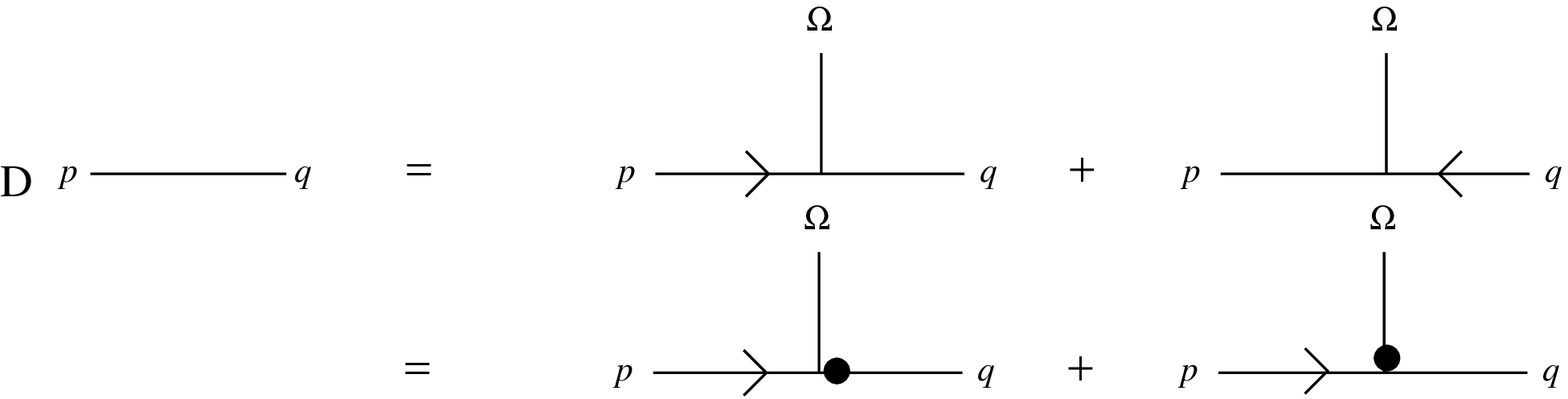}}} $$
and
$$ \mbox{\epsfxsize=12cm{\epsfbox{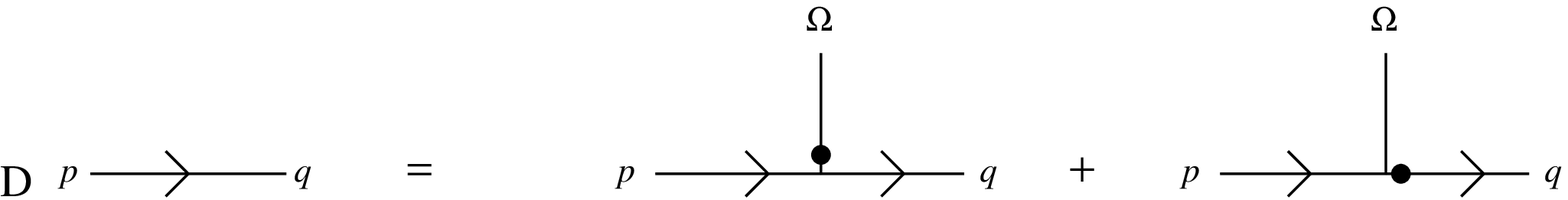}}} .$$

From this last variation, one can compute the covariant variations of the correlation functions and free energies through
the following lemma:

\bl \label{lemDOmega} For any symmetric bilinear form $f(q,p)=f(p,q)$:
\bea
D_\Omega \left( \Res_{q\to {\bf a}} {dE_{q}(p)\over \om(q)}\, f(q,\qbar) \right)_{x(p)}
&=& 2    \sum_{i,j} \Res_{r\to a_i}\Res_{q\to a_j} { dE_{r}(p) \over \om(r)}\, \Omega(r)\, {dE_{q}(r) \over \om(q)}\, f(q,\qbar) \cr
 && + \sum_j \Res_{q\to a_j} {dE_{q}(p)\over \om(q)} \,\, D_\Omega \left(f(q,\qbar)\right)_{x(q)} . \cr
\eea
\el

$$ \mbox{\epsfxsize=12cm{\epsfbox{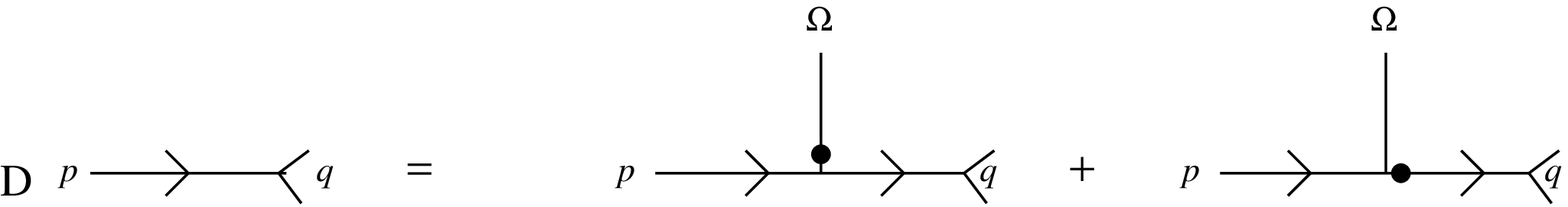}}} $$

Graphically, this means that taking the variation of a diagram just consists in adding a leg $\Omega$ in all possible edges of the graph.
In particular if $\Omega$ can be written as:
\beq
\Omega(p) = \int_{\partial\Omega} B(p,q) \Lambda(q)
\eeq
where the path ${\partial\Omega}$ does not intersect small circles around branch-points\footnote{This excludes the case where $\Omega$ corresponds to the variation of an hard edge, cf \cite{eynhardedges,  chekhovhe, Marco2}.}, then we have:

\bt \label{variat} Variations of correlation functions and free energies:
For $g+k>1$ we have
\beq\encadremath{
\left. D_\Omega W_k^{(g)}(p_1,\dots,p_k)\right|_{x(p_i)} =  \int_{\partial\Omega} W_{k+1}^{(g)}(p_1,\dots,p_k,q) \Lambda(q)
}\eeq
and,
for $g \geq 1$,
\beq\label{deltaFgintW1L}\encadremath{
D_\Omega F^{(g)} = - \int_{\partial\Omega} W_1^{(g)}(p) \L(p).
}\eeq

\et

This theorem is proved in appendix \ref{sectproofvariat}. It follows directely from lemma \ref{lemgraphaddleg} and lemma \ref{lemDOmega}.

\subsection{Loop insertion operator}

In particular for any point $q$ lying away from the branch-points, if we choose
\beq
\Omega(p) = B(p,q)
\eeq
we call $D_{B(.,q)}$ the {\bf loop insertion operator}, by analogy with matrix models \cite{ACKM,courseynard}.

It acts on the correlation functions and free energies as follows:
\bt
\beq
D_B \, W_{k}^{(g)}(p_1,\dots,p_k) = W_{k+1}^{(g)}(p_1,\dots,p_k,q)
\eeq
\beq
D_B \, F^{(g)} = - W_{1}^{(g)}(q)
\eeq
and
\beq
D_B \, F^{(0)} = y(q)dx(q) +{1 \over 4 i \pi} \left( \kappa \oint_{\bcycle}ydx \right)^t \oint_{\bcycle} \oint_{\bcycle} W_{3,0}
\kappa \oint_{\bcycle}ydx
.
\eeq
\et
Thus, the loop insertion operator, adds one leg to correlation functions.

\subsection{Variations with respect to the moduli}

Let us consider canonical variations of the curve corresponding to each moduli of the curve defined in section \ref{sectmoduliofcurve}.
We use \eq{ydxmodulitka}
\beq\label{ydxmodulitkabis}
ydx = \sum_{\alpha,k} t_{\alpha,k} B_{\alpha,k} + \sum_\alpha t_\alpha dS_{\alpha,o} + 2i\pi \sum_i \epsilon_i du_i(p)
\eeq
to identify the $\Omega$'s corresponding to varying only one modulus.

\subsubsection*{Variation of filling fractions}

Consider the variation of the curve
\beq
\Omega(p) = -2i\pi du_i(p)  = -\oint_{\bcycle_i} B(p,q).
\eeq
i.e. $\partial\Omega=\bcycle_i$ and $\L=-1$.
It is such that:
\beq
\delta_\Omega \epsilon_j =  \delta_{ij}
\virg
\delta_\Omega t_\alpha= 0
\virg
\delta_\Omega V_\alpha= 0.
\eeq
Therefore it is equivalent to varying only the filling fraction $\epsilon_i={1\over 2i\pi}\oint_{\acycle_i} y dx$:
\beq
D_{-2i\pi du_i} = {\partial \over \partial \epsilon_i}
\eeq

Theorem \ref{variat} gives
\beq
{\partial \over \partial \epsilon_i} \, W_{k}^{(g)}(p_1,\dots,p_k) = - \oint_{\bcycle_i} W_{k+1}^{(g)}(p_1,\dots,p_k,q),
\eeq
and
\beq
{\partial \over \partial \epsilon_i} \, F^{(g)} =   \oint_{\bcycle_i} W_{1}^{(g)}(q),
\eeq
and
\beq
 {\partial \over \partial \epsilon_i} \, F^{(0)}= - \oint_{\bcycle_j} ydx + {1 \over 4 i \pi} \left( \kappa  \oint_{\bcycle} ydx\right)^t \; \delta_{- 2 i \pi du_i} (\tau) \;  \kappa  \oint_{\bcycle} ydx.
\eeq

\subsubsection*{Variation of temperatures}

Let $\alpha$ and $\alpha'$ be two distinct poles of $y dx$.
Consider the variation of the curve
\beq
\Omega(p) = - dS_{\alpha,\alpha'}(p)  = \int_{\alpha}^{\alpha'} B(p,q)
\virg{\rm i.e.}\,\,\,
\partial\Omega = [\alpha,\alpha'] \,\,\, , \,\, \L=1
\eeq
It is such that:
\beq
\delta_\Omega \epsilon_j =0
\virg
\delta_\Omega t_\beta= \delta_{\alpha, \beta} - \delta_{\alpha', \beta}
\virg
\delta_\Omega V_\beta= 0.
\eeq
Therefore it is equivalent to varying only the temperatures $t_\alpha$ and $t_{\alpha'}$:
\beq
D_{-dS_{\alpha,\alpha'}} = {\partial \over \partial t_\alpha} - {\partial \over \partial t_\alpha'}
\eeq
Notice that it is impossible to vary only one temperature $t_\alpha$ since we have $\sum_\beta t_\beta=0$.

Theorem \ref{variat} gives
\beq
\left( {\partial \over \partial t_\alpha} - {\partial \over \partial t_\alpha'} \right) \, W_{k}^{(g)}(p_1,\dots,p_k) = \int_\alpha^{\alpha'} W_{k+1}^{(g)}(p_1,\dots,p_k,q),
\eeq
\beq
\left({\partial \over \partial t_\alpha} - {\partial \over \partial t_\alpha'}\right) \, F^{(g)} =  \int_{\alpha'}^{\alpha} W_{1}^{(g)}(q)
\eeq
and
\beq
\left({\partial \over \partial t_\alpha} - {\partial \over \partial t_\alpha'}\right)  F^{(o)} =  \mu_\alpha-\mu_{\alpha'} + {1 \over 4 i \pi} \left( \kappa  \oint_{\bcycle} ydx\right)^t \; \delta_{-dS_{\alpha,\alpha'}} (\tau) \;  \kappa  \oint_{\bcycle} ydx.
\eeq

\subsubsection*{Variation of the moduli of the poles}

Let $\alpha$ be a pole of $y dx$.
Consider the variation of the curve
\beq
\Omega(p) = - B_{\alpha,k}  =  \Res_\alpha B(p,q) z_\alpha^k(q),
\eeq
i.e. $\partial\Omega$ is a small circle around $\alpha$ and $\L={1\over 2i\pi}\, z_\alpha^k$.
It is such that:
\beq
\delta_\Omega \epsilon_j = 0
\virg
\delta_\Omega t_\beta= 0
\virg
\delta_\Omega t_{\beta,k'}= \delta_{\alpha,\beta} \delta_{k,k'}
\eeq
Therefore it is equivalent to varying only the coefficient $t_{\alpha,k}$:
\beq
D_{-B_{\alpha,k}} = {\partial \over \partial t_{\alpha,k}}
\eeq

Theorem \ref{variat} gives
\beq
{\partial \over \partial t_{\alpha,k}} \, W_{k}^{(g)}(p_1,\dots,p_k) =  \Res_\alpha z_\alpha^k(q) W_{k+1}^{(g)}(p_1,\dots,p_k,q),
\eeq
\beq
{\partial \over \partial t_{\alpha,k}} \, F^{(g)} = - \Res_\alpha z_\alpha^k(q) W_{1}^{(g)}(q)
\eeq
and
\beq
{\partial \over \partial t_{\alpha,k}} F^{(o)} =   \Res_{\alpha} ydx z_\alpha^k + {1 \over 4 i \pi} \left( \kappa  \oint_{\bcycle} ydx\right)^t \; \delta_{B_{\alpha,k}} (\tau) \;  \kappa  \oint_{\bcycle} ydx.
\eeq

\subsection{Homogeneity}

\bt \label{homogeneity} For $g>1$, we have the homogeneity property:
\beq
(2-2g)F^{(g)}=
\sum_{\alpha,k} t_{\alpha,k} {\partial\over \partial t_{\alpha,k}} F^{(g)}
+ \sum_{\alpha} t_{\alpha} {\partial\over t_{\alpha}} F^{(g)}
+ \sum_{i} \epsilon_i {\partial\over \partial \epsilon_i} F^{(g)}
\eeq
i.e. $F^{(g)}$ is homogeneous of degree $2-2g$.
\et
The proof is given in appendix \ref{sectproofvariat}.

\subsection{Variations of  \texorpdfstring{$F^{(0)}$}{F0} with respect to the moduli.}

In this section we compute the first and second deriatives of $F^{(0)}$ with respect to the moduli of the
curve. This paragraph is only for bookkeeping since those expressions have been known for some time \cite{Kri, Marco2,MarcoF}. 
Here we set $\kappa=0$.

\subsubsection*{First derivatives of $F^{(0)}$.}

\beq
{\partial {F}^{(0)}\over \partial t_{\alpha,k}} = \Res_\alpha z_\alpha^k \, y dx
\eeq

\beq
{\partial {F}^{(0)}\over \partial t_{\alpha,\beta}}=\left({\partial\over \partial t_{\alpha}}-{\partial\over \partial t_{\beta}}\right)  F^{(0)} = \mu_\alpha-\mu_\beta
\eeq

\beq
{\partial F^{(0)}\over \partial \epsilon_i} = - \oint_{\bcycle_i} ydx
\eeq

\bigskip

\subsubsection*{Second derivatives of $F^{(0)}$.}

\beq
{\partial^2 F^{(0)}\over \partial t_{\alpha,k} \partial t_{\beta,l} } = (\delta_{\alpha,\beta}-1)\, \Res_{p\to \alpha}\Res_{q\to \beta} z_\alpha(p)^{k} B(p,q) z_\beta(q)^{l}
\eeq
\beq
{\partial^2 F^{(0)}\over \partial t_{\alpha,k} \partial t_{\gamma,\beta}} = \Res_\alpha z_\alpha^k dS_{\gamma,\beta}
\eeq
\beq
{\partial^2 F^{(0)}\over \partial t_{\alpha,k} \partial \epsilon_i } = 2i\pi \Res_\alpha z_\alpha^{k} du_i = - \oint_{\bcycle_i} B_{\alpha,k}
\eeq
\beq
{\partial^2 F^{(0)}\over \partial \epsilon_i \partial t_{\alpha,\beta}} = 2i\pi (u_i(\beta)-u_i(\alpha))
\eeq
\beq
{\partial^2 F^{(0)}\over \partial \epsilon_i \partial \epsilon_j} = 0 \;\; (-2i\pi \tau_{ij} \; for \; \underline{F}^{(0)} )
\eeq
\beq
{\partial^2 F^{(0)}\over \partial t_{\alpha,\beta}^2} = \ln{(d\zeta_\alpha(\alpha)d\zeta_\beta(\beta)\primef(\alpha,\beta)^2)}
\eeq
\beq
{\partial^2 F^{(0)}\over \partial t_{\alpha,\beta} \partial t_{\alpha,\gamma}} = \ln{\left(d\zeta_\alpha(\alpha)\primef(\alpha,\beta)\primef(\alpha,\gamma)\over \primef(\beta,\gamma)\right)}
\eeq
\beq
{\partial^2 F^{(0)}\over \partial t_{\alpha,\beta} \partial t_{\delta,\gamma}} = \ln{\left(\primef(\delta,\beta)\primef(\alpha,\gamma)\over \primef(\alpha,\delta)\primef(\beta,\gamma)\right)}
\eeq
where $\zeta_\alpha = {1 \over z_\alpha}$ is a local coordinate around the pole $\alpha$.

\br
The definition of $F^{(0)}$ given in \eq{defF0} is nothing but the homogeneity property since it is written in terms
of the first derivatives. One can also write a formula focusing more on the second order derivatives of $F^{(0)}$:
\bea
F^{(0)}&=&  -{1\over 2}\sum_{\alpha,\beta} \Res_{p\to \alpha}\Res_{q\to \beta} V_\alpha(p) B(p,q) V_\beta(q)
+ \sum_{\alpha,\beta} t_\beta \Res_\alpha V_\alpha dS_{\beta,o}  \cr
&& \; - {1\over 2}\sum_{\alpha, \beta} t_\alpha t_\beta \ln{(\gamma_{\alpha,\beta})}
- {1\over 2}\epsilon^t \oint_{\bcycle} ydx  \cr
\eea
where 
\beq
\ln{\gamma_{\alpha,\alpha}} = -\int_\alpha^o (dS_{\alpha,o'} + {dz_\alpha\over z_\alpha}) +  \ln{(z_\alpha(o))}
\eeq
and
\beq
\ln{\gamma_{\alpha,\beta}} = \ln{\left(\primef(\alpha,\beta)\primef(o,o')\over \primef(\alpha,o')\primef(\beta,o)\right)}.
\eeq

One can notice that, in these terms,
\beq
{\partial^2 F^{(0)}\over \partial t_{\alpha,\beta}^2} = \ln(\gamma_{\alpha,\alpha}\gamma_{\beta,\beta}).
\eeq

\er

\section{Variations with respect to  \texorpdfstring{$\kappa$}{k} and modular transformations}

\subsection{Variations with respect to  \texorpdfstring{$\kappa$}{k}}\label{sectvarikappa}

We have introduced the matrix $\kappa$ in order to easily compute modular transformations of our functions. Somehow variations of $\kappa$ play the role of infinitesimal modular transformations.
Therefore it is important to compute ${\partial/\partial \kappa}$, and we will use this result in section \ref{sectmodular}.

First, notice that $W^{(g)}_k(p_1,\dots,p_k)$ is a polynomial in $\kappa$ of degree $3g+2k-3$, and $F^{(g)}$ is a polynomial in $\kappa$ of degree $3g-3$ for $g>1$ (number of propagators in a graph of ${\cal G}_k^{(g)}$).

\bt\label{thdWdkappa}
\beq\encadremath{
\begin{array}{lcl}
2i\pi {\partial \over \partial \kappa_{ij}}\,W^{(g)}_{k}({\bf p_K})
&=&  {1\over 2}\,\oint_{r\in\bcycle_j}\oint_{s\in\bcycle_i} W^{(g-1)}_{k+2}({\bf p_K},r,s) \cr
&& + {1\over 2}\,\sum_h \sum_{L\subset K} \oint_{r\in\bcycle_i} W^{(h)}_{|L|+1}({\bf p_L},r) \oint_{s\in\bcycle_j} W^{(g-h)}_{k-|L|+1}({\bf p_{K/L}},s) \cr
\end{array}
}\eeq
and in particular for $g \geq 2$:
\beq\encadremath{
- 2i\pi {\partial \over \partial \kappa_{ij}}\,F^{(g)}
=  {1\over 2}\,\oint_{r\in\bcycle_j}\oint_{s\in\bcycle_i} W^{(g-1)}_{2}(r,s) + {1\over 2}\,\sum_{h=1}^{g-1} \oint_{r\in\bcycle_i} W^{(h)}_{1}(r) \oint_{s\in\bcycle_j} W^{(g-h)}_{1}(s)
}\eeq

and
\beq
 {\partial \over \partial \kappa_{ij}}\,F^{(1)} =  {1 \over \kappa_{ji}}.
\eeq

\et
This theorem is proved in Appendix \ref{sectproofvariat}

Notice that these equations are to be compared with the Kodaira--Spencer theory \cite{ABK,ADKMV,BCOV, EMO}.

\subsection{Modular transformations} \label{sectmodular}

Consider a modular change of cycles:
\beq
\pmatrix{\underline\acycle \cr \underline\bcycle} = \pmatrix{\delta_{\acycle \acycle'} & \delta_{\acycle \bcycle'}\cr \delta_{\bcycle \acycle'} & \delta_{\bcycle \bcycle'}} \pmatrix{\underline\acycle' \cr \underline\bcycle'}
\virg
\pmatrix{\underline\acycle' \cr \underline\bcycle'} = \pmatrix{\delta_{\acycle' \acycle} & \delta_{\acycle' \bcycle}\cr \delta_{\bcycle' \acycle} & \delta_{\bcycle' \bcycle}} \pmatrix{\underline\acycle \cr \underline\bcycle}
\eeq
where $\delta_{\acycle' \acycle} = \delta_{\bcycle \bcycle'}^t$, $\delta_{\acycle' \bcycle} = - \delta_{\acycle \bcycle'}^t$,
$\delta_{\bcycle' \bcycle} = \delta_{\acycle \acycle'}^t$, $\delta_{\bcycle' \acycle} = - \delta_{\bcycle \acycle'}^t$ and the matrices $\delta_{\acycle \acycle'},\delta_{\acycle \bcycle'},\delta_{\bcycle \acycle'},\delta_{\bcycle \bcycle'}$ have integer coefficients
and satisfy $\delta_{\acycle \acycle'} \delta_{\bcycle \bcycle'}^t - \delta_{\bcycle \acycle'} \delta_{\acycle \bcycle'}^t = Id$.

Under this transformation of the cycle homology basis, the Abel map and the matrix of period change like:
\beq
du' = {\cal{J}} du \virg du = {\cal{J}}^{-1} du'
\eeq
with ${\cal{J}} = 
(\delta_{\acycle \acycle'}^t+\tau' \delta_{\acycle \bcycle'}^t) = (\delta_{\bcycle \bcycle'}-\tau \delta_{\acycle \bcycle'})^{-1}$
and
\beq
\tau' = (\delta_{\bcycle \bcycle'}-\tau \delta_{\acycle \bcycle'})^{-1} (-\delta_{\bcycle \acycle'}+\tau \delta_{\acycle \acycle'})
\virg
\tau = (\delta_{\acycle \acycle'}^t+\tau' \delta_{\acycle \bcycle'}^t)^{-1}\,(\delta_{\bcycle \acycle'}^t+\tau' \delta_{\bcycle \bcycle'}^t).
\eeq

Let us define the following symmetric matrix:
\beq
\widehat{\kappa}=\widehat{\kappa}^t=(\delta_{\bcycle \bcycle'} \delta_{\acycle \bcycle'}^{-1}-\tau )^{-1}
= 
\delta_{\acycle \bcycle'} {\cal{J}}
\eeq
it is such that the Bergmann kernel changes like:
\beq\label{modularchB}
\underline{B}' = \underline{B} + 2i\pi\, du^t \widehat{\kappa} du .
\eeq
The $\kappa$-modified Bergmann kernel changes like:
\bea
B'(p,q) &=& \underline{B}(p,q) + 2 i \pi \left[ du'^t(p) \kappa du'(q) + du^t(p) \widehat{\kappa} du(q) \right] \cr
&=& \underline{B}(p,q) + 2 i \pi du^t(p) \left( \widehat{\kappa} + {\cal{J}}^t
\kappa {\cal{J}} \right) du(q).
\eea

In other words, the effect of a modular change of cycles is equivalent to a change $\kappa\to \widehat{\kappa} + {\cal{J}}^t \kappa {\cal{J}} $ in the definition of the kernel $B(p,q)$.

Thus, the modular variations of the free energies satisfy the following theorem:
\bt
\label{thmodular}
For $g\geq 2$ the modular transformation of the free energies $F^{(g)}$ consists in changing $\kappa$ to
$\widehat{\kappa} + {\cal{J}}^t \kappa {\cal{J}}$ in the definitions of the modified Bergmann kernel and Abelian differential.

The first correction $F^{(1)}$ changes like:
\beq
F^{(1)'} = F^{(1)} - {1\over 2} \ln{(\delta_{\bcycle \bcycle'} -\tau \delta_{\acycle \bcycle'})}.
\eeq

An equivalent way of saying the same thing, is that:
if we change the basis of cycles and change the matrix $\kappa\to {\cal J}^{t\,-1}(\kappa-\widehat\kappa){\cal J}^{-1}$, then the $F^{(g)}$'s are unchanged for $g>1$.
\et
\proof{
The result for $g\geq 2$ comes directly from the variation of the Bergmann kernel.

$F^{(1)}$ depends on the cycles only through the Bergmann tau function.
Since,
\bea
{\d \ln{(\tau_{B'x})}\over \d x(a_i)} - {\d \ln{(\tau_{Bx})}\over \d x(a_i)}
&=&  \Res_{p\to a_i} {\underline{B}'(p,\pbar)-\underline{B}(p,\pbar)\over dx(p)} \cr
&=& 2i\pi\, \Res_{p\to a_i} {du^t(p) \widehat\kappa du(\pbar)\over dx(p)} \cr
&=& - 2i\pi\, \Res_{p\to a_i} {du^t(p) \widehat\kappa du(p)\over dx(p)} \cr
&=& - \Tr \kappa {\d \tau \over \d x(a_i)} \cr
&=& - \Tr (\delta_{\bcycle \bcycle'}-\tau \delta_{\acycle \bcycle'})^{-1} {\d \tau \delta_{\acycle \bcycle'} \over \d x(a_i)} \cr
&=& {\d  \ln{\det (\delta_{\bcycle \bcycle'}-\tau \delta_{\acycle \bcycle'})}  \over \d x(a_i)} \cr
\eea
and this characterizes the Bergmann tau function totally (up to a general multiplicative factor),
one obtains the second result of the theorem.
}

\br
The transformation of leading order $F^{(0)}$ is more complicated and its computation is more involved
as the final result depends explicitely on the position of the poles $\alpha$ in the fundamental domain.
Let us just mention that it depends on all the parameters of the modular transformation explicitely.
\er

\bt
If one chooses $\kappa={i\over 2\, \Im\tau}$, then $F^{(g)}(\kappa)$ is modular invariant.
\et

\proof{

It is well known that for that value of $\kappa$, the modified Bergmann kernel is the Schiffer kernel and is modular invariant.
Indeed, it is easy to check that if $\kappa={i\over 2\,\Im\tau}$ one has
$\hat\kappa+{\cal J}^t \kappa {\cal J} = {i\over 2\,\Im\tau'}$.

Since the only modular dependence of $F^{(g)}$ for $g\geq 1$ is in the Bergmann kernel, this proves the modular invariance of the $F^{(g)}$'s.

}

\section{Symplectic invariance}

The following theorem is mostly the reason why we call $F^{(g)}$'s invariants of the curve.
This theorem seems to be rather important and it has beautiful applications as we will see in section \ref{sectkontse}.

\bt
\label{symplinv}
The following transformations of $\curve$ leave the $F^{(g)}$'s unchanged:

$\bullet$ \vspace{0.3mm} $x\to {ax+b\over cx+d}$ and $y\to {(cx+d)^2\over ad-bc} y$.

$\bullet$ \vspace{0.3mm} $y\to y +R(x)$ where $R$ is any rational function.

$\bullet$ \vspace{0.3mm} $y\to  y$ and $x\to -x$.

$\bullet$ \vspace{0.3mm} $y\to x$ and $x\to y$.

\et
Notice that these are transformations which conserve the symplectic  form
\beq|
dx\wedge dy|
\eeq
In particular we have the $PSL_2(\C)$ invariance:
\beq
\pmatrix{x\cr y} \longrightarrow \pmatrix{a & b \cr c & d}\,\pmatrix{x\cr y} \qquad\virg (ad-bc)^2=1
\eeq

That symplectic invariance seems to be a very powerful tool to see if different matrix models have the same topological expansion. For example, in section \ref{sectkontse} we show how symplectic invariance can be used to provide a new and very easy proof of some properties of the Kontsevitch integral.

\medskip

\proof{
Invariance under the first 2 transformations is obvious from the definitions, because the only $x$ and $y$ dependance of the $W_k^{(g)}$'s is in $\om(q)=(y(q)-y(\qbar))dx(q)$ which is clearly unchanged under the first 2 transformations (notice that the transformation $x\to {ax+b\over cx+d}$ conserves the branchpoints). In fact the first 2 transformations leave $W_k^{(g)}$ unchanged.

In the 3rd transformation the only thing which changes is the sign of $\om$, and it is easy to see that $W_k^{(g)}$ is multiplied by $(-1)^{2g-2+k}=(-1)^k$, and $F^{(g)}$ is multiplied by $(-1)^{2g-2}=1$.

The 4th transformation is the difficult one. 
The proof consists in building some ''mixed correlation functions'', and seeing that their definition by inverting the roles of $x$ and $y$ lead to the same objetcs.
Since it is long and involves new results in the framework of matrix model, it is written in a separate paper
\cite{EOSym}.

The case of $F^{(0)}$ and $F^{(1)}$ are done separately in appendix \ref{appsymplinvF0F1}.
}

\section{Singular limits}

\label{sectsingular}

Consider a 1-parameter family of algebraic curves:
\beq
\curve(x,y,t)
\eeq
such that the curve at $t=0$ has a singular branchpoint $a$ with a $p/q$ rational singularity,
i.e. in some local coordinate $z$ near $a$ we have:
\beq
\left\{\begin{array}{l}
t=0\cr
x(z) \sim x(a) + (z-a)^q \cr
y(z) \sim y(a) + (z-a)^p \cr
\end{array}\right.
\eeq

At $t\neq 0$, the singularity is smoothed, and we have (the local parameter is now $\zeta = z\, t^{-\nu}$):
\beq
\left\{\begin{array}{l}
x(z,t) \sim x(a) + t^{q\nu}\,\,Q(\zeta) + o(t^{q\nu}) \cr
y(z,t) \sim y(a) + t^{p\nu}\,\,P(\zeta) +o(t^{p\nu}) \cr
\end{array}\right.
\eeq
where $Q$ is  a polynomial of degree $q$ and $P$ is a polynomial of degree $p$, and where $\nu$ is some exponent which depends on the choice of the parameter $t$.

The curve
\beq
\curve_{\rm sing}(\xi,\eta)=\left\{\begin{array}{l}
\xi(\zeta) = Q(\zeta)  \cr
\eta(\zeta) = P(\zeta)  \cr
\end{array}\right.
= {\rm Resultant}(Q-\xi,P-\eta)
\eeq
is called the singular spectral curve.

One observes that $F^{(g)}(\curve)$ is singular in the small $t$ limit, and it behaves like
\beq
F^{(g)}(\curve(t)) \sim t^{\gamma_g} F^{(g)}_{\rm sing} + o(t^{\gamma_g}) \qquad ,\,\, {\rm for}\, g\geq 2
\eeq
\beq
F^{(1)}(\curve(t)) \sim -{1\over 24}\,(p-1)(q-1)\nu\,\ln{(t)}  + O(1) \qquad ,\,\, {\rm for}\, g=1
\eeq
$F^{(g)}_{\rm sing}$ is called the double scaling limit of $F^{(g)}$.
The exponent $\gamma_g$ and $F^{(g)}_{\rm sing}$ are given by the following theorem:

\bt\label{thsinglimit} Singular limits:
\beq\encadremath{
F^{(g)}_{\rm sing}(\curve) = F^{(g)}(\curve_{\rm sing}) \qquad ,\,\, {\rm for}\, g\geq 2
}\eeq
and
\beq
\gamma_g = (2-2g)(p+q)\nu .
\eeq
In other words, our construction of $F^{(g)}$ commutes with the singular limit.

\et

\proof{
It is easy to see that the most singular term in the limit of the Bergmann kernel in that regime, behaves like $t^{0}$, and thus $dE_{z'}(z)$ as well.
The denominator $((y(z)-y(\overline{z}))dx(z)$ in the recursion behaves like $t^{\nu(p+q)} (P(\zeta)-P(\ovl\zeta))Q'(\zeta)d\zeta$,
and by recursion on $k$ and $g$, we easily see that:
\beq
W_{k}^{(g)}(z_1,\dots,z_k) \sim t^{(2-2g-k)(p+q)\nu}\,\, \om_k^{(g)}(\zeta_1,\dots,\zeta_k)
\eeq
if all $z_i$'s are close to $a$, and is subdominant if some $z_i$'s are not in the vicinity of $a$.
The leading contribution to $W_{k}^{(g)}$ is thus obtained by taking $z'$ and $\overline{z}'$ in the vicinity of $a$ only in eq.\ref{defWkgrecursive},
i.e. $\om_k^{(g)}(\zeta_1,\dots,\zeta_k)$ obey the same recursion formula as eq.\ref{defWkgrecursive}, with the curve $\curve_{\rm sing}$.
The same holds for the free energy.

}

\section{Integrability}
\label{sectintegrability}

Here, we prove that $Z$ is a tau-function, because it satisfies some Hirota equation.

\subsection{Baker-Akhiezer function}

Given two points $\xi$ and $\eta$ in the fundamental domain, we define the following kernel as a formal series in $1/N$:
\bea
K_N(\xi,\eta) &=& { \ee{- N\int_\eta^\xi y dx}\over E(\xi,\eta)\,\sqrt{dx(\xi)dx(\eta)}} \cr
&& \; \exp{\left( - \sum_{g=0}^\infty \sum_{l=1, 2-2g-l<0}^\infty {1\over l!}\,\,N^{2-2g-l}\,\,\int_\eta^\xi\int_\eta^\xi\dots\int_\eta^\xi W^{(g)}_l(p_1,\dots,p_l) \right)}
\cr
\eea
where the integration path lies in the fundamental domain.

This kernel has the following properties:
\begin{itemize}

\item Notice that $(x(\xi)-x(\eta))K_N(\xi,\eta) \to 1$ when $\eta\to\xi$.

\item We have:
\beq
\mathop{{\rm lim}}_{\eta\to\xi}\, \Big( K_N(\xi,\eta) - {1\over (x(\xi)-x(\eta))} \Big) = -Ny(\xi) + {W_1(\xi)\over dx(\xi)} 
\eeq
where $W_1=\sum_{g=1}^\infty N^{1-2g} W_1^{(g)}$.

\item we have:
\beq
K_N(\xi,\eta) = K_{-N}(\eta,\xi)
\eeq


\item One may think that $K_N$ is singular at branchpoints because $\ln{K_N}$ has poles, however, using the singular limit theorem \ref{thsinglimit}, we see that the leading behavior of all $W_l^{(g)}$'s is given by the $W_l^{(g)}$'s of the Airy curve $y=\sqrt{x}$ described in section \ref{sectairy}.
Therefore, near a branchpoint $a$, when $\xi,\eta \to a$, to leading order $K_N$ is the Tracy-Widom kernel \cite{TWlaw}:
\bea
K_N(\xi,\eta) \sim {Ai(\hat\xi)Ai'(\hat\eta)-Ai'(\hat\xi)Ai(\hat\eta)\over \hat\xi-\hat\eta} \cr
\hat\xi=N^{2/3}(x(\xi)-x(a)) \virg \hat\eta=N^{2/3}(x(\eta)-x(a))
\eea
In other words, $K_N$ is not singular near branchpoints.

\item The only singularities of $K_N(\xi,\eta)$ are essential singularities at all the poles of $ydx$, with a singular part equal to $\exp{(-N\int^\xi_\eta ydx)}$, as well as a simple pole at $\xi=\eta$.

\end{itemize}

\bigskip

Then, given a pole $\alpha$ of $ydx$, we define for $\xi$ in the vicinity of $\alpha$:
\bea
\psi_{\alpha,N}(\xi) &=& {\ee{-N\, V_\alpha(\xi)} \ee{-N\int_\alpha^\xi (ydx-d V_\alpha+t_\alpha {dz_\alpha\over z_\alpha})} \over E(\xi,\alpha)\,\sqrt{dx(\xi) d\zeta_\alpha(\alpha)}}\,\left(z_\alpha(\xi)\right)^{N t_\alpha}  \cr
&& \exp{\left( - \sum_{g=0}^\infty \sum_{l=1, 2-2g-l<0}^\infty {1\over l!}\,\,N^{2-2g-l}\,\,\int_\alpha^\xi\int_\alpha^\xi\dots\int_\alpha^\xi W^{(g)}_l(p_1,\dots,p_l)\right)} \cr
&=&\mathop{{\rm lim}}_{\eta\to\alpha} \left( K(\xi,\eta)\,\sqrt{dx(\eta)\over d\zeta_\alpha(\eta)}\,\ee{-NV_\alpha(\eta)}\,\left(z_\alpha(\eta)\right)^{Nt_\alpha}\,\right)
\eea
where $\zeta_\alpha$ is the local parameter near $\alpha$,
$\zeta_\alpha={1\over z_\alpha}$,
and
$
\phi_{\alpha,N}(\xi) = \psi_{\alpha,-N}(\xi) 
$.

They have the following properties:
\begin{itemize}


\item $\psi_{\alpha,N}$ was defined only in the vicinity of $\alpha$, but it can be easily analytically continued to the whole curve, by choosing an arbitrary point $o$ in the vicinity of $\alpha$ and writing:
\bea
&& \int_{\alpha}^\xi (ydx-dV_\alpha+t_\alpha {dz_\alpha\over z_\alpha}) + V_\alpha(\xi) - t_\alpha\ln{(z_\alpha(\xi))} \cr
&=&
\int_o^\xi ydx + \int_{\alpha}^o (ydx-dV_\alpha+t_\alpha {dz_\alpha\over z_\alpha}) + V_\alpha(o) - t_\alpha\ln{(z_\alpha(o))}
\eea

\item Using the singular limit theorem \ref{thsinglimit} near branchpoints, we see that the leading behavior of all $W_l^{(g)}$'s is given by the $W_l^{(g)}$'s of the Airy curve $y=\sqrt{x}$ described in section \ref{sectairy}.
Therefore, near a branchpoint $a$, when $\xi,\eta\to a$ we have:
\bea
\psi_{\alpha,N}(\xi) \sim  C\, Ai(\hat\xi) \virg \hat\xi=N^{2/3}(x(\xi)-x(a)) 
\eea
where $C$ is some normalization constant ($C={\psi_{\alpha,N}(a)/ Ai(0)}$).
In other words, $\psi_{\alpha,N}$ is not singular near branchpoints.

\item The only singularities of $\psi_{\alpha,N}$ are essential singularities at all the poles of $ydx$, with a singular part equal to $\exp{(-N\int^\xi ydx)}$.

\end{itemize}

This is why we call those formal functions Baker-Akhiezer functions (cf \cite{BBT}).

\br
In fact, those functions are exactly Baker-Akhiezer functions only when the curve has genus $\genus=0$.
In the general case, the Baker-Akhiezer functions must also have the property that they take the same value after going around a non-trivial cycle.
It is not difficult to multiply $\psi_{\alpha,N}$ by the appropriate $\theta$ function in order to fulfill that property. However, if we do that, we destroy the $1/N^2$ expansion.

This is why the $\psi_{\alpha,N}$ defined above can be called a "formal Baker-Akhiezer function".
This definition is sufficient to find a formal Hirota equation, valid only order by order in $1/N^2$.

\er

\subsection{Sato relation}

Given two points $\xi$ and $\eta$ on $\overline{\Sigma}$, and a complex number $r$, we define the curve:
\beq
\curve + r[\xi,-\eta] = \left\{ \left(x(p),y(p)+r{dS_{\xi,\eta}(p)\over dx(p)}\right)\quad , \,\, p\in \Sigma  \right\}
\eeq
The differential $ydx + rdS_{\xi,\eta}$ has  the same $\acycle$-contour integrals as $ydx$, the same poles with the same singular part, plus two additional simple poles, one located at $p=\xi$ with residue $r$,
and one at $p=\eta$ with residue $-r$.

We have Sato's relation:
\bt\label{thSato}
\beq
K_N(\xi,\eta) = {Z_N(\curve+{1\over N}[\xi,-\eta])\over Z_N(\curve)}
\virg
\psi_{\alpha,N}(\xi) = {Z_N(\curve+{1\over N}[\xi,\alpha])\over Z_N(\curve)}
\eeq
\et
Indeed, the definition of $K_N$ is the formal Taylor expansion in powers of $r=1/N$ of the RHS.

\subsection{Hirota equation}

Consider two algebraic curve $\curve(x,y)$ and $\td\curve(x,y)$
 with the same conformal structure (i.e. the same compact Riemann surface $\overline\Sigma$), 
then we have an Hirota bilinear relation:

\bt\label{thHirota}
We have the bilinear relation:
\beq
\Res_{\eta\to\zeta} dx(\eta)\,K_N(\xi,\eta)\,\td{K}_{\td{N}}(\eta,\zeta) = K_N(\xi,\zeta)
\eeq
and also:
\beq\label{hirotapsi}
\Res_{\xi\to \alpha} dx(\xi)\,\psi_{\alpha,N}(\xi)\,\td{\psi}_{\alpha,-\td{N}}(\xi) = 0 \qquad \qquad {\rm if}\,\, \td{N}\td{t}_\alpha >  N t_\alpha+1
\eeq
which takes exactly the form of the Hirota equation \cite{BBT, kostovhirota}.
\et

It is important to notice that this Hirota equation makes sense only order by order in its $1/N^2$ expansion. In the way we have obtained it, it is meaningless for finite $N$ (appart maybe from the genus zero case $\genus=0$, under the condition that the $1/N^2$ series is convergent).
Therefore, we have a "formal Hirota equation".


\section{Application: topological expansion of matrix models}
\label{sectmatrixmodel}

In this section, we show how the objects defined in section.\ref{sectdefspfree} (i.e. $\kappa=0$) correspond to the terms of the topological expansion of the free energy and correlation functions of various matrix models when one considers appropriate curves $\curve(x,y)$. Notice that in all this section we consider $\kappa=0$.

\subsection{Formal 1-matrix model} \label{sect1MM}

The formal 1-matrix model (cf \cite{eynform}), is known to be the generating function which enumerates maps of given topology since the work of \cite{BIPZ}, then \cite{ambjornrmt, davidRMT, KazakovRMT}. Its topological expansion was computed in several steps.
The authors of \cite{ACKM} introduced a recursive algorithm to compute the $F^{(g)}$'s, only in the 1-cut case (i.e. $\genus=0$), and then the method was extended to other cases \cite{Ak96, AkAm}.
The computation of the subleading term $F^{(1)}$ was done in general by Chekhov \cite{Chekh}.
The computation to all orders of the correlation functions was found in \cite{eynloop1mat}, and the free energies in \cite{ec1loopF}.

\subsubsection{Definition}

\bd {\bf Formal  1-matrix model.}

Consider a ''semi-classical'' potential $V(x)$, i.e. such that $V'(x)$ is a rational function.
Let $D(x)$ be its denominator, i.e. $D(x)V'(x)$ is a polynomial of degree $d$, and let $X_1,\dots,X_d$ be its zeroes:
\beq
D(x)V'(x) = \prod_{i=1}^d (x-X_i)
\eeq
We write:
\beq
\delta V_i(x) = V(x) - V(X_i) - {1\over 2} V''(X_i) (x-X_i)^2
\eeq

Choose an integer $n$, and a $d-$partition of $n$, $\vec{n}=\{n_1,\dots,n_{d}\}$, such that
\beq
\sum_{j=1}^d n_j=n.
\eeq

The following gaussian integral (where each matrix $M_i$ is of size $n_i$) is a polynomial in $T$ of the form:
\bea
&& {(-1)^l n^l\over l!\, T^l}\,e^{-{n \over T} \sum_i n_i V(X_i)} \;\int dM_1 \dots dM_d \,\, \prod_{i=1}^{d} \ee{-{n V''(X_i)\over 2T}\Tr (M_i-X_i\,{\bf 1}_{n_i})^2} \cr
&& \qquad \,\,\prod_{i>j} \det(M_i\otimes {\bf 1}_{n_j} - {\bf 1}_{n_i} \otimes M_j)^2 \,\,(\sum_i \Tr \delta V_i(M_i))^l \cr
&=& \sum_{k=l/2}^{dl/2} A_{k,l} T^k
\eea
We define the formal matrix integral as the formal power series in $T$:
\beq
Z_{\rm 1MM} = \sum_{k=0}^\infty T^k \big( \sum_{j=0}^{2k} A_{k,j} \big).
\eeq
One can also define its formal logarithm, i.e. the free energy
\beq
F_{\rm 1MM}=- \ln{Z_{\rm 1MM}} = \sum_{k=0}^\infty T^k B_k
\eeq

It is a standard computation discovered by 't Hooft (\cite{thooft, eynform}), that, for fixed $\epsilon_i={T n_i\over n}$, for every $k$, $n^{-2} B_{k}$ is a polynomial in $1/n^2$:
\beq
B_k(n_1,\dots,n_d) = \sum_{g=0}^{g_{\rm max}(k)} B_{k,g}(\epsilon_1,\dots,\epsilon_d)\, \left(n\over T\right)^{2-2g}
\eeq
Thus we define the following formal power series in $T$:
\beq
F_{\rm 1MM}^{(g)}= \sum_{k=0}^\infty T^k B_{k,g}(\epsilon_1,\dots,\epsilon_d)
\eeq

\ed

{\bf Remark:} Here, the question of convergence of those series is not relevant.
It is well known that each $F_{\rm 1MM}^{(g)}$ is a convergent series (because it is written in terms of algebraic functions of $T$), but $F_{\rm 1MM}$ is not.

\subsubsection{Loop equations and classical spectral curve}

It is easy to see (this property holds for each power of $T$ because it holds for gaussian integrals) that the formal matrix integral satisfies, order by order in powers of $T$, the loop equations (i.e. Virasoro constraints), which can be written (see \cite{ZJDFG,Virasoro}):
\beq
y(x)^2 + {T^2\over n^2}\om_2(x,x) = {V'(x)^2\over 4} - {T\over n}\left<\Tr {V'(x)-V'(M)\over x-M}\right>
\eeq
where $y(x)={V'(x)\over 2}-{T\over n}\left<\Tr{1\over x-M}\right>$ and $\om_2(x,x')=\left<\Tr{1\over x-M}\,\Tr{1\over x'-M}\right>_c$.
where the expectation value $<.>$ is defined in a formal way similar to $F$.

If one identifies the coefficients of $n^0$ in each side,
 one gets an algebraic equation (here hyperelliptical), which is called the "classical spectral curve":
\beq
\curve_{\rm 1MM}(x,y) = D(x)^2\,(y^2 - {1\over 4}V'^2(x)+P(x))
\eeq
where $D(x)P(x)$ is a polynomial of degree at most $\deg(D(x)V'(x))-1$, and completely determined by the condition that the polynomial $P(x)$ is a formal power series in powers of $T$ such that at $T=0$:
\beq
P(x,T=0) = \sum_{i=1}^d \epsilon_i {V'(x)\over x-X_i}
\eeq
It is such that there exist some contours $\acycle_i$, $i=1,\dots,d$ such that:
\beq
{1\over 2i\pi}\oint_{\acycle_i} y dx = \epsilon_i
\eeq
Notice that the genus $\genus$ of the curve $\curve_{\rm 1MM}$ is the number of non vanishing $\epsilon_i$'s minus one.

Most often in the litterature, $V$ is chosen polynomial such that $V'(0)=0$, and only the 1-cut case is considered, with only one non-vanishing filling fraction at $X=0$. The resulting curve has genus $\genus=0$. This is the case which is relevant for enumerating polygonal surfaces.

\bigskip

It was proved in \cite{eynloop1mat, ec1loopF}, in the case of polynomial potentials only (but it is clear that it can be extended to the semiclassical case), that one has:

\bt
\beq
\encadremath{F^{(g)}_{\rm 1MM} = \underline{F}^{(g)}(\curve_{\rm 1MM})}
\eeq
\et

This proves that $F^{(g)}$ (which we recall is a formal power series in $T$) genericaly has a finite radius of convergence $T<T_c$.

We also have:

\bea
\left<\Tr{1\over x(p_1)-M}\,\dots\,\Tr{1\over x(p_k)-M}\right>_{\rm c}
&=& \sum_{g=0}^{\infty} N^{2-k-2g} {\underline{W}_k^{(g)}(p_1,\dots,p_k)\over dx(p_1)\dots dx(p_k)} \cr
&& + \delta_{k,1}\, N({1\over 2}V'(x(p_1))-y(p_1)) \cr
&& - {\delta_{k,2} \over (x(p_1)-x(p_2))^2} \cr
\eea

\subsection{2-matrix model}

The formal 2-matrix model is known to be the generating function which counts bicolored maps (let us 
say the 2 colors are + or -, thus it is an Ising model on a random map), it was introduced by Kazakov \cite{Kazakov}. The loop equations were first written in \cite{staudacher}.
$F^{(0)}$ was computed in \cite{Kri, MarcoF, Marco2, eynmultimat}.
$F^{(1)}$ was first found in \cite{eynm2m} for the $\genus=0$ case, the in \cite{eynm2mg1} for $\genus=1$, then in \cite{EKK} for arbitrary $\genus$.
Then higher orders for the correlation functions were first derived in \cite{eyno}, and the $F^{(g)}$'s for $g\geq 2$ were first found in \cite{CEO}.
During the same time it became clear that matrix models topological expansion was closely related to algebraic geometry \cite{KazMar, DV, DW, DReview}.

\subsubsection{Definition}

\bd
Similarly, consider $V'_1$ and $V'_2$ two rational functions with respective denominators $D_1(x)$ and $D_2(x)$.
The equation:
\beq
\left\{
\begin{array}{l}
V'_1(X_i)=Y_i\cr
V'_2(Y_i)=X_i
\end{array}\right. \qquad \quad i=1,\dots, d
\eeq
has $d= \deg(V'_1 D_1) * \deg(V'_2 D_2)$ solutions.
We then write:
\beq
\delta V_{1,i}(x) = V_1(x) - V_1(X_i) - Y_i (x-X_i) - {V_1''(X_i)\over 2}(x-X_i)^2
\eeq
et
\beq
\delta V_{2,i}(y) = V_2(y) - V_2(Y_i) - X_i (y-Y_i) - {V_2''(Y_i)\over 2}(y-Y_i)^2
\eeq

We then choose an integer $n$, and a $d-$partition of $n$:
\beq
n=\sum_{i=1}^d n_i .
\eeq

The following gaussian integral (where each matrix $M_i$ or $\td{M}_i$ is of size $n_i$) is a polynomial in $T$ of the form:
\bea
&& {(-1)^l n^l\over l!\, T^l}\, e^{-{n \over T} \sum_i n_i \Tr\left( V_1(X_i) + V_2(Y_i) - X_i Y_i\right)}
\int dM_1 \dots dM_d d\td{M}_1 \dots d\td{M}_d \cr
&& \,\, \prod_{i=1}^{d} \ee{-{n \over T}\Tr \left( {V_1''(X_i)\over 2}(M_i-X_i\,{\bf 1}_{n_i})^2+{V_2''(Y_i)\over 2}(\td{M}_i-Y_i\,{\bf 1}_{n_i})^2 - (M_i-X_i\,{\bf 1}_{n_i})(\td{M}_i-Y_i\,{\bf 1}_{n_i})\right)} \cr
&& \qquad \,\,\prod_{i>j} \det(M_i\otimes {\bf 1}_{n_j} - {\bf 1}_{n_i} \otimes M_j) 
\,\,\prod_{i>j} \det(\td{M}_i\otimes {\bf 1}_{n_j} - {\bf 1}_{n_i} \otimes \td{M}_j) \cr
&& \,\,(\sum_i \Tr \delta V_{1,i}(M_i)+ \delta V_{2,i}(\td{M}_i))^l \cr
&=& \sum_{k=l/2}^{dl/2} A_{k,l} T^k
\eea
Similarly to the 1-matrix case, we can define the formal 2-matrix model as a formal power series in powers of $T$ (see \cite{eynform}):
\beq
Z_{\rm 2MM} = \sum_{k=0}^\infty T^k \big( \sum_{j=0}^{2k} A_{k,j} \big).
\eeq
One can also define its formal logarithm, i.e. the free energy
\beq
F_{\rm 2MM}=- \ln{Z_{\rm 2MM}} = \sum_{k=0}^\infty T^k B_k
\eeq

Again, it is a standard computation (\cite{thooft, eynform}), that, for fixed $\epsilon_i={T n_i\over n}$, for every $k$, $n^{-2} B_{k}$ is a polynomial in $1/n^2$:
\beq
B_k(n_1,\dots,n_d) = \sum_{g=0}^{g_{\rm max}(k)} B_{k,g}(\epsilon_1,\dots,\epsilon_d)\, \left(n\over T\right)^{2-2g}
\eeq
Thus we define the following formal power series in $T$:
\beq
F_{\rm 2MM}^{(g)}= \sum_{k=0}^\infty T^k B_{k,g}(\epsilon_1,\dots,\epsilon_d)
\eeq

\ed

\subsubsection{Loop equations and classical spectral curve}

Again, the formal 2-matrix integral satisfies, order by order in powers of $T$, the loop equations (i.e. $W$-algebra constraints), which can be written (see \cite{ZJDFG,Virasoro}):
\beq
(y-y(x)) U(x,y) + {T^2\over n^2} U(x,y;x) = E(x,y)
\eeq
where $y(x)= V_1'(x)-{T\over n}\left<\Tr{1\over x-M_1}\right>$, $U(x,y)= x-V_2'(y)+{T\over n}\left<\Tr{1\over x-M_1}\,{V'_2(y)-V'_2(M_2)\over y-M_2}\right>$ and $U(x,y;x')= \left<\Tr{1\over x-M_1}\,{V'_2(y)-V'_2(M_2)\over y-M_2}\,\Tr{1\over x'-M_1}\right>_{c}$, and $E(x,y)=(y-V'_1(x))(x-V_2'(y))-{T\over n}\left<\Tr{V'_1(x)-V'_1(M_1)\over x-M_1}\,{V'_2(y)-V'_2(M_2)\over y-M_2}\right>+T$,
and where the expectation value $<.>$ is defined in a formal way similar to $F$.

If one chooses $y=y(x)$ and identifies the coefficients of $n^0$ in each side,
 one gets an algebraic equation $\curve_{\rm 2MM}(x,y)=0$, which is called the "classical spectral curve" \cite{eynm2m,eynm2mg1}:
\beq
\curve_{\rm 2MM}(x,y) = D_1(x)\,D_2(y)\,((V'_1(x)-y)(V'_2(y)-x)-P(x,y)+T)
\eeq
where $D_1(x)\,D_2(y)P(x,y)$ is a polynomial of degree $\leq \deg(D_1 V'_1)$ in $x$ and a polynomial of degree $\leq \deg(D_2 V'_2)$ in $y$, and 
with fixed filling fractions:
\beq
{1\over 2i\pi}\oint_{\acycle_I} y dx = T{n_I\over N}= \epsilon_I \virg \sum_I \epsilon_I = T
\eeq
and such that in the limit $T\to 0$ and $\forall I\,\,\epsilon_{I}\to 0$, one has:
\beq
y \sim V'_1(x) - \sum_i {\epsilon_{i}\over x-X_i} + O(T^2)
\virg
x \sim V'_2(y) - \sum_i {\epsilon_{i}\over y-Y_i} + O(T^2).
\eeq

Then one has:
\bt
\beq\encadremath{
F^{(g)}_{\rm 2MM} = \underline{F}^{(g)}(\curve_{\rm 2MM})
}\eeq
\et

\proof{This theorem was proved in \cite{eyno,CEO}.}

Again, this proves a posteriori, that $F_{\rm 2MM} ^{(g)}$ (which we recall is a formal power series in $T$) genericaly has a finite radius of convergence $T<T_c$.

We also have:
\bea
\left<\Tr{1\over x(p_1)-M_1}\,\dots\,\Tr{1\over x(p_k)-M_1}\right>_{\rm c}
&=& \sum_{g=0}^{\infty} N^{2-k-2g} {\underline{W}_k^{(g)}(p_1,\dots,p_k)\over dx(p_1)\dots dx(p_k)} \cr
&& + \delta_{k,1}\, N(V'(x(p_1))-y(p_1)) \cr
&& - {\delta_{k,2} \over (x(p_1)-x(p_2))^2} .\cr
\eea

Remark that the 1-Matrix model is a special case of the 2-matrix model with $V_2$ a quadratic polynomial.

\subsection{Double scaling limits of matrix models, minimal CFT}

It is well known that double scaling limits of the 1-matrix model, or 2-matrix models are in relationship with $(p,q)$ minimal models of conformal field theory \cite{Kos, ZJDFG, DKK}.

We have seen that as long as the curve is regular, all the $F^{(g)}$'s can be computed.
This shows that the radius of convergence in $T$ of $F^{(g)}(T)$ is reached for singular curves.
So far, only rational singularities have been studied in detail.

Thus, consider the case where the potentials $V_1$ and $V_2$ are fine-tuned so that the curve $\curve_{\rm 2MM}$ has a $p/q$ singularity at $T=T_c$ (notice that for the one matrix model one necessarily has $q=2$):
\beq
\left\{\begin{array}{l}
T=T_c\cr
x(z) \mathop{\sim}_{p \to a}\, x(a) + (z-z(a))^q \cr
y(z) \mathop{\sim}_{p \to a}\, y(a) + (z-z(a))^p \cr
\end{array}\right.
\eeq
We can use the notation of section \ref{sectsingular} with $t=T-T_c$, and thus at $t\neq 0$, the singularity is resolved, and we have (the local parameter is now $\zeta = z t^{-\nu}$):
\beq\label{dsllimitcurve2MM}
\left\{\begin{array}{l}
x(z,t) \sim x(a) + t^{q\nu}\,\,Q(\zeta) + o(t^{q\nu}) \cr
y(z,t) \sim y(a) + t^{p\nu}\,\,P(\zeta) +o(t^{p\nu}) \cr
\zeta = z t^{-\nu}
\end{array}\right.
\eeq
where $Q$ is  a polynomial of degree $q$ and $P$ is a polynomial of degree $p$.

The curve
\beq
\curve_{\rm sing}(\xi,\eta)=\left\{\begin{array}{l}
\xi = Q(\zeta)  \cr
\eta = P(\zeta)  \cr
\end{array}\right.
= {\rm Resultant}(Q-\xi,P-\eta)
\eeq
is called the singular spectral curve.

The dependence on $T$ of the 2-Matrix model is such that:
\beq
\left. {d\over dT} ydx\right|_{x} = dS_{\infty_x,\infty_y}
\eeq
where $\infty_x$ and  $\infty_y$ are the 2 common poles of $x$ and $y$. They are far away from branch-points, and in particular from the singularity.
This means that in the vicinity of the singularity we have:
\beq
\left. {d\over dT} ydx\right|_{x} \sim C\, t^\nu \,d\zeta + O(t^{2\nu})
\eeq
where $C$ is some constant of order $1$.
After substitution with the limit eq.\ref{dsllimitcurve2MM} this implies the Poisson relation\footnote{This Poisson relation is well known and can be found in \cite{DKK,ZJDFG}}:
\beq\label{Poissonpq1}
pP(\zeta)Q'(\zeta) - q Q(\zeta)P'(\zeta) = {C\over\nu} \, t^{1-(p+q-1)\nu}
\eeq
and therefore
\beq
\nu = {1\over p+q-1}.
\eeq

Thereom \ref{thsinglimit} proves that:
\beq
F_{\rm 2MM}^{(g)}(T) \mathop{\sim}_{T\to T_c} (T-T_c)^{(2-2g)(p+q)/(p+q-1)} F^{(g)}(\curve_{\rm sing})
 \qquad ,\,\, {\rm for}\, g\geq 2 
 \eeq
We also have:
\beq
F_{\rm 2MM}^{(0)}(T) \mathop{\sim}_{T\to T_c} {C^2\over 2} {(p+q-1)^2\over (p+q)(p+q+1)}\,(T-T_c)^{2+2\nu} + {\rm reg} \qquad ,\,\, {\rm for}\, g=0
\eeq
\beq
F_{\rm 2MM}^{(1)}(T) \mathop{\sim}_{T\to T_c}  -{1\over 24}\,(p-1)(q-1)\nu\,\ln{(T-T_c)}  + O(1) \qquad ,\,\, {\rm for}\, g=1
\eeq

We thus have a way to compute explicitely the double scaling limit of $F_{\rm 2MM}^{(g)}$.

\subsubsection{(p,q) minimal models} \label{sectpq}

Let us study in more details the curve $\curve_{(p,q)}$ (cf \cite{DKK}).

As we have seen above, the curve for the $(p,q)$ minimal model is of the form:
\beq\label{curvepq}
\curve_{(p,q)}(x,y)=
\left\{\begin{array}{l}
x = Q(\zeta) \cr
y = P(\zeta)  \cr
\end{array}\right.
= {\rm Resultant}(Q-x,P-y)
\eeq
where $P$ and $Q$ are polynomials of respective degrees $p$ and $q$, satisfying:
\beq\label{Poissonpq}
 p PQ'-q QP'  = {t_1\over \nu}
\eeq
The solution of which can be written \cite{ZJDFG}:
\beq
P= (Q^{p/q})_+
\eeq
and
\beq
(Q^{p/q})_- = {t_1\over q} \zeta^{1-q} + O(\zeta^{-q}),
\eeq
where we have used the notations $f= f_+ + f_-$, with $f_+$ and $f_-$ denoting respectively the positive and the negative
part of the Laurent series of $f$.
This last equation implies $q-2$ equations for the coefficients of $Q$.

The curve $\curve_{(p,q)}$ has genus zero $\genus=0$, and is such that $x$ and $y$ have only one pole $\alpha=\infty$.
The Bergmann kernel is:
\beq
B(\zeta_1,\zeta_2) = {d\zeta_1 \, d\zeta_2\over (\zeta_1-\zeta_2)^2}.
\eeq

The moduli (of the pole) of that curve are the $Q_k$ and $P_k$ such that:
\beq
Q(\zeta) = \sum_{k=0}^q Q_k \zeta^k
\virg
P(\zeta) = \sum_{k=0}^p P_k \zeta^k
\eeq
by a translation on $\zeta$, we can assume that $Q_{q-1}=0$, and by a rescaling of $\zeta$ we can assume that $Q_{q-2}=-q Q_q$, and the Poisson \eq{Poissonpq} implies that $P_{p-1}=0$ and $P_{p-2}=-p P_p$, thus:
\beq
Q_{q-1}=P_{p-1} = 0 \virg {Q_{q-2}\over Q_q}=-q \virg {P_{p-2}\over P_p}=-p.
\eeq

We find:
\beq
F^{(0)}(\curve_{(p,q)})= 0 .
\eeq

\subsubsection{Other times}

More generaly \cite{ZJDFG}, we can deform the $(p,q)$ minimal model with $p+q-2$ times $t_1,\dots, t_{p+q-2}$. 
For this purpose, one considers  $Q(\zeta)$ a degree $q$ monic polynomial,
\beq
Q(\zeta) = \zeta^q + \sum_{j=0}^{q-2} u_{q-j} \zeta^j
\eeq
whose coefficients $u_2,\dots, u_{q}$ are determined as functions of $q-1$ parameters, $t_1,\dots, t_{q-1}$, by the following requirement:
\beq
(Q^{p/q})_- = \sum_{j=1}^{q-2} {q-j\over q} t_{q-j} Q^{-j/q} + {t_1\over q} \zeta^{1-q} + O(\zeta^{-q}).
\eeq
Then we define the degree $p$ monic polynomial $P(\zeta)$ by:
\beq
P = \zeta^p + \sum_{j=0}^{p-2} v_{p-j} \zeta^j =   Q^{p/q}_+ - \sum_{j=1}^{p-1} {j+q\over q} t_{q+j-1} Q^{j/q}_+
\eeq
which depends on $p-1$ other times $t_q, \dots , t_{q+p-2}$.

The corresponding classical spectral curve is:
\beq
\curve_{(p,q)}(x,y) = {\rm Resultant}(x-Q,y-P).
\eeq
and depends on times $t_1,\dots, t_{p+q-2}$. One can check that if $t_2=t_3=\dots=t_{p+q-1}=0$, one recovers the $(p,q)$ minimal model.

It is well knwon that this curve is the spectral curve of the dispersionless Witham hierarchy \cite{krichwitham}.

\subsubsection{Examples of minimal models}

$\bullet$ {\bf Airy curve $(2,1)$}

The classical spectral curve for the $(2,1)$ minimal model is
\beq
\curve_{(2,1)}(x,y) = y^2 +t_1 - x .
\eeq
\beq
Q(\zeta)=\zeta^2+t_1 \virg P(\zeta)=\zeta
\eeq
It is studied with particular care in section \ref{sectairy} since it describes the behaviour of a generic curve around the branch points, and thus coincides with Tracy-Widom law \cite{TWlaw}.

\vs

\noindent $\bullet$ {\bf Pure gravity $(3,2)$}
\beq
Q(\zeta) = \zeta^2 - 2 v
\virg
P(\zeta) = \zeta^3 - 3v \zeta
\virg
t_1=3 v^2
\eeq
The classical spectral curve is:
\beq
\curve_{(3,2)}(x,y) = x^3-3v^2 x - y^2 + 2 v^3
\eeq
and is studied in details in section \ref{sectpure} .

\vs

\noindent $\bullet$ {\bf Ising model $(4,3)$}
\beq
Q(\zeta) = \zeta^3 - 3 v \zeta - 3 w
\virg
P(\zeta) = \zeta^4 - 4v \zeta^2 - 4w\zeta +2v^2 - {5\over 3}t_5(\zeta^2-2v)
\eeq
with:
\beq
t_1 = 4v^3+6w^2  \virg t_2=6vw.
\eeq
The classical spectral curve is
\bea
\curve_{(4,3)}(x,y)
&=& x^4 - y^3 - 4 v^3 x^2 + 3 v^4 y +   2 v^6 \cr
&& +12 w v ( - x y + v^2  x)
+6 w^2 (-   x^2 + 2 v  y - 4 v^3 )
+ 8 w^3 x
- 3 w^4.
 \cr
&& + 5 t_5(-   x^2 y-    v^2 x^2+ 2   v^3 y+ 2 v^5
- 2   w y x+ 2   v^2 w x+ 3   w^2 y-  17 v^2 w^2) \cr
&& + {25\over 3} t_5^2 ( v^2 y+ 2 v^4- 4 v w x- 12 v w^2) \cr
&& + {125\over 27} t_5^3 (-x^2+ 2 v^3- 6  w x- 9 w^2)
\eea

The variations with respect to the moduli $t_5$, $t_2$ and $t_1$ correspond respectively to the variations of the form $ydx$
\beq
\Omega_5 = -d\L_5 = d(\zeta^5 - 5 v \zeta^3 + 10 v^2 \zeta),
\eeq
\beq
\Omega_1 = -d\L_1 = d(\zeta +{5\over 12}\,{t_5\over v^3-w^2}\,(2v^2\zeta + 2 v w - w \zeta^2)),
\eeq
and
\beq
\Omega_2 = -d\L_2 = d(\zeta^2 +{5\over 6}\,{t_5\over v^3-w^2}\,(v^2\zeta^2 - 2 v w \zeta  -2 w^2))
\eeq
in the notations of section \ref{sectcompstruct}.

\noindent $\bullet$ {\bf Unitary models $(q+1,q)$}
\beq
\curve_{(q+1,q)}(x,y)=T_{q+1}(x)-T_q(y)
\eeq

\beq
Q(\zeta) = T_q(\zeta)
\virg
P(\zeta) = T_{q+1}(\zeta)
\eeq
where $T_p$ is the $p^{\rm th}$ Tchebychev's polynomial.

\subsection{Matrix model with external field}\label{sectGK}

Define the formal matrix model with external field \cite{PZinnmatext}:
\beq
Z_{\rm Mext} = \int dM\,\,\ee{-N\Tr (V(M)-M \hat\L)}
\eeq
where $V'(x)$ is a rational function with denominator $D(x)$, and $\hat\L$ is a fixed $N\times N$ matrix.
Consider its topological expansion (in the sense of a formal integral as  in the 1-matrix case above \cite{eynform}):
\beq
F_{\rm Mext}=-\ln{Z_{\rm Mext}} = \sum_{g=0}^\infty N^{2-2g} \,F^{(g)}_{\rm Mext}.
\eeq

Let us assume that $\hat\L$ has $s$ distinct eigenvalues $\hat\l_1,\dots,\hat\l_s$ with respective multiplicities $m_1,\dots, m_s$ such that $\sum_i m_i=N$.
The minimal polynomial of $\hat\L$ is:
\beq
S(y)=\prod_i (y-\hat\l_i).
\eeq

Define the classical curve obtained by ``removing'' the $1/N^2$ connected term in the loop equations:
\beq\label{defcurveMext}
\curve_{\rm Mext}(x,y) = ((V'(x)-y)S(y)-P(x,y))D(x)
\eeq
where
\beq\label{defP}
P(x,y) = {1\over N}\left<\Tr {V'(x)-V'(M)\over x-M}\,{S(y)-S(\hat\Lambda)\over y-\hat\Lambda}\right>
\eeq
so that $P(x,y)D(x)$ is a  polynomial in both $x$ and $y$.

Then one has:
\bt\label{thMext}
\beq
\encadremath{
F^{(g)}_{\rm Mext} = \underline{F}^{(g)}(\curve_{\rm Mext})
}\eeq
\et

\proof{This theorem is proved in appendix \ref{proofthGKontsevitch}, and the proof is very similar to that of the 2-matrix case above, see \cite{CEO}.}

\subsubsection{Application to Kontsevitch integral}
\label{sectkontse}

The Kontsevitch integral is known to be the generating function which computes intersection numbers of moduli space of Riemann surfaces (see \cite{kontsevitch, DReview}).
It is defined as:
\beq
Z_{\rm Kontsevitch} = \int dM\,\,\ee{-N\Tr ({M^3\over 3}-M (\L^2+t_1))} \virg t_1={1\over N}\Tr {1\over \L}
\eeq
where $\L$ has eigenvalues $\l_1,\dots,\l_N$.
Thus  it is $Z_{\rm Mext}$ with $V(x)=x^3/3$ and $\hat\L=\L^2 + t_1$.
Its classical curve is:
\beq
\curve_{\rm Kontsevitch}(x,y) = (x^2-y)S(y)-xS_1(y)-S_2(y)
\eeq
where $S_1(y)$ and $S_2(y)$ are polynomials in $y$ of degree at most $s-1$.

If we assume that the curve has genus zero (which is the case if we want the $F_{\rm Kontsevitch}^{(g)}$ to be the generating functions for intersection numbers), then we can find explicitely a rational parametrization:
\beq
\curve_{\rm Kontsevitch}(x,y) = 
\left\{
\begin{array}{l}
x(z) = z + {1\over 2N}\,\Tr {1\over \L}{1\over z-\L}\cr
y(z) = z^2 + {1\over N}\,\Tr{1\over \L}
\end{array} 
\right. .
\eeq
Using the symplectic invariance of theorem \ref{symplinv}, we may exchange the roles of $x$ and $y$.
There is a unique branch point in $y$, solution of $y'(z)=0$, located at $z=0$.
Since the formulae for $F^{(g)}$ consist in taking residues of rational functions at the branch point, we may consider the Taylor expansion of $x(z)$ near $z=0$, i.e.
\beq
\curve_{\rm Kontsevitch}(x,y) = 
\left\{
\begin{array}{l}
x(z) = z - {1\over 2}\sum_{k=0}^\infty t_{k+2} z^k  \cr
y(z) = z^2 + t_1\cr
\end{array}
\right.
\eeq
where we have defined the Kontsevitch times:
\beq
t_k = {1\over N}\Tr \L^{-k}
\eeq
so that:
\beq
\encadremath{
F^{(g)}_{\rm Kontsevitch} = F^{(g)}(\curve_{\rm Kontsevitch})
}\eeq

Again, using symplectic invariance of theorem \ref{symplinv}, we may add to $x(z)$ any rational function of $y(z)$, i.e. we immediately get a 1-line proof of the following well-known theorem:
\bt
$F_{\rm Kontsevitch}^{(g)}$ depends only on the odd times $t_{2k+1}$, with $k\leq 3g-2$:
\beq
\encadremath{
F_{\rm Kontsevitch}^{(g)}=F_{\rm Kontsevitch}^{(g)}(t_1,t_3,t_5,\dots,t_{6g-3})
}\eeq
\et

And, if we assume that $t_k=0$ for $k\geq p+2$, the curve is:
\beq
\left\{
\begin{array}{l}
x(z) = z - {1\over 2}\sum_{k=0}^p  t_{k+2} z^k  \cr
y(z) = z^2 +t_1 \cr
\end{array}
\right.
\eeq
which is identical to the curve of the $(p,2)$ model of section \ref{sectpq}, and which is well known to satisfy KdV hierarchy.
Thus, again we have a 1-line proof of the well known result:
\bt
$Z_{\rm Kontsevitch}$ is a KdV hierarchy tau function.
\et

\bigskip

\noindent $\bullet$ {\bf Examples: the first few correlation functions}

For Kontsevitch's curve we have:
\beq
B(z,z')={dz \, dz'\over (z-z')^2}
\virg
dE_z(z') = {z \,dz'\over z^2-z'^2}
\;
, \;
\om(z)=2z^2 dz \,(2 - \sum_j t_{2j+3}\, z^{2j})
\eeq
and the only branchpoint is located at $z=0$.

From the def.\ref{defloopfctions} we easily get the first correlation functions:
\bea
W_1^{(1)}(z)
&=& - {dz\over 8 (2-t_3)} \,\left({1\over z^4}+{t_5\over (2-t_3)z^2}\right), \cr
\eea

\bea
W_3^{(0)}(z_1,z_2,z_3)  
&=& - {1 \over 2-t_3}\, {dz_1\,dz_2\,dz_3\over z_1^2 z_2^2 z_3^2},\cr
\eea

\bea
 {W_2^{(1)}(z_1,z_2) }
&=& {dz_1 \, dz_2 \over 8 (2-t_3)^4 z_1^6 z_2^6} \Big[ (2-t_3)^2 ( 5 z_1^4 + 5 z_2^4+3z_1^2 z_2^2) \cr
&& \qquad + 6 t_5^2 z_1^4 z_2^4 + (2-t_3) \big(6 t_5 z_1^4 z_2^2 +6 t_5 z_1^2 z_2^4 + 5 t_7 z_1^4 z_2^4\big) \Big] \cr
\eea
and
\bea
{W_1^{(2)}(z) } &=& - { dz \over 128 (2-t_3)^7 z^{10}} \Big[ 252\, t_5^4 z^8 + 12\, t_5^2 z^6 (2-t_3) (50\, t_7 z^2 + 21\, t_5) \cr
&& \quad + z^4 (2-t_3)^2 ( 252\, t_5^2 + 348\, t_5 t_7 z^2 + 145\, t_7^2 z^4 + 308\, t_5 t_9 z^4) \cr
&& \qquad + z^2 (2-t_3) (203\, t_5 +145\, z^2 t_7 + 105\, z^4 t_9 +105\, z^6 t_{11}) \cr
&& \qquad \quad + 105\, (2 -t_3)^4 \Big] .\cr
\eea

\vs

The first and second order free energies are found:
\beq
F_{\rm Kontsevitch}^{(1)} = - {1 \over 24} \ln\left(1-{t_3\over 2} \right)
\eeq
and
\beq
F_{\rm Kontsevitch}^{(2)} = {1 \over 1920}\, {252\, t_5^3 + 435\, t_5 t_7 (2-t_3) + 175\, t_9 (2-t_3)^2 \over (2 -t_3)^5 }.
\eeq
which coincide with expressions previously found in the litterature \cite{IZK}.

\subsection{Example: Airy curve}
\label{sectairy}

The curve $y=\sqrt{x}$ is particularly important, because it corresponds to the leading behaviour of any generic curve near its branch points.
It is also the minimal model $(1,2)$ (cf \cite{ZJDFG, BookPDF}), also called Tracy-Widom law \cite{TWlaw}.

\bigskip

Consider the curve
\beq
\curve(x,y)=y^2-x.
\eeq
We chose the uniformization $p=y$:
\beq
\left\{
\begin{array}{lll}
x(p)&=& p^2 \cr
y(p)&=& p
\end{array}
\right.
\eeq
There is only one pole $\alpha=\infty$, and there is only one branch point located at $a=0$, the conjugated point is $\overline{p}=-p$.
The Bergmann kernel is the Bergmann kernel of the Riemann sphere:
\beq
B(p,q)={dp \, dq\over (p-q)^2}
\virg
dE_q(p) = {q \,dp\over q^2-p^2}
\virg
\om(q)=4q^2 dq
\eeq

It is easy to see that all correlation functions with $2g+k\geq 3$ are of the form:
\beq
W_k^{(g)}(p_1,\dots,p_k) = \om_k^{(g)}(p_1^2,\dots,p_k^2)\,dp_1\dots dp_k
\eeq
Moreover, the diagrammatic rules are clearly homogenous, so that the function $W_1^{(g)}(p)$ must be a homogeneous function of $p$. It is easy to find that:
\beq
W_1^{(g)}(p) = {c_g\, dp\over p^{6g-2}}
\eeq
and the total $1-$point function is
\beq\label{Adefiry1ptfc}
W_1(p,N) = - N ydx + \sum_{g=1}^\infty N^{1-2g} W^{(g)}_1(p) = W_1(N^{1\over 3}p,1).
\eeq
Similarly, the total $2-$point function is
\beq\label{Adefiry2ptfc}
W_2(p,q,N) = \sum_{g=0}^\infty N^{-2g} W^{(g)}_2(p,q) = W_2(N^{1\over 3}p,N^{1\over 3}q,1)
\eeq
and in general
\beq\label{Adefirykptfc}
W_k(p_1,\dots,p_k,N) = \sum_{g=0}^\infty N^{2-2g-k} W^{(g)}_k(p_1,\dots,p_k) = W_k(N^{1\over 3}p_1,\dots,N^{1\over 3}p_k,1).
\eeq

\medskip

The solution of the recursion def.\ref{defloopfctions} can be found explicitely in terms of the Airy function.

Consider $g(x)=Ai'(x)/Ai(x)$ where $Ai(x)$ is the Airy function, i.e. $g'(x)+g^2(x)=x=p^2$. In terms of the variable $p$ we write:
\beq
f(p)=g(p^2)
\virg
f^2 + {f'\over 2p} = p^2 .
\eeq
It can be expanded for large $p$:
\beq
f(p) = \sum_{k=0}^\infty f_k p^{1-3k}
= p - {1\over 4 p^2} - {9\over 32 p^5} + \dots
\eeq
where the coefficients in the expansion satisfy
\beq
{4-3k\over 2}f_{k-1}+\sum_{j=0}^k f_j f_{k-j} = \delta_{k,0}.
\eeq

The solution of the recursion def.\ref{defloopfctions} for the $1-$point function is:
\beq
W_1(p,1) = -2 {p^2-f(p)f(-p)\over f(p)-f(-p)}\, p dp
= -2p^2dp + {dp\over (2p)^4} + {9!!\,dp\over 3^2\,(2p)^{10}}+{15!!\over 3^4\,(2p)^{16}}\dots
\eeq

\beq
W_2(p,p',1) = - 4\,{(f(p)-f(p'))\,(f(-p)-f(-p'))\over (p^2-p'^2)^2\,(f(p)-f(-p))\,(f(p')-f(-p'))}\, pdp\, p'dp'
\eeq
In particular
\beq
W_2(p,p,1) = {f'(p) f'(-p)\over (f(p)-f(-p))^2}\, dp^2
\eeq
so that:
\beq
W_2(p,p,1)+W_1(p,1)^2  = 4p^4\, dp^2 = x\, dx^2
\eeq


Similarly we find for instance (with obvious cyclic conventions for the indices):
\bea
&& W_3(p_1,p_2,p_3,1) \cr
&= &{dx_1 dx_2 dx_3 \over (p_3^2-p_2^2)(p_3^2-p_1^2)(p_2^2-p_1^2)} \times \cr
&& \;\; \times {\sum_{i=1}^3 f(p_i)f(-p_i)(f(p_{i-1})+f(-p_{i-1})-f(p_{i+1})-f(-p_{i+1})) 
\over (f(p_1)-f(-p_1))(f(p_2)-f(-p_2))(f(p_3)-f(-p_3))} \cr
&=& {dp_1 \,dp_2 \,dp_3 \over 2\, p_1^2\, p_2^2\, p_3^2} + {dp_1 \,dp_2 \,dp_3 \over 2^6\, p_1^8\, p_2^8\, p_3^8} + \dots  \cr
\eea
and one can easily find similar expressions for all $W_k$'s.

\medskip

In fact all correlation functions can be written with a determinantal formula \cite{Dysondet, eynmetha},
with the Tracy-Widom kernel \cite{TWlaw}:
\beq
K(x,x')= {Ai(x)Ai'(x')-Ai'(x)Ai(x')\over x-x'}
\eeq
The fact that the Baker--Akhiezer function is $Ai(x)$ and satisfies the differential equation $Ai''=x Ai$ can be seen as a consequence of the Hirota equation theorem.\ref{thHirota}.

\bigskip

\br 
To large $N$ leading order the first term $W^{(0)}_k$ can be written in terms of Ferrer diagrams (Young diagrams):
\beq
W_k^{(0)}(p_1,\dots,p_k) = {k-3 !\over 2^{k-2}}\prod_j {dp_j\over p_j^2}\,\sum_{|\l|=k-3} M_\l(1/p_i^2)\,\prod_{j} \,\,{2\l_j+1!!\over \l_j!}\,\,{1\over n_j(\l)!}
\eeq
where $\l$ is a Ferrer diagram, $n_i(\l)=\#\{j\,/\,\, \l_j=i\}$, and $M_\l$ are the elementary monomial symetric polynomials:
\beq
M_\l(z_i) = \sum_{i_1\neq i_2\neq\dots\neq i_{k-3}} z_{i_1}^{\l_1}\dots z_{i_{k-3}}^{\l_{k-3}}.
\eeq

For instance with $k=4$ there is only one diagram $(1)$, and $M_{(1)}(z_1,z_2,z_3,z_4) = z_1+z_2+z_3+z_4$, and:
\beq
W_{4}^{(0)}(p_1,p_2,p_3,p_4)
= {3\over 4} \, {dp_1\, dp_2\, dp_3\, dp_4\over p_1^2\,p_2^2\,p_3^2\,p_4^2}\,\left({1\over p_1^2}+{1\over p_2^2}+{1\over p_3^2}+{1\over p_4^2}\right).
\eeq
\er

\medskip

\br
The free energies are all vanishing for that curve:
\beq
\forall g\quad F^{(g)}=0 .
\eeq
\er

\subsection{Example: pure gravity (3,2)}\label{sectpure}

In this section we study in details the $(3,2)$ minimal model, also called pure gravity \cite{ZJDFG}.



It corresponds to the curve
\beq
\curve_{(3,2)}= \; 
\left\{
\begin{array}{l}
x(z) = z^2-2v \cr
y(z) = z^3-3v z \cr
t_1=3 v^2
\end{array}\right. .
\eeq
We recognize Tchebychev's polynomials $T_2$ and $T_3$, which satisfy the Poisson relation \eq{Poissonpq} .
Up to a rescaling $z=\sqrt{v}\, p$, the curve reads:
\beq
\curve_{(3,2)} = \; 
\left\{
\begin{array}{l}
x(p) = v(p^2-2) \cr
y(z) = v^{3/2} (p^3-3 p) \cr
t_1=3 v^2
\end{array}\right. .
\eeq
There is only one $x$-branch point at $p=0$, and the conjugated point is $\overline{p}=-p$.
The Bergmann kernel is the Bergmann kernel of the Riemann sphere:
\beq
B(p,q)={dp \, dq\over (p-q)^2}
\virg
dE_q(p) = {q \,dp\over q^2-p^2}
\eeq
\beq
\om(q) = (y(q)-y(\qbar))dx(q) = 4v^{5/2}\, (q^2-3)\, q^2\, dq
\eeq
\beq
\Phi(q) = v^{5/2}({2\, q^5\over 5} - 2 q^3)
\eeq
Under a variation of $t_1$ we have:
\beq
\Omega_1(p) = - \left.{\partial y(p)dx(p)\over \partial t_1}\right|_{x(p)} = v^{1/2}\,dp = -v^{1/2}\,\Res_\infty q B(p,q) 
\virg \L_1(q)= -v^{1/2}\,q
\eeq
so the effect of $\partial/\partial t_1$ is equivalent to:
\beq\label{ddt32}
\left.{\partial \over \partial t_1} W_k^{(g)}\right|_{x} = - v^{1/2}\, \Res_{q\to \infty} q  W_{k+1}^{(g)}
\eeq




%

\subsubsection{Some correlation functions}

Using def.\ref{defloopfctions}, we find:
\beq\label{W30pq32}
W_{3}^{(0)}({p}_1,{p}_2,{p}_3) = - {v^{-5/2}\over 6}\,\,{d{p}_1 d{p}_2 d{p}_3\over {p}_1^2 {p}_2^2 {p}_3^2}
\eeq

\beq\label{W11pq32}
W_{1}^{(1)}(p) = -{v^{-5/2}\over (12)^2}\,{p^2+3\over p^4}\,dp
\eeq

\beq\label{W21pq32}
W_{2}^{(1)}({p},{q}) = v^{-5}\,\,{15{q}^4 + 15{p}^4 + 6{p}^4{q}^2 + 2{p}^4{q}^4 + 9{p}^2{q}^2 + 6{p}^2{q}^4\over 2^5\,3^3\,\,{p}^6\,{q}^6}\,dp\,dq
\eeq

\beq\label{W12pq32}
W_{1}^{(2)}(p) = - v^{-15/2}\,\,
7 {135 + 87 p^2 + 36 p^4 + 12 p^6 + 4 p^8\over 2^{10}\, 3^5\,p^{10}}\, dp
\eeq

\beq
W_{4}^{(0)}({p}_1,{p}_2,{p}_3,{p}_4) = {v^{-5}\over 9 {p}_1^2 {p}_2^2 {p}_3^2 {p}_4^2}\,\left(1+3\sum_i {1\over {p}_i^2}\right)\,dp_1 dp_2 dp_3 dp_4
\eeq

\beq
{W_{5}^{(0)}({p}_1,{p}_2,{p}_3,{p}_4,{p}_5) \over dp_1 dp_2 dp_3 dp_4 dp_5} = {v^{-15/2}\over 9 {p}_1^2 {p}_2^2 {p}_3^2 {p}_4^2 {p}_5^2}\,\left(1+3\sum_i {1\over {p}_i^2}+6\sum_{i<j} {1\over {p}_i^2{p}_j^2}+5\sum_i {1\over {p}_i^4}\right)
\eeq
etc...

Using def.\ref{defFg}, and eq.\ref{ddt32}, we find from eq.\ref{W30pq32}:
\beq
{\partial^3 F^{(0)}\over \partial t_1^3} = -{1\over 6v} = -{1\over 2\sqrt{3 t_1}}
\longrightarrow
{\partial^2 F^{(0)}\over \partial t_1^2} = -{t_1^{1/2}\over \sqrt{3}}
\eeq
and using eq.\ref{W11pq32}:
\beq
{\partial F^{(1)}\over \partial t_1} = -{1\over (12)^2 v^2}=-{1\over 48 t_1}
\longrightarrow
{\partial^2 F^{(1)}\over \partial t_1^2} = {1\over 48 \,t_1^2}
\eeq
as well as using eq.\ref{W12pq32}:
\beq
{\partial F^{(2)}\over \partial t_1} =  - v^{-7}\,\,{ 7 \over 2^8\, 3^5} = - { 7\over 2^8\, 3^{3/2}\, t_1^{7/2}} 
\longrightarrow
{\partial^2 F^{(2)}\over \partial t_1^2} = {49\over 2^9\,3^{3/2}\,t_1^{9/2}} .
\eeq
We may thus verify that the second derivative of the free energy:
\beq
u = {\partial^2 F\over \partial t_1^2} = \sum_{g=0}^\infty t_1^{(1-5 g)/2} u^{(g)}
\virg
F^{(g)}={4 u^{(g)}\over 5(1-g)(3-5g)}\, .
\eeq
satisfies the Painlev\'e equation to the first orders:
\beq
u^2+{1\over 6}u'' = {1\over 3}t_1 .
\eeq
It is well known that this equation is satisfied to all orders \cite{ZJDFG}, and here this can be seen as a consequence of the Hirota equation theorem. \ref{thHirota}.



\section{Conclusion}

In this paper, we have constructed an infinite family of invariants of algebraic curves.
By construction, these invariants coincide with the topological expansion of matrix integrals in the special case where the algebraic curve is the large $N$ part of the matrix integral's spectral curve.
But we emphasize again that the construction presented here goes beyond matrix models.

Our invariants are defined only in terms of algebraic geometry, and they have many interesting properties, like homogeneity, and integrability (they obey some Hirota equation).

The problem of computing the $F^{(g)}$'s for matrix models is an old problem which was addressed many times, and which found more and more elaborate answers \cite{ACKM}.
We claim that ours is more efficient, because it contains all multicut cases, and various types of matrix models at once. Also, even in the simplest cases (1-matrix model, 1 cut), our expressions are simpler than what existed before \cite{ACKM}.
Our $F^{(g)}$'s are defined recursively, like those of \cite{ACKM}, but the recursions are much easier to handle, and it is much easier to deduce properties to any order $g$ from our construction.

The efficiency of our method becomes striking when one wants to compare different models (Kontsevitch and KdV for instance), or when one wants to take singular limits.
Another important application of our method, is to prove that our $F^{(g)}$'s provide a solution to the holomorphic anomaly equations of \cite{BCOV} in topological string theory, thus confirming the Dijkgraaf Vafa correspondence. We claim that this can be proved easily from our work, and we present it in a coming paper \cite{EMO}.
It would also be interesting to compare our free energies with the $D$-Modules considered in \cite{Mth1,Mth2} as partition functions of a unified matrix M-theory by checking
that they indeed satisfy the equations of \cite{Mth2}.

One of the reasons our method is very efficient also, is that it can be represented diagrammatically, without equations, and thus very easy to remember.

\bigskip

{\bf Perspectives and generalizations:}

\begin{itemize}

\item The first thing one could think about is to understand what our $F^{(g)}$'s compute in algebraic geometry. There has been some attempts to recognize the first few of them $F^{(1)}$ as the Dedekind function \cite{Dubrovin1, Dubrovin2}, $F^{(2)}$ as the Eisenstein series \cite{klemm}, but the answer for higher $g$ is still obscure.
Also a combinatoric interpretation is missing.

\item Beyond that, it would be important to understand the combinatorics behind our diagrammatic construction. Since all diagrams are obtained from trees, it seems to be related to Schaeffer's method for counting maps \cite{schaeffer}, but this issue needs to be investigated further.

\item Double scaling limits of matrix models are in relationship with conformal field theory (CFT), and in particular minimal models. It would be interesting to compare our formulae with those obtained directely from CFT \cite{BookPDF, Zamozamo, Zamo}, and, since higher genus CFT is far less known, maybe our method can bring something new to CFT.

\item Also, the link between formal matrix models (i.e. those defined as combinatorial generating functions, for which it makes sense to consider a topological expansion, cf. \cite{eynform}), and actual convergent matrix integrals needs to be better understood. The difficulty lies beyond any order in perturbation, and integrability could play a key role in understanding the relationship in more details.

\item It would be interesting to check that the topological expansion for the chain of matrices \cite{eynmultimat},  is also given by the same $F^{(g)}$'s. This is an open question, but there are strong evidences that the answer is positive. For instance it is easy to see from \cite{eynmultimat} that $F^{(0)}$ and some of the first few correlation functions are indeed the same.

\item It would be interesting also to extend our construction to other types of matrix models, for instance non-hermitian (real symmetric or quaternionic). A first attempt was done in \cite{ChEynbeta}.
Another possible extension is towards the $O(n)$ matrix model, whose large $N$ limit is known in terms of an agebraic curve \cite{ECK1,ECK2}. To leading order, the $O(n)$ matrix model solution looks very similar to that of the 1-matrix model, except that correlation functions are no longer meromorphic functions on the curve. Instead they gain a phase shift after going around a non-trivial cycle. This would probably allow to define a twisted version of our construction.

\end{itemize}

\subsection*{Acknowledgements}

We would like to thank Mina Aganagic, Michel Berg\`ere, Leonid Chekhov, Philippe Di Francesco, Robert Dijkgraaf, Aleix Prats Ferrer, Ivan Kostov, Amir Kashani-Poor, Marcos Mari\~no, Andrei Okounkov, Pierre Vanhove, Jean-Bernard Zuber, for fruitful discussions on this subject.
This work is partly supported by the Enigma European network MRT-CT-2004-5652, by the ANR project G\'eom\'etrie et int\'egrabilit\'e en physique math\'ematique ANR-05-BLAN-0029-01,
 by the Enrage European network MRTN-CT-2004-005616, 
 by the European Science foundation through the Misgam program,
 by the French and Japaneese governments through PAI Sakurav, by the Quebec government with the FQRNT.

\vfill\eject


\appendix
\section{}
\section{Properties of correlation functions} 
\label{appprop}

In this section we prove the theorems stated in section \ref{propcorrel}.

We use very much the following obvious properties:

\beq\label{sumBdxdx}
\sum_i B(p^i,q) = {dx(p)dx(q)\over (x(p)-x(q))^2},
\eeq
\beq\label{eqsymdEom}
{dE_{\qbar}(p)\over \om(\qbar)} = {dE_{q}(p)\over \om(q)},
\eeq
and
\beq\label{eqdEtoB}
\mathop{{\rm lim}}_{q\to a_i} {dE_{q}(p)\over \om(q)}\,dx(q) = - {1\over 2}\,\mathop{{\rm lim}}_{q\to a_i} {B(p,q)\over dy(q)}\,dx(q).
\eeq
The third one is nothing but De L'H\^opital's rule.

\vs

{\bf Theorem \ref{thW30} }\proof{
From the definition \ref{defloopfctions}, we have:
\bea
W_{3}^{(0)}(p,p_1,p_2)
&=&  \Res_{q\to \bfa} {dE_{q}(p)\over \om(q)}\,\Big(B(q,p_1)B(\qbar,p_2)+B(q,p_2)B(\qbar,p_1) \Big) \cr
&=& 2 \Res_{q\to \bfa} {dE_{q}(p)\over \om(q)}\,B(q,p_1)B(\qbar,p_2) \cr
&=& 2 \Res_{q\to \bfa} \Res_{r\to \overline{q}} {dE_{q}(p)\over (y(q)-y(\qbar))(x(r)-x(q))}\,B(q,p_1)B(r,p_2) \cr
&=& - 2 \Res_{q\to \bfa} \Res_{r\to q} {dE_{q}(p)\over (y(q)-y(\qbar))(x(r)-x(q))}\,B(q,p_1)B(r,p_2) \cr
&=& - 2 \Res_{q\to \bfa} {dE_{q}(p)\over \om(q)}\,B(q,p_1)B(q,p_2) \cr
&=& \Res_{q\to \bfa} {B(q,p)B(q,p_1)B(q,p_2)\over dx(q) dy(q)}
\eea
where we have used \eq{eqsymdEom}, \eq{sumBdxdx} and \eq{eqdEtoB}.
}

\vs

{\bf Theorem \ref{thpolesWkgbp}} \proof{
It is  obvious from the definition that if $p$ is away from branchpoints, the residues are finite integrals, and $W_{k+1}^{(g)}$ is finite.
The only poles can be obtained when $p$ pinches an integration contour, i.e. at branch points.

Then, it is easy to see by recursion on $k$ and $g$ that it holds for any $p_i$.

}
\vs

{\bf Theorem \ref{thWkcycle}} \proof{The first one is a property of $dE_{q}(p)$, and the second one follows from recursion on $k$ and $g$, and it holds for $W_{2}^{(0)}=B$.}

\vs

{\bf Theorem \ref{thsumWk}} \proof{
The case $k=1,g=0$ comes from \eq{sumBdxdx}, and by integration:
\beq
\sum_i dE_{q}(p^i) = 0
\eeq
which proves equation \ref{thsumWk1}.
Then, it is clear from the iterative definition of $W_{k+1}^{(g)}$ for $k\geq 1$,
that the dependance of $W_{k+1}^{(g)}(p_1,p,p_2,\dots,p_k)$  in $p$ is a sum of integrals involving only the following quantities:
\beq
\sum_j \Res_{q\to a_j}\,{dE_{j,q}(q_1)\over \om_j(q)} B(q,p)\, W_{l+1}^{(g)}(q, q_2,\dots,q_l)
\eeq
for some $q_1,\dots,q_l$\footnote{Since it may not be clear which branch point $dE$ and $\omega$ refers to,
one indicates it with an index.}.
Using equation \ref{sumBdxdx} we have to compute:
\bea
&& \sum_i \sum_j \Res_{q\to a_j}\,{dE_{j,q}(q_1)\over \om_j(q)} B(q,p^i)\, W_{l+1}^{(g)}(q, q_2,\dots,q_l) \cr
&=&  \sum_j \Res_{q\to a_j}\,{dE_{j,q}(q_1)\over \om_j(q)} {dx(q)dx(p)\over (x(q)-x(p))^2}\, W_{l+1}^{(g)}(q, q_2,\dots,q_l) \cr
&=& {1\over 2}\, \sum_j \Res_{q\to a_j}\,{dE_{j,q}(q_1)\over \om_j(q)} {dx(q)dx(p)\over (x(q)-x(p))^2}\, \left( W_{l+1}^{(g)}(q, q_2,\dots,q_l)+W_{l+1}^{(g)}(\qbar, q_2,\dots,q_l)\right) \cr
&=& - {1\over 2}\,\sum_{q^i\neq q,\qbar}\, \sum_j \Res_{q\to a_j}\,{dE_{j,q}(q_1)\over \om_j(q)} {dx(q)dx(p)\over (x(q)-x(p))^2}\,  W_{l+1}^{(g)}(q^i, q_2,\dots,q_l) \cr
&=& 0
\eea
where the second equality holds due to \ref{eqsymdEom}, the third equality holds due to equation \ref{thsumWk1}, and the last equality holds because that last expression has no poles at the branchpoints.
This proves equation \ref{thsumWk2}.
}

\vs

{\bf Theorem \ref{thPkpol}} \proof{
It is clearly a rational function of $x(p)$ because it is a symmetric sum on all sheets.
From theorem \ref{thpolesWkgbp}, the RHS may have poles at branch points, and/or at the poles of some $y(p^i)$, and/or when $x(p)=x(p_l)$ for some $l$.
Let us prove that the poles at branch points actualy cancel.

Let us denote in this section:
\beq
U_k^{(g)}(q,q',p_K) =
\sum_{m=0}^g \sum_{J\subset K} W_{j+1}^{(m)}(q,p_J)W_{k-j+1}^{(g-m)}(q',p_{K/J})
+ W_{k+2}^{(g-1)}(q,q',p_K).
\eeq
From theorem \ref{thsumWk}, we have:
\bea
&& \sum_i U_k^{(g)}(q^i,q^i,p_K) \cr
&=& -\sum_i\sum_{l\neq i} U_k^{(g)}(q^l,q^i,p_K) \cr
&=& - U_k^{(g)}(q,\qbar,p_K) - U_k^{(g)}(\qbar,q,p_K) - \sum_{q^i\neq q,\qbar} (U_k^{(g)}(q,q^i,p_K)+U_k^{(g)}(\qbar,q^i,p_K)) \cr
&& - \sum_{q^i\neq q,\qbar} (U_k^{(g)}(q^i,q,p_K)+U_k^{(g)}(q^i,\qbar,p_K))  - \sum_{q^i\neq q,\qbar\,}\,\, \sum_{l\neq i, q^l\neq, q,\qbar} U_k^{(g)}(q^l,q^i,p_K) \cr
\eea
and from theorem \ref{thsumWk} and theorem \ref{thpolesWkgbp}, only the first two terms have poles at branch points, and thus:
\bea
&& - \sum_i \sum_j \Res_{q\to a_j} {dE_{j,q}(p)\over \om_j(q)}\, U_k^{(g)}(q^i,q^i,p_K) \cr
&=&  \sum_j \Res_{q\to a_j} {dE_{j,q}(p)\over \om_j(q)}\, \left(U_k^{(g)}(q,\qbar,p_K)+U_k^{(g)}(\qbar,q,p_K)\right) \cr
&=&  2 \sum_j \Res_{q\to a_j} {dE_{j,q}(p)\over \om_j(q)}\, U_k^{(g)}(q,\qbar,p_K) \cr
&=& 2 W_{k+1}^{(g)}(p,p_K)
\eea
where the last equality holds from the definition of $W_{k+1}^{(g)}$.
Thus we have:
\bea
W_{k+1}^{(g)}(p,p_K)
&=& - {1\over 2}\,\sum_j \Res_{q\to a_j} {dE_{j,q}(p)\over \om_j(q)}\, \left( \sum_i U_k^{(g)}(q^i,q^i,p_K)  \right). \cr
\eea
Then we rewrite it in terms of $P_k^{(g)}$ above:
\bea
&& W_{k+1}^{(g)}(p,p_K) \cr
&=& - {1\over 2}\,\sum_j \Res_{q\to a_j} {dE_{j,q}(p)\over \om_j(q)}\, \Big( P_k^{(g)}(x(q),p_K)\,dx(q)^2  \cr
&& \qquad \quad \qquad + 2\sum_i y(q^i)dx(q)W_{k+1}^{(g)}(q^i,p_K)  \Big) \cr
&=& - {1\over 2}\,\sum_j \Res_{q\to a_j} {dE_{j,q}(p)\over \om_j(q)}\, \Big( P_k^{(g)}(x(q),p_K)\,dx(q)^2  + 2y(q)dx(q)W_{k+1}^{(g)}(q,p_K) \cr
&&  \qquad \quad \qquad + 2y(\qbar)dx(q)W_{k+1}^{(g)}(\qbar,p_K) \Big) \cr
&=& - {1\over 2}\,\sum_j \Res_{q\to a_j} {dE_{j,q}(p)\over \om_j(q)}\, \Big( P_k^{(g)}(x(q),p_K)\,dx(q)^2  \cr
&& \qquad \quad \qquad + 2(y(q)-y(\qbar))dx(q)W_{k+1}^{(g)}(q,p_K) \Big) \cr
\eea
where we have used theorem \ref{thsumWk} again. That gives
\bea
&& W_{k+1}^{(g)}(p,p_K) \cr
&=& - {1\over 2}\,\sum_j \Res_{q\to a_j} {dE_{j,q}(p)\over \om_j(q)}\,  P_k^{(g)}(x(q),p_K)\,dx(q)^2 \cr
&& \qquad \quad - \sum_j \Res_{q\to a_j} dE_{j,q}(p)\,W_{k+1}^{(g)}(q,p_K).  \cr
\eea
Let us compute that last integral:
\bea\label{proofthPkpolRbilin}
&& \sum_j \Res_{q\to a_j} dE_{j,q}(p)\,W_{k+1}^{(g)}(q,p_K)  \cr
&=& - \Res_{q\to p} dE_{j,q}(p)\,W_{k+1}^{(g)}(q,p_K) + {1\over 2i\pi}\,\sum_i \oint_{q'\in\underline\bcycle_i}B(p,q')\, \oint_{\underline\acycle_i} W_{k+1}^{(g)}(q,p_K)  \cr
&& -  {1\over 2i\pi}\,\sum_i \oint_{{q'\in\underline\acycle_i}}B(p,q')\, \oint_{\underline\bcycle_i} W_{k+1}^{(g)}(q,p_K)  \cr
&=& - \Res_{q\to p} dE_{j,q}(p)\,W_{k+1}^{(g)}(q,p_K) + \sum_i  du_i(p)\, \oint_{\acycle_i} W_{k+1}^{(g)}(q,p_K)  \cr
&=& - \Res_{q\to p} dE_{j,q}(p)\,W_{k+1}^{(g)}(q,p_K)   \cr
&=& - W_{k+1}^{(g)}(p,p_K)
\eea
where we have deformed the contour of integration using Riemann bilinear identity, and then we have used theorem \ref{thWkcycle}.
Thus we have:
\beq
\sum_j \Res_{q\to a_j} {dE_{j,q}(p)\over \om_j(q)}\,  P_k^{(g)}(x(q),p_K)\,dx(q)^2  = 0 .
\eeq
Since this holds for any $p$, we can write for any $m\geq 0$:
\bea
0
&=& \Res_{p\to a_i} (y(p)-y(a_i))\,(x(p)-x(a_i))^m\, \sum_j \Res_{q\to a_j} {dE_{j,q}(p)\over \om_j(q)}\,  P_k^{(g)}(x(q),p_K)\,dx(q)^2  \cr
&=& - \Res_{q\to a_i}   \Res_{p\to q,\qbar} {dE_{j,q}(p)\over \om_j(q)}\,  (y(p)-y(a_i))\,(x(p)-x(a_i))^m\,P_k^{(g)}(x(q),p_K)\,dx(q)^2  \cr
&=& - \Res_{q\to a_i} dx(q)\,  (x(q)-x(a_i))^m\,P_k^{(g)}(x(q),p_K)  \cr
\eea
which proves that $P_k^{(g)}(x(q),p_K)$ can have no pole at $q=a_i$.

}

\vs

{\bf Theorem \ref{thsymWk}} \proof{
Assume this is proved for $h<g$, and at $g$, it is proved for $l<k$.
Then we have:
\bea
&& W_{2}^{(g)}(p_1,p_2) \cr
&=&  \sum_i \Res_{q\to a_i} {dE_{q}(p_1)\over \om(q)}\,\Big[2\sum_{m=0}^g  W_{2}^{(m)}(q,p_2)W_{1}^{(g-m)}(\qbar)
+ W_{3}^{(g-1)}(q,\qbar,p_2) \Big] \cr
&=&  \sum_i \Res_{q\to a_i} {dE_{q}(p_1)\over \om(q)}\,B(q,p_2)\,W_{1}^{(g)}(\qbar) \cr
&& + \sum_i \Res_{q\to a_i} \sum_{i'} \Res_{q'\to a_{i'}} {dE_{q}(p_1)\over \om(q)}\,{dE_{i',q'}(p_2)\over \om_{i'}(q')}\,  \Big[ 2  \sum_{m'=0}^{g-1}  W_{2}^{(m')}(q',q)W_{2}^{(g-m'-1)}(\qbar',\qbar) \cr
&& \quad + 2 \sum_{m=1}^g W_{1}^{(g-m)}(\qbar)W_{3}^{(m-1)}(q',\qbar',q)  +  2  \sum_{m'=0}^{g-1}  W_{3}^{(m')}(q',q,\qbar)W_{1}^{(g-m'-1)}(\qbar') \cr
&& \qquad + 4\sum_{m=1}^g\sum_{m'=0}^m  W_{2}^{(m')}(q',q)W_{1}^{(m-m')}(\qbar')W_{1}^{(g-m)}(\qbar) + W_{4}^{(g-2)}(q,q',\qbar,\qbar') \Big] \cr
\eea
and $W_{2}^{(g)}(p_2,p_1)$ is given by the same integral except that the order for computing residues is reversed, the residue in $q$ is computed before $q'$.
The difference is thus obtained by pushing the contour of $q'$ through the contour of $q$, and is obtained as the residue at $q=q'$.
Notice that only $W_2^{(0)}(q,q')$ has a pole at $q=q'$, all the other $W_k^{(g)}$ have no poles at $q=q'$ from theorem \ref{thpolesWkgbp}. Thus we have:
\beq
\begin{array}{l}
W_{2}^{(g)}(p_1,p_2)-W_{2}^{(g)}(p_2,p_1)
- \sum_i \Res_{q\to a_i} {dE_{q}(p_1)\,B(q,p_2)-dE_{q}(p_2)\,B(q,p_1)\over \om(q)}\,W_{1}^{(g)}(\qbar) \cr
=  \sum_i \Res_{q\to a_i} \Res_{q'\to q} {dE_{q}(p_1)\over \om(q)}\,{dE_{q'}(p_2)\over \om_{i}(q')}\, B(q,q') \cr
 \qquad \Big[4\sum_{m=0}^g  W_{1}^{(m)}(\qbar')W_{1}^{(g-m)}(\qbar)  + 2  W_{2}^{(g-1)}(\qbar',\qbar)  \Big] \cr
= 2 \sum_i \Res_{q\to a_i} \Res_{q'\to q} {dE_{q}(p_1)\over \om(q)}\,{dE_{q'}(p_2)\over \om_{i}(q')}\, B(q,q')\, U_{0}^{(g)}(\qbar,\qbar') \cr
=  \sum_i \Res_{q\to a_i} \Res_{q'\to q} {dE_{q}(p_1)\over \om(q)}\,{dE_{q'}(p_2)\over \om_{i}(q')}\, B(q,q')\, \left( U_{0}^{(g)}(q,q') + U_{0}^{(g)}(\qbar,\qbar')\right) \cr
\end{array}\eeq
where we have used the notation of theorem \ref{thPkpol}.

Similarly, for higher values of $k$, we find:
\bea
&& W_{2+k}^{(g)}(p_1,p_2,p_K)-W_{2+k}^{(g)}(p_2,p_1,p_K) \cr
&& \qquad - \sum_i \Res_{q\to a_i} {dE_{q}(p_1)\,B(q,p_2)-dE_{q}(p_2)\,B(q,p_1)\over \om(q)}\,W_{k+1}^{(g)}(\qbar,p_K) \cr
&=&  \sum_i \Res_{q\to a_i} \Res_{q'\to q} {dE_{q}(p_1)\over \om(q)}\,{dE_{q'}(p_2)\over \om_{i}(q')}\, B(q,q') \,\left(U_{k}^{(g)}(q,q',p_K)+U_{k}^{(g)}(\qbar,\qbar',p_K)\right) \cr
\eea
Then it gives:
\bea
&& W_{2+k}^{(g)}(p_1,p_2,p_K)-W_{2+k}^{(g)}(p_2,p_1,p_K) \cr
&& \qquad - \sum_i \Res_{q\to a_i} {dE_{q}(p_1)\,B(q,p_2)-dE_{q}(p_2)\,B(q,p_1)\over \om(q)}\,W_{k+1}^{(g)}(\qbar,p_K) \cr
&=&  \sum_i \Res_{q\to a_i} {dE_{q}(p_1)\over \om(q)}\,d_{q'}\left({dE_{q'}(p_2)\over \om_{i}(q')}\, \left(U_{k}^{(g)}(q,q',p_K)+U_{k}^{(g)}(\qbar,\qbar',p_K)\right)\right)_{q'=q} \cr
\eea
and by integrating half of it by parts we get:
\beq
\begin{array}{l}
 W_{2+k}^{(g)}(p_1,p_2,p_K)-W_{2+k}^{(g)}(p_2,p_1,p_K) \cr
= - \sum_i \Res_{q\to a_i} {B(q,p_1) dE_{q}(p_2)-B(q,p_2) dE_{q}(p_1)\over \om(q)^2}\, \left(U_{k}^{(g)}(q,q,p_K)+U_{k}^{(g)}(\qbar,\qbar,p_K)\right) \cr
 +  \sum_i \Res_{q\to a_i} {dE_{q}(p_1)\,B(q,p_2)-dE_{q}(p_2)\,B(q,p_1)\over \om(q)}\,W_{k+1}^{(g)}(\qbar,p_K) \cr
\end{array}
\eeq

Now we use the following Lemma:
\bl\label{LemmathsymWk}
If $f(q,q')$ is localy a bilinear differential near a branchpoint $a_i$, with no poles, and symmetric in $q$ and $\qbar$, and in $q'$ and $\qbar'$, then:
\beq
\Res_{q\to a_i} {B(q,p_1) dE_{q}(p_2)-B(q,p_2) dE_{q}(p_1)\over \om(q)^2} f(q,q) =0.
\eeq
\el
{\bf Proof of the Lemma:}

The residue is a simple pole, and we can use formula \ref{eqdEtoB}, that gives:
\bea
&& \Res_{q\to a_i} {B(q,p_1) dE_{q}(p_2)-B(q,p_2) dE_{q}(p_1)\over \om(q)^2} f(q,q) \cr
&=& \Res_{q\to a_i} {B(q,p_1) B(q,p_2)-B(q,p_2) B(q,p_1)\over \om(q) dy(q) dx(q)} f(q,q) \cr
&=& 0
\eea
$\square$.
\medskip

Using this Lemma, as well as theorem \ref{thPkpol}, we get:
\beq
\begin{array}{l}
W_{2+k}^{(g)}(p_1,p_2,p_K)-W_{2+k}^{(g)}(p_2,p_1,p_K) \cr
= - \sum_i \Res_{q\to a_i} {B(q,p_1) dE_{q}(p_2)-B(q,p_2) dE_{q}(p_1)\over \om(q)^2}\, (y(q)-y(\qbar))dx(q)W_{k+1}^{(g)}(q,p_K) \cr
 +  \sum_i \Res_{q\to a_i} {dE_{q}(p_1)\,B(q,p_2)-dE_{q}(p_2)\,B(q,p_1)\over \om(q)}\,W_{k+1}^{(g)}(\qbar,p_K) \cr
= 0.\cr
\end{array}
\eeq

}

\vs

{\bf Corollary \ref{corResxyWk}} \proof{

For any rational function $R(x)$ with no pole at $x(a_i)$ we have:
\bea
&& \Res_{a_i} R(x(p)) W^{(g)}_{k+1}(p,p_1,\dots,p_k)  \cr
&=& {1\over 2} \Res_{a_i} R(x(p)) \left(W^{(g)}_{k+1}(p,p_1,\dots,p_k)+W^{(g)}_{k+1}(\pbar,p_1,\dots,p_k)\right)  \cr
&=& 0
\eea
due to theorem \ref{thsumWk}.

\medskip

For $m=0,1$ compute
\bea
&& \sum_\alpha \Res_{p\to \alpha} x(p)^m y(p) W^{(g)}_{k+1}(p,p_K) \cr
&=& -{1\over 2}\sum_\alpha \Res_{p\to \alpha} {x(p)^m\over dx(p)}\, \Big( -2y(p)dx(p) W^{(g)}_{k+1}(p,p_K)+ \cr
&& \qquad +\sum_{h=0}^g \sum_{I\subset K} W^{(h)}_{|I|+1}(p,p_I)W^{(g-h)}_{k-|I|+1}(p,p_{K/I})+W^{(g-1)}_{k+2}(p,p,p_K)\Big)\cr
&=& {1\over 2} \Res_{p\to a_i,p_K} {x(p)^m\over dx(p)}\, \Big( -2y(p)dx(p) W^{(g)}_{k+1}(p,p_K)+ \cr
&& \qquad +\sum_{h=0}^g \sum_{I\subset K} W^{(h)}_{|I|+1}(p,p_I)W^{(g-h)}_{k-|I|+1}(p,p_{K/I})+W^{(g-1)}_{k+2}(p,p,p_K)\Big)\cr
&=& {1\over 2}\sum_{j=1}^k \Res_{p\to p_j} {x(p)^m\over dx(p)}\, \Big( B(p,p_j) W^{(g)}_{k}(p,p_{K/\{j\}})\Big)\cr
&& +{1\over 2}\sum_i \Res_{p\to a_i} {x(p)^m\over dx(p)}\, \Big( -2y(p)dx(p) W^{(g)}_{k+1}(p,p_K)+ \cr
&& \qquad +\sum_{h=0}^g \sum_{I\subset K} W^{(h)}_{|I|+1}(p,p_I)W^{(g-h)}_{k-|I|+1}(p,p_{K/I})+W^{(g-1)}_{k+2}(p,p,p_K)\Big)\cr
&=& {1\over 2}\sum_{j=1}^k d_{p_j}\, \Big( {x(p_j)^m\,W^{(g)}_{k}(p_{K})\over dx(p_j)}  \Big)\cr
&& +{1\over 4}\sum_i \Res_{p\to a_i} {x(p)^m\over dx(p)}\, \Big( -2y(p)dx(p) W^{(g)}_{k+1}(p,p_K)+ \cr
&& \qquad +\sum_{h=0}^g \sum_{I\subset K} W^{(h)}_{|I|+1}(p,p_I)W^{(g-h)}_{k-|I|+1}(p,p_{K/I})+W^{(g-1)}_{k+2}(p,p,p_K)\Big)\cr
&& +{1\over 4}\sum_i \Res_{p\to a_i} {x(p)^m\over dx(p)}\, \Big( -2y(\pbar)dx(p) W^{(g)}_{k+1}(\pbar,p_K)+ \cr
&& \qquad +\sum_{h=0}^g \sum_{I\subset K} W^{(h)}_{|I|+1}(\pbar,p_I)W^{(g-h)}_{k-|I|+1}(\pbar,p_{K/I})+W^{(g-1)}_{k+2}(\pbar,\pbar,p_K)\Big)\cr
&=& {1\over 2}\sum_{j=1}^k d_{p_j}\, \Big( {x(p_j)^m\,W^{(g)}_{k}(p_{K})\over dx(p_j)}  \Big)\cr
&& +{1\over 4}\sum_i \Res_{p\to a_i} {x(p)^m\over dx(p)}\, \Big( P_k^{(g)}(x(p),p_K) dx^2(p) \Big) \cr
&=& {1\over 2}\sum_{j=1}^k d_{p_j}\, \Big( {x(p_j)^m\,W^{(g)}_{k}(p_{K})\over dx(p_j)}  \Big)\cr
\eea
due to theorem \ref{thPkpol}.
}

\vs

{\bf Theorem \ref{thintPhi}} \proof{
The case $k=1$, $g=0$ is easy:
\beq
\Res_{p_2\to p_1} \Phi(p) B(p_1,p_2) = d\Phi(p_2) = y(p_1)dx(p_1).
\eeq
We prove the theorem by recursion on $g$ and $k$. Suppose it is proved for all $k'$ for $g'\leq g-1$, and for $k'\leq k-1$ if $g'=g$.
We write $K=\{1,\dots,k\}$ and $K'=\{1,\dots,k-1\}$.
Then we have from \eq{defWkgrecursive}:
\bea
&& \Res_{p_{k}\to \bfa} \Phi(p_{k}) W_{k+1}^{(g)}(p,p_1,\dots,p_k) \cr
&=& \Res_{p_{k}\to \bfa} \Res_{q\to \bfa} \Phi(p_{k}) \,{dE_{q}(p)\over \om(q)}\,\Big(
\qquad \sum_{m=0}^g \sum_{J\subset K'} W_{j+2}^{(m)}(q,p_J,p_k)W_{k-j}^{(g-m)}(\qbar,p_{K'/J})  \cr
&& \qquad +\sum_{m=0}^g \sum_{J\subset K'} W_{j+1}^{(m)}(q,p_J)W_{k-j+1}^{(g-m)}(\qbar,p_{K'/J},p_k)
+ W_{k+2}^{(g-1)}(q,\qbar,p_{K'},p_k) \Big) .\cr
\eea
Then we exchange the contours of integration:
\beq
\Res_{p_{k}\to \bfa} \Res_{q\to \bfa} = \Res_{q\to \bfa} \Res_{p_{k}\to \bfa} + \Res_{q\to \bfa} \Res_{p_{k}\to q,\qbar}.
\eeq
Thus
\bea
&& \Res_{p_{k}\to \bfa} \Phi(p_{k}) W_{k+1}^{(g)}(p,p_1,\dots,p_k) \cr
&=& \Res_{q\to \bfa} \Res_{p_{k}\to \bfa} \Phi(p_{k}) \,{dE_{q}(p)\over \om(q)}\,\Big(
\qquad \sum_{m=0}^g \sum_{J\subset K'} W_{j+2}^{(m)}(q,p_J,p_k)W_{k-j}^{(g-m)}(\qbar,p_{K'/J})  \cr
&& \qquad +\sum_{m=0}^g \sum_{J\subset K'} W_{j+1}^{(m)}(q,p_J)W_{k-j+1}^{(g-m)}(\qbar,p_{K'/J},p_k)
+ W_{k+2}^{(g-1)}(q,\qbar,p_{K'},p_k) \Big) \cr
&& + \Res_{q\to \bfa} \Res_{p_{k}\to q,\qbar} \Phi(p_{k}) \,{dE_{q}(p)\over \om(q)}\,\Big(
\qquad \sum_{m=0}^g \sum_{J\subset K'} W_{j+2}^{(m)}(q,p_J,p_k)W_{k-j}^{(g-m)}(\qbar,p_{K'/J})  \cr
&& \qquad +\sum_{m=0}^g \sum_{J\subset K'} W_{j+1}^{(m)}(q,p_J)W_{k-j+1}^{(g-m)}(\qbar,p_{K'/J},p_k)
+ W_{k+2}^{(g-1)}(q,\qbar,p_{K'},p_k) \Big) .\cr
\eea
The first term is computed from the recursion hypothesis, and the second term can exist only if the correlation function containing $p_k$ has poles at $p_k=q$ or $p_k=\qbar$, and  from theorem \ref{thpolesWkgbp},
this can happen only if the correlation function containing $p_k$  is a Bergmann kernel. That gives:
\beq
\begin{array}{l}
 \Res_{p_{k}\to \bfa} \Phi(p_{k}) W_{k+1}^{(g)}(p,p_1,\dots,p_k) \cr
=  \Res_{q\to \bfa} {dE_{q}(p)\over \om(q)}\,\Big(
\qquad \sum_{m=0}^g \sum_{J\subset K'} (2m+(j+1)-2) W_{j+1}^{(m)}(q,p_J) W_{k-j}^{(g-m)}(\qbar,p_{K'/J})  \cr
 \qquad +\sum_{m=0}^g \sum_{J\subset K'} (2(g-m)+(k-j)-2) W_{j+1}^{(m)}(q,p_J) W_{k-j}^{(g-m)}(\qbar,p_{K'/J}) \cr
 \qquad + (2(g-1)+k+1-2) W_{k+1}^{(g-1)}(q,\qbar,p_{K'}) \Big) \cr
 + \Res_{q\to \bfa} \Res_{p_{k}\to q,\qbar} \Phi(p_{k}) \,{dE_{q}(p)\over \om(q)}\,\Big(
\qquad  B(q,p_k)W_{k+1}^{(g)}(\qbar,p_{K'})  +  W_{k}^{(g)}(q,p_{K'}) B(\qbar,p_k) \Big) \cr
=  (2g+k-3)\, \Res_{q\to \bfa} {dE_{q}(p)\over \om(q)}\,\Big( \sum_{m=0}^g \sum_{J\subset K'}  W_{j+1}^{(m)}(q,p_J) W_{k-j}^{(g-m)}(\qbar,p_{K'/J})  \cr
\qquad +  W_{k+1}^{(g-1)}(q,\qbar,p_{K'}) \Big) \cr
 + \Res_{q\to \bfa} {dE_{q}(p)\over y(q)-y(\qbar)}\,\Big( y(q) W_{k+1}^{(g)}(\qbar,p_{K'})  +  y(\qbar) W_{k}^{(g)}(q,p_{K'})  \Big) \cr
=  (2g+k-3)\, W_k(p,p_1\dots,p_{k-1}) \cr
 + \Res_{q\to \bfa} {dE_{q}(p)\over y(q)-y(\qbar)}\,\Big( y(q) (W_{k+1}^{(g)}(\qbar,p_{K'})+W_{k}^{(g)}(q,p_{K'}))  +  (y(\qbar)-y(q)) W_{k}^{(g)}(q,p_{K'})  \Big) \cr
=  (2g+k-3)\, W_k(p,p_1\dots,p_{k-1}) \cr
 + \Res_{q\to \bfa} {dE_{q}(p)\over y(q)-y(\qbar)}\, (y(\qbar)-y(q)) W_{k}^{(g)}(q,p_{K'})   \cr
=  (2g+k-3)\, W_k(p,p_1\dots,p_{k-1})  - \Res_{q\to \bfa} dE_{q}(p)\, W_{k}^{(g)}(q,p_{K'})   \cr
=  (2g+k-3)\, W_k(p,p_1\dots,p_{k-1})  + {1\over 2}\,\Res_{q\to \bfa} dS_{q,o}(p) \, W_{k}^{(g)}(q,p_{K'}) \cr
 \qquad  -  {1\over 2}\,\Res_{q\to \bfa} dS_{\qbar,o}(p) \, W_{k}^{(g)}(q,p_{K'})  \cr
=  (2g+k-3)\, W_k(p,p_1\dots,p_{k-1})  + {1\over 2}\,\Res_{q\to \bfa} dS_{q,o}(p) \, W_{k}^{(g)}(q,p_{K'}) \cr
 \qquad -  {1\over 2}\,\Res_{q\to \bfa} dS_{q,o}(p) \, W_{k}^{(g)}(\qbar,p_{K'})  \cr
=  (2g+k-3)\, W_k(p,p_1\dots,p_{k-1})  + \Res_{q\to \bfa} dS_{q,o}(p) \, W_{k}^{(g)}(q,p_{K'}) \cr
=  (2g+k-3)\, W_k(p,p_1\dots,p_{k-1})  - \Res_{q\to p} dS_{q,o}(p) \, W_{k}^{(g)}(q,p_{K'}) \cr
 -{1\over 2i\pi} \sum_j \left( \oint_{\underline{\bcycle}_j} B(q,p) \oint_{\underline{\acycle}_j} W_{k}^{(g)}(q,p_{K'}) - \oint_{\underline{\acycle}_j} B(q,p) \oint_{\underline{\bcycle}_j} W_{k}^{(g)}(q,p_{K'})\right) \cr
=  (2g+k-3)\, W_k(p,p_1\dots,p_{k-1})  - \Res_{q\to p} dS_{q,o}(p) \, W_{k}^{(g)}(q,p_{K'}) \cr
 - du^t(p)\,\left( (1+\kappa\tau) \oint_{\underline{\acycle}} W_{k}^{(g)}(q,p_{K'}) - \kappa \oint_{\underline{\bcycle}} W_{k}^{(g)}(q,p_{K'})\right) \cr
=  (2g+k-3)\, W_k(p,p_1\dots,p_{k-1})  - \Res_{q\to p} dS_{q,o}(p) \, W_{k}^{(g)}(q,p_{K'}) \cr
 \qquad - du^t(p)\,\oint_{\acycle} W_{k}^{(g)}(q,p_{K'})  \cr
=  (2g+k-3)\, W_k(p,p_1\dots,p_{k-1})  - \Res_{q\to p} dS_{q,o}(p) \, W_{k}^{(g)}(q,p_{K'})   \cr
=  (2g+k-2)\, W_k(p,p_1\dots,p_{k-1})
\end{array}
\eeq

}

\section{Variation of the curve} 
\label{sectproofvariat}
In this appendix, we proove the theorems stated in sections (\ref{sectcompstruct})
and section (\ref{sectvarikappa}).

{\bf Lemma \ref{lemDOmega} }\proof{
\beq
\begin{array}{l}
D_\Omega\, \left( \sum_j \Res_{q\to a_j} {dE_{q}(p)\over \om(q)}\, f(q,\qbar) \right)_{x(p)}
\cr
=   \sum_j \Res_{q\to a_j} {dE_{q}(p)\over \om(q)} \,\, D_\Omega \left(f(q,\qbar)\right)_{x(q)} -  \sum_j \Res_{q\to a_j} {dE_{q}(p)\over (\om(q))^2}\,(\Omega(q)-\Omega(\qbar))\, f(q,\qbar)  \cr
 +2   \sum_j \Res_{q\to a_j} \sum_i \Res_{r\to a_i} { dE_{r}(p) \over \om(r)}\, \Omega(r)\, {dE_{q}(r) \over \om(q)}\, f(q,\qbar) \cr
 
=   \sum_j \Res_{q\to a_j} {dE_{q}(p)\over \om(q)} \,\, D_\Omega \left(f(q,\qbar)\right)_{x(q)} - 2  \sum_j \Res_{q\to a_j} {dE_{q}(p)\over (\om(q))^2}\,\Omega(q)\, f(q,\qbar)  \cr
 +2   \sum_j \Res_{q\to a_j} \sum_i \Res_{r\to a_i} { dE_{r}(p) \over \om(r)}\, \Omega(r)\, {dE_{q}(r) \over \om(q)}\, f(q,\qbar) \cr
=   \sum_j \Res_{q\to a_j} {dE_{q}(p)\over \om(q)} \,\, D_\Omega \left(f(q,\qbar)\right)_{x(q)} - 2  \sum_j \Res_{q\to a_j} \Res_{r\to q} {dE_{q}(r) dE_r(p)\over \om(q) \om(r)}\,\Omega(r)\, f(q,\qbar)  \cr
 +2   \sum_j \Res_{q\to a_j} \sum_i \Res_{r\to a_i} { dE_{r}(p) \over \om(r)}\, \Omega(r)\, {dE_{q}(r) \over \om(q)}\, f(q,\qbar) \cr
=   \sum_j \Res_{q\to a_j} {dE_{q}(p)\over \om(q)} \,\, D_\Omega \left(f(q,\qbar)\right)_{x(q)} \cr
 +2    \sum_i \Res_{r\to a_i}\sum_j \Res_{q\to a_j} { dE_{r}(p) \over \om(r)}\, \Omega(r)\, {dE_{q}(r) \over \om(q)}\, f(q,\qbar) \cr
\end{array}
\eeq
}

\vs

{\bf Theorem \ref{variat}} \proof{
This theorem straightforwardly comes from the diagrammatic rules described in the preceding paragraph except for
the variation of $F^{(1)}$.

Let us prove it for $F^{(1)}$ with $
\Omega(p) = \int_{\cal C} B(p,q) \Lambda(q)$.
One has
\beq
\begin{array}{rcl}
- \int_{\cal C} W_1^{(g)}(p) \L(p) &=&
- \Res_{q \to {\bf a}} {\int_{\cal C} dE_q(p) \L(p) \over \omega(q)} B(q,\qbar) \cr
&=& - \Res_{q \to {\bf a}} {\int_{\cal C} dE_q(p) \L(p) dz_i(q) dz_i(\qbar) \over \omega(q)} \left[ {1 \over (z(q)-z(\qbar))^2} + {1 \over 6} S_B(q) \right] \cr
\end{array}
\eeq
where $z_i$ is a local variable near the branch point $a_i$ and $S_B$ is the corresponding Bergmann projective connection.
Since the last term has a simple pole at the branch point $a_i$, one can write
\bea
- \Res_{q \to {\bf a}} {\int_{\cal C} dE_q(p) \L(p) dz_i(q) dz_i(\qbar) \over \omega(q)} {1 \over 6} S_B(q)
&=& - {1 \over 2} \sum_i {\Omega(a_i) \over dy(a_i)}  \Res_{q \to a_i} {B(q,\qbar) \over dx(q)} \cr
&=& - {1 \over 2} \delta_{\Omega} ln \tau_{Bx}. \cr
\eea

On the other hand, one can express the first term thanks to the local variables $z_i$ and compute
\bea
- \Res_{q \to {\bf a}} {\int_{\cal C} dE_q(p) \L(p) dz_i(q) dz_i(\qbar) \over \omega(q) (z(q)-z(\qbar))^2} 
&=& - {1 \over 24} \delta_\Omega (y'(a_i))\cr
\eea
provided that $z_i(q) - z_i(\qbar)= 2 z_i(q)$.
}

\vs

{\bf Theorem (\ref{homogeneity}) }\proof{
Using theorem \ref{thintPhi}, we have:
\beq
(2-2g)F^{(g)} = -\Res_\bfa \Phi W_1^{(g)}
\eeq
and using eq.\ref{ydxmoduli}, we can choose for any arbitrary $o'$:
\beq
\Phi(p)=
-\sum_\alpha \Res_{\alpha} V_\alpha dS_{p,o'}
+ \sum_\alpha t_\alpha \int_{o}^\alpha dS_{p,o'}
+ \sum_i \epsilon_i \oint_{\bcycle_i} dS_{p,o'}
\eeq
which implies:
\beq
\begin{array}{l}
 (2-2g)F^{(g)}
= -\Res_\bfa \Phi W_1^{(g)} \cr
= \sum_\alpha \Res_{p\to\bfa}  \Res_{q\to\alpha} V_\alpha(q) dS_{p,o'}(q) W_1^{(g)}(p)  - \sum_\alpha t_\alpha \Res_{p\to\bfa}  \int_{q=o}^\alpha dS_{p,o'}(q) W_1^{(g)}(p) \cr
 - \sum_i \epsilon_i \Res_{p\to\bfa} \oint_{q\in\bcycle_i} dS_{p,o'}(q) W_1^{(g)}(p) .\cr
\end{array}
\eeq
Since the poles $\alpha$ and the branch points $a_i$ do not coincide, one can exchange the order of integration. Then, one can
move the integration contours for $p$ in order to integrate only around the last pole $p \to q$:
\beq
\begin{array}{l}
 (2-2g)F^{(g)}
\cr
= \sum_\alpha t_\alpha  \int_{q=o}^\alpha \Res_{p\to q}  dS_{p,o'}(q) W_1^{(g)}(p) - \sum_\alpha  \Res_{q\to\alpha} \Res_{p\to q}  V_\alpha(q) dS_{p,o'}(q) W_1^{(g)}(p)   \cr
 + \sum_i \epsilon_i \oint_{q\in\bcycle_i} \Res_{p\to q}  dS_{p,o'}(q) W_1^{(g)}(p) \cr
=  \sum_\alpha  \Res_{q\to\alpha} V_\alpha(q) W_1^{(g)}(q)  - \sum_\alpha t_\alpha  \int_{q=o}^\alpha W_1^{(g)}(q)  - \sum_i \epsilon_i \oint_{q\in\bcycle_i} W_1^{(g)}(q) .\cr
\end{array}
\eeq
}

\vs

{\bf Theorem (\ref{thdWdkappa})} \proof{
We have
\bea
2i\pi {\partial \over \partial \kappa_{ij}}\,W^{(0)}_{2}(p_1,p_2)
&=& {1\over 2}(2i\pi)^2 (du_i(p_1) du_j(p_2)+du_i(p_2) du_j(p_1))\cr
&=& {1\over 2}\,\oint_{r\in\bcycle_j}\oint_{s\in\bcycle_i} (B(p_1,r)B(p_2,s) + B(p_2,r)B(p_1,s)). \cr
\eea
Thus the theorem holds for $k=2$ and $g=0$.
And by integration we get 
\bea
- 2i\pi {\partial \over \partial \kappa_{ij}} dE_q(p)
&=&  2 (i\pi)^2 (du_i(p) (u_j(q)-u_j(\qbar))+du_j(p) (u_i(q)-u_i(\qbar))) \cr
&=& - {1\over 2}\,\oint_{r\in \bcycle_i}\oint_{s\in \bcycle_j}  (B(p,r) dE_q(s)+B(p,s) dE_q(r)) .\cr
\eea

\beq
\begin{array}{rl}
& 2i\pi {\partial \over \partial \kappa_{ij}} W_{k+1}^{(g)}(p,p_K) \cr
=& 2i\pi {\partial \over \partial \kappa_{ij}} \Res_{q\to\bfa} {dE_q(p)\over (y(q)-y(\qbar))dx(q)}\, \Big(
W_{k+2}^{(g-1)}(q,\qbar,p_K) + \sum W_{j+1}^{(h)}(q,p_J) W_{k-j+1}^{(g-h)}(\qbar,p_{K/J}) \Big) \cr
=& {1 \over 2}\,\oint_{r\in \bcycle_i}\oint_{s\in \bcycle_j}  \Res_{q\to\bfa} {B(p,r) dE_q(s)+B(p,s) dE_q(r)\over (y(q)-y(\qbar))dx(q)}\, \Big(
W_{k+2}^{(g-1)}(q,\qbar,p_K) \cr
& \hspace{8cm} + \sum W_{j+1}^{(h)}(q,p_J) W_{k-j+1}^{(g-h)}(\qbar,p_{K/J}) \Big) \cr
&  + \Res_{q\to\bfa} {dE_q(p)\over (y(q)-y(\qbar))dx(q)}\, \Big( 2i\pi {\partial \over \partial \kappa_{ij}} W_{k+2}^{(g-1)}(q,\qbar,p_K) \Big) \cr
&+ \Res_{q\to\bfa} {dE_q(p)\over (y(q)-y(\qbar))dx(q)}\, \Big( \sum 2i\pi {\partial \over \partial \kappa_{ij}} W_{j+1}^{(h)}(q,p_J) W_{k-j+1}^{(g-h)}(\qbar,p_{K/J}) \Big) \cr
& + \Res_{q\to\bfa} {dE_q(p)\over (y(q)-y(\qbar))dx(q)}\, \Big( \sum  W_{j+1}^{(h)}(q,p_J)  2i\pi {\partial \over \partial \kappa_{ij}}W_{k-j+1}^{(g-h)}(\qbar,p_{K/J}) \Big) \cr
=&   {1 \over 2}\,\oint_{r\in \bcycle_i}\oint_{s\in \bcycle_j}  \Big( B(p,r) W^{(g)}_{k+1}(s,p_K)+B(p,s)  W^{(g)}_{k+1}(r,p_K) \Big) \cr
& + {1\over 2}\,\oint_{r\in \bcycle_i}\oint_{s\in \bcycle_j} \Res_{q\to\bfa} {dE_q(p)\over (y(q)-y(\qbar))dx(q)}\,
\Big( W_{k+4}^{(g-2)}(q,\qbar,p_K,r,s) \cr
& + 2 W_{j+3}^{(h)}(q,\qbar,p_J,r) W_{k-j+1}^{(g-1-h)}(p_{K/J},s) + 2 W_{j+2}^{(h)}(q,p_J,r) W_{k-j+2}^{(g-1-h)}(\qbar,p_{K/J},s)
\Big) \cr
& + {1\over 2}\,\oint_{r\in \bcycle_i}\oint_{s\in \bcycle_j} \Res_{q\to\bfa} {dE_q(p)\over (y(q)-y(\qbar))dx(q)}\,  \sum_J W_{k-j+1}^{(g-h)}(\qbar,p_{K/J}) \times \cr
&\hspace{2cm} \times  \Big( W_{j+3}^{(h-1)}(q,p_J,r,s) + 2 \sum_L W_{l+2}^{(m)}(q,p_L,r) W_{k-j-l+1}^{(h-m)}(p_{K/(J\cup L)},s) \Big) \cr
& + {1\over 2}\,\oint_{r\in \bcycle_i}\oint_{s\in \bcycle_j} \Res_{q\to\bfa} {dE_q(p)\over (y(q)-y(\qbar))dx(q)}\,  \sum_J W_{k-j+1}^{(g-h)}(q,p_{K/J}) \times \cr
& \hspace{2cm} \times \Big( W_{j+3}^{(h-1)}(\qbar,p_J,r,s) + 2 \sum_L W_{l+2}^{(m)}(\qbar,p_L,r) W_{k-j-l+1}^{(h-m)}(p_{K/(J\cup L)},s) \Big)   \cr
\end{array}
\eeq
We regroup together all terms with 2 $W$'s and all terms with 3 $W$'s:
\beq
\begin{array}{rl}
& 2i\pi {\partial \over \partial \kappa_{ij}} W_{k+1}^{(g)}(p,p_K) \cr
=&   {1 \over 2}\,\oint_{r\in \bcycle_i}\oint_{s\in \bcycle_j}  \Big( B(p,r) W^{(g)}_{k+1}(s,p_K)+B(p,s)  W^{(g)}_{k+1}(r,p_K) \Big) \cr
& + \oint_{r\in \bcycle_i}\oint_{s\in \bcycle_j} \Res_{q\to\bfa} {dE_q(p)\over (y(q)-y(\qbar))dx(q)}\,  \sum_J  W_{j+3}^{(h-1)}(q,p_J,r,s) W_{k-j+1}^{(g-h)}(\qbar,p_{K/J}) \cr
& + \oint_{r\in \bcycle_i}\oint_{s\in \bcycle_j} \Res_{q\to\bfa} {dE_q(p)\over (y(q)-y(\qbar))dx(q)}\,   \Big(  W_{j+2}^{(h)}(q,p_J,r) W_{k-j+2}^{(g-1-h)}(\qbar,p_{K/J},s) \Big) \cr
& + {1\over 2}\,\oint_{r\in \bcycle_i}\oint_{s\in \bcycle_j} \Res_{q\to\bfa} {dE_q(p)\over (y(q)-y(\qbar))dx(q)}\, W_{k+4}^{(g-2)}(q,\qbar,p_K,r,s) \cr
& + \oint_{r\in \bcycle_i}\oint_{s\in \bcycle_j} \Res_{q\to\bfa} {dE_q(p)\over (y(q)-y(\qbar))dx(q)} \times \cr
& \hspace{2cm} \times \Big( \sum_J  \sum_L W_{l+2}^{(m)}(\qbar,p_L,r) W_{k-j-l+1}^{(h-m)}(p_{K/(J\cup L)},s)  W_{k-j+1}^{(g-h)}(q,p_{K/J}) \Big) \cr
& + \oint_{r\in \bcycle_i}\oint_{s\in \bcycle_j} \Res_{q\to\bfa} {dE_q(p)\over (y(q)-y(\qbar))dx(q)} \times \cr
& \hspace{2cm} \times \Big( \sum_J   \sum_L W_{l+2}^{(m)}(q,p_L,r) W_{k-j-l+1}^{(h-m)}(p_{K/(J\cup L)},s)  W_{k-j+1}^{(g-h)}(\qbar,p_{K/J}) \Big) \cr
& + {1\over 2}\,\oint_{r\in \bcycle_i}\oint_{s\in \bcycle_j} \Res_{q\to\bfa} {dE_q(p)\over (y(q)-y(\qbar))dx(q)}\,  \Big( 2 W_{j+3}^{(h)}(q,\qbar,p_J,r) W_{k-j+1}^{(g-1-h)}(p_{K/J},s)  \Big) \cr
\end{array}
\eeq
We recognize the recursion  relation  eq.\ref{defWkgrecursive} in lines 2 to 5, and in lines 6 to 8, and this gives the theorem.

The theorem for the free energies, is easy obtained using theorem  \ref{thintPhi}.
}

\section{Proof of the symplectic invariance of \texorpdfstring{$F^{(0)}$}{F0} and \texorpdfstring{$F^{(1)}$}{F1}}
\label{appsymplinvF0F1}

\subsection{\texorpdfstring{$F^{(1)}$}{F1}}
Let us study how $F^{(1)}$ is changed under the exchange of the roles of $x$ and $y$.
For this purpose, we define the images of $F^{(1)}$ and $W_1^{(1)}(p)$ under this
transormation, i.e.
\beq
\widehat{W}_1^{(1)}(p):= - \Res_{q \to {\bf b}} {\int_{\tilde{q}}^q B(p,\xi) \over 2 (x(q)-x(\tilde{q})) dy(q)} B(q, \tilde{q})
\eeq
and
\beq
\widehat{F}^{(1)} = -{1 \over 2 } ln \tau_{By} - {1 \over 24} ln \prod_j x'(b_j)
\eeq
where ${\bf b}$ denotes the set of $y$-branch points,
$\tilde{q}$ is the only point satisfying $y(q) = y(\tilde{q})$ and approaching a branch point $b_j$
when $q \to b_j$, $\tau_{By}$ is the Bergmann tau function associated to $y$ and $x'(b_j)= {dx(b_j) \over d\tilde{z}_j(b_j)}$.
According to theorem \ref{variat}, for any variation $\Omega$ of the curve, the variation of the free energy reads
\beq
\delta_\Omega \widehat{F}^{(1)} = \int_{\cal{C}} \widehat{W}_1^{(1)}(p) \L(p).
\eeq
Thus, the variation of the difference between the two "free energies" reads
\beq
\delta_\Omega \left( F^{(1)} - \widehat{F}^{(1)} \right) = \int_{\cal{C}} \left( W_1^{(1)}(p) + \widehat{W}_1^{(1)}(p) \right) \L(p) .
\eeq

In order to evaluate this quantity, one needs the following lemma:
\bl
For any choice of variable $z$:
\beq\label{sumW}
W_1^{(1)}(p) + \widehat{W}_1^{(1)}(p) = {1 \over 24} d_p \left[{1 \over x' y'} \left( 2  S_{Bz}(p)
+ {x'' y'' \over x' y'} + {x''^2 \over x'^2} - {x''' \over x'} + {y''^2 \over y'^2} - {y''' \over y'} \right) \right]
\eeq
where the derivatives are taken with respect to the variable $z$ and $S_{Bz}$ denotes
the Bergmann projective connection associated to $z$.
\el

\proof{
From the definitions , one easily derives that
\beq\label{sumW1}
W_1^{(1)}(p) + \widehat{W}_1^{(1)}(p) = \Res_{q \to p} {B(p,q) \over (x(p)-x(q)) (y(p)-y(q))}.
\eeq

Let us now expand the integrand in terms of an arbitrary local variable $z$ when $p \to q$. The different factors
read
\beq
{B(p,q) \over dz(p) dz(q)}= {1 \over (z(p)-z(q))^2} + {1 \over 6} S_{Bz}(p) - {z(p)-z(q) \over 12} S_{Bz}'(p)+ O((z(p)-z(q))^2)
\eeq
and
\bea
{1 \over y(p)-y(q)} &=& {1 \over (z(p)-z(q)) y'(p)} \Big[ 1 - (z(p)-z(q)) {y'' \over 2 y'} \cr 
&& \; \; + (z(p)-z(q))^2 \left( {y''^2 \over 4 y'^2} - {y'''\over 6 y'}\right) \cr
&& \; \; \; \;   + {(z(p)-z(q))^3 \over 24} \left(- {y'''' \over y'} +4 {y''' y'' \over y'^2} - {3 y''^3 \over y'^3} \right) \cr
&& \qquad + O((z(p)-z(q))^4) \Big] .\cr
\eea
Inserting it altogether inside \eq{sumW1}, one can explicitly compute the residues and one recognizes the
formula \ref{sumW}.
}

We can now prove the following theorem stating the symmetry of $F^{(1)}$ under the exchange of $x \leftrightarrow y$:
\bl\label{symF1}
The two free energies transform in the same way under any variation of the moduli of the curve:
\beq
\delta_\Omega  F^{(1)} = \delta_\Omega \widehat{F}^{(1)}
\eeq
\el

\proof{
We already know that
\beq
\delta_\Omega \left( F^{(1)} - \widehat{F}^{(1)} \right) = \int_{p \in \cal{C}}
{1 \over 24} d_p \left[{1 \over x' y'} \left( 2  S_{Bz}(p)
+ {x'' y'' \over x' y'} + {x''^2 \over x'^2} - {x''' \over x'} + {y''^2 \over y'^2} - {y''' \over y'} \right) \right]
\L(p)
\eeq
for an arbitrary variable $z$.

We just have to check that this quantity vanishes for the transformations corresponding to varying the moduli
of the curve. Because the function inside the differential wrt $p$ vanishes at the poles of $ydx$, one can
check that it is indeed the case.
}

Thus the first correction to the free energy satisfies the following variation under symplectic transformations:
\bt
$F^{(1)}$ does not change under the following transformations of $\curve$:

$\bullet$ \vspace{0.3mm} $y\to y +R(x)$ where $R$ is a rational function.

$\bullet$ \vspace{0.3mm} $y\to c y$ and $x\to {1\over c}x$ where $c$ is a non-zero complex number.

$\bullet$ \vspace{0.3mm} $x\to x +R(y)$ where $R$ is a rational function.

$\bullet$ \vspace{0.3mm} $y\to x$ and $x\to y$.

$F^{(1)}$ is shifted by a multiple of $ i \pi/12$ when $\curve$ is changed by $x \to - x$.

\et

\proof{The first transformation is obvious from the definition since neither $ln \tau_{Bx}$ nor $y'(a_i)$ changes.

The second one follows from theorem 2 in \cite{EKK} for $f=x$ and $g={x \over c}$ which shows that
the variations of $ln(\tau_{Bx})$ and $y'(a_i)$ compensate.

The fourth one is nothing but lemma (\ref{symF1}) and it gives the third one when combined with the first.

The last transformation holds because $ln \tau_{Bx}$ is left unchanged and $y'(a_i)$ changes sign.

}

\subsection{\texorpdfstring{$F^{(0)}$}{F0}}

\bt
$F^{(0)}$ does not change under the following transformations of $\curve$:

$\bullet$ \vspace{0.3mm} $y\to y +{\cal{P}}(x)$ where ${\cal{P}}$ is a polynomial.

$\bullet$ \vspace{0.3mm} $y \to y$ and $x \to -x$.

$\bullet$ \vspace{0.3mm} $y\to c y$ and $x\to {1\over c}x$ where $c$ is a non-zero complex number.

$\bullet$ \vspace{0.3mm} $x\to x +{\cal{P}}(y)$ where ${\cal{P}}$ is a polynomial.

$\bullet$ \vspace{0.3mm} $y\to x$ and $x\to y$.

\et

\proof{
The second and third transformations come from the structure of $F^{(0)}$ which is bilinear in objects proportionnal to $ydx$.

One can obtain the last one following the same method as for $F^{(1)}$: we compute the variation of the difference
of the two free energies under the changes of the moduli of $\curve$ and show that they vanish.

Let us now show the first invariance. In this case, the variation of the free energy is
$\delta_{{\cal{P}}(x)dx} F^{(0)}$. Since ${\cal{P}}(x)$ is a polynomial one can write it
\beq
{\cal{P}}(x(p)) = \Res_{q \to p} B(p,q) {\cal{Q}}(x(q)) = - \sum_\alpha \Res_{q \to \alpha} B(p,q) {\cal{Q}}(x(q)),
\eeq
where ${\cal{Q}}(x)$ is also a polynomial in $x$. Then, theorem (\ref{variat}) implies the invariance
of $F^{(0)}$ under this transformation.

The fourth transformation is a combination of the first and the last one.

}


\section{Matrix model with an external field}
\label{proofthGKontsevitch}

We consider the matrix model with external field defined in section
(\ref{sectGK}):
\beq\label{defZLambda} Z(\Lambda):=\int_{H_n} dM \, e^{-N
Tr(V(M) - M\widehat{\L} )} \eeq

where we assume that $\widetilde\L$ is the diagonal matrix:
\beq
\widehat{\L}={\rm diag}\,(\,\mathop{\overbrace{\widehat{\lambda}_1,\dots,\widehat{\lambda}_1}}^{n_1},\mathop{\overbrace{\widehat{\lambda}_2,\dots,\widehat{\lambda}_2}}^{n_2},\dots,\mathop{\overbrace{\widehat{\lambda}_s,\dots,\widehat{\lambda}_s}}^{n_s}\,)
\eeq
and $V'(x)$ is a rational fraction with denominator $D(x)$:
$V'(x) = {\sum_{k=0}^{d} g_k x^k \over D(x)}$.

In particular, the polynomial
$S(y):=\prod_{i=1}^s (y-\widehat{\lambda}_i)$ is the minimal polynomial of $\widetilde\L$.

We define the correlation functions
$\overline{w}_{k}(x_1,\dots,x_k) := N^{k-2}\left< \prod_{i=1}^k \tr{1\over x_{i}-M}\right>_c$
and their $1/N^2$ expansion
\beq\label{Mextdefwklh}
\overline{w}_{k}(\bfx_K) = \sum_{h=0}^\infty {1\over N^{2h}}\,\overline{w}_{k}^{(h)}(\bfx_K).
\eeq

We also define the auxiliary functions
\beq\label{Mextdefubarkl}
\overline{u}_{k}(x,y;\bfx_K) := N^{|K|-1} \left< \tr {1\over x-M} {S(y)-S(\L)\over y-\L}\,\,
\prod_{r=1}^{|K|} \tr{1\over x_{i_r}-M}\right>_c
\eeq
and
\beq\label{Mextdefpkl}
P_{k}(x,y;\bfx_K) := N^{|K|-1} \left< \tr {V'(x)-V'(M_1)\over x-M} {S(y)-S(\L)\over y-\L}\,\,
\prod_{r=1}^{|K|} \tr{1\over x_{i_r}-M}\right>_c.
\eeq

Notice that $\overline{u}_{k,}(x,y;\bfx_K)$ is a polynomial in $y$ of degree $s-1$, and
$D(x) P_{k}(x,y;\bfx_K)$ is a polynomial in $x$ of degree $d-1$ and in $y$ of degree $s-1$ (note that
$P_{0}$ corresponds to $P$ in \eq{defP}).

It is convenient to renormalize those functions, and define:
\beq\label{defukl}
u_{k}(x,y;\bfx_K):=\overline{u}_{k}(x,y;\bfx_K)  -\delta_{k,0}S(y)
\eeq
and
\beq\label{defwkl}
w_{k}(\bfx_K):=\overline{w}_{k}(\bfx_K)  + {\delta_{k,2}  \over (x_1-x_2)^2} .
\eeq

\subsection{Loop equations}

Consider the change of variables
\beq
\delta M = {1\over x-M}{S(y)-S(\L)\over y-\L}.
\eeq
You get the loop equation
\beq
\overline{w}_1(x)\overline{u}_0(x,y)+{1\over N^2}\overline{u}_1(x,y;x)
=
V'(x)\overline{u}_0(x,y) - P_0(x,y)-y\overline{u}_0(x,y) + S(y)\overline{w}_1(x)
\eeq
i.e.
\beq
(y+\overline{w}_1(x)-V'(x))(\overline{u}_0(x,y)-S(y))+{1\over N^2}\overline{u}_1(x,y;x)
= (V'(x)-y)S(y) - P_0(x,y).
\eeq
We define the polynomial both in $x$ and $y$
\beq
E_{Mext}(x,y):= \left((V'(x)-y)S(y) - P_0(x,y)\right) D(x)
\eeq
and
\beq
Y(x):=V'(x)-\overline{w}_1(x).
\eeq
The loop equation thus implies:
\beq
(y-Y(x))u_0(x,y)D(x)+{1\over N^2}u_1(x,y;x)D(x) = E_{Mext}(x,y)
\eeq
and in particular
\beq\label{Mextloop1}
E_{Mext}(x,Y(x))={1\over N^2}u_1(x,Y(x);x)D(x).
\eeq

The leading order of the topological expansion reads
\beq
E_{Mext}^{(0)}(x,Y(x))
= 0
\eeq
which defines an algebraic curve.

\subsection{Leading order algebraic curve}

Let us study the curve $\curve_{Mext}(x,y)=E_{Mext}^{(0)}(x,y)=0$ defining a compact Riemann surface $\overline{\Sigma}$ and
two functions $x$ and $y$ defined on it.

Because $y$ is a solution of a degree $s+1$ equation, $\curve_{Mext}(x,y)$  has $s+1$ $x$-sheets.
The sheets can be identified by their large $x$ behavior:

$\bullet$ in the physical sheet, we have $Y(x)\sim V'(x)-1/x+O(1/x^2)$

$\bullet$ in the other sheets, $Y(x)\sim \widehat{\lambda}_i + {n_i\over N}\,{1\over x}+O(1/x^2)$

Let us note by $p^i \in \overline{\Sigma}$ with $i= 0 \dots s$ the different points of the curve whose $x$-projection are $x(p)$, i.e.
\beq
\forall i,j \;\;\; x(p^i) = x(p^j).
\eeq
The superscript $0$ corresponds to the point in the physical sheet.

From the correlation functions previously defined on the $x$ and $y$ projections, one defines the corresponding
meromorphic 1-forms on the curve as follows:
\beq
W_k({\bf p_K}) :=w_k({\bf x(p_K)})\, dx(p_1)\dots dx(p_k)
\eeq
and
\beq
U_k(p,y;{\bf p_K}) :=u_k(x(p),y;{\bf x(p_K)})\, dx(p)dx(p_1)\dots dx(p_k)
\eeq
as well as there topological expansions
\beq
W_k({\bf p_K}) = \sum_{h=0}^\infty N^{2-2h}\,W_k^{(h)}({\bf p_K})
\;\;\; \hbox{and} \;\;\;
U_k(p,y;{\bf p_K}) = \sum_{h=0}^\infty N^{2-2h}\,U_k^{(h)}(p,y;{\bf p_K}).
\eeq

\subsubsection{Filling fractions and genus}

The curve has a genus $g\leq d s -1$ and we work with fixed filling fractions
\beq
\epsilon_I:={1\over 2i\pi}\oint_{{\cal A}_I} y dx.
\eeq

%
%
%
%
%

\subsubsection{Subleading loop equations}
Consider the topological expansion of the loop equation \eq{Mextloop1}. It reads, for $h\geq 1$:
\bea\label{Mextmastloop2} E^{(h)}(x,y) &=&
D(x) (y-Y(x)) u_0^{(h)}(x,y) + D(x) w_{1,0}^{(h)}(x) u_0^{(0)}(x,y) \cr && +
D(x) \sum_{m=1}^{h-1} w_{1,0}^{(m)}(x) u_0^{(h-m)}(x,y) + D(x) u_1^{(h-1)}(x,y;x), \cr \eea
where $E^{(h)}(x,y)$ is the $h$'th term in the $\hbar^2$-expansion
of the spectral curve.

\subsection{Diagrammatic rules for the correlation functions and the free energy}
In this section, one proves that the correlation functions' and the free energy's topological expansion of this model do coincide
with the $\underline{W}_k^{(h)}$'s and $\underline{F}^{(h)}$'s defined following the definitions of \eq{defspcorr} and \eq{defspfree} for the classical spectral
curve ${\cal{E}}_{Mext}(x,y)=0$.

\subsubsection{The semi-classical spectral curve}
Let us reexpress the semi-classical spectral curve (i.e. the whole formal series $E_{Mext}(x,y)$) in terms of the
classical one $E_{Mext}^{(0)}(x,y)$.

\bt
\beq \label{MextdefE}
\begin{array}{rcl}
E_{Mext}(x,y) &=& -D(x)``\left<\prod_{i=0}^s (y-V'(x(p))+{1 \over N} \Tr{1 \over x(p^{(i)})-M}) \right>'' \cr
&=& D(x)\left[(V'(x)-y)S(y) - P_0(x,y)\right]
\end{array}
\eeq
and
\beq
U_0(p,y) = - ``\left<\prod_{i=1}^s (y-V'(x(p))+{1 \over N} \Tr{1 \over x(p^{(i)})-M}) \right>''
\eeq
where $``<.>''$ means that one replace $\overline{w}_2$ by $w_2$ in the expansion.
\et

\proof{
One proves that the ${1 \over N^2}$-expansions of 
\beq
\widetilde{E}(x,y) = -D(x) \left<\prod_{i=0}^s (y-V'(x(p))+{1 \over N} \Tr{1 \over x(p^{(i)})-M}) \right>
\eeq
and 
\beq
\widetilde{U}(p,y) = -D(x) \left<\prod_{i=1}^s (y-V'(x(p))+{1 \over N} \Tr{1 \over x(p^{(i)})-M}) \right>
\eeq
coincide with the expansion of $E_{Mext}(x,y)$ and $U_0(p,y)$.

Let the topological expansions be
\beq
\widetilde{E}(x,y) = \sum_g N^{-2g} \widetilde{E}^{(g)}(x,y)
\;\;\; ,\;\;\; \widetilde{U}(p,y) = \sum_g N^{-2g} \widetilde{U}^{(g)}(p,y).
\eeq

Expanding the expressions of $\widetilde{E}(x,y)$ and $\widetilde{U}(p,y)$ into cumulants, one recovers
\bea \widetilde{E}^{(h)}(x,y) &=&
(y-Y(x)) D(x) \widetilde{U}_0^{(h)}(x,y) + D(x) w_{1}^{(h)}(x) \widetilde{U}_0^{(0)}(x,y) \cr && +
D(x) \sum_{m=1}^{h-1} w_{1}^{(m)}(x) \widetilde{U}_0^{(h-m)}(x,y) +  D(x) \widetilde{U}_1^{(h-1)}(x,y;x), \cr \eea
which coincides with \eq{Mextmastloop2}.

One easily proves that this system of equations admits a unique solution thanks to the polynomial
properties of $\widetilde{U}(p,y)$ and that the leading orders $h=0$ coincide.
The proof is extremely similar to that for the 2-matrix model (cf. theorem 1 in \cite{CEO}).
}

\subsubsection{Diagrammatic solution}

One has:
\beq
W_2^{(0)}(p_1,p_2) = \underline{B}(p_1,p_2)
\eeq
where $\underline{B}$ is the Bergmann kernel of the algebraic curve ${\cal{E}}_{Mext}$.

The coefficient of $y^s$ of \eq{MextdefE}, divided by $D(x)$, is:
\beq
V'(x) +\sum_i \L_i = \sum_i Y(p^i).
\eeq
It implies that
\beq
{dx(p)dx(q)\over (x(p)-x(q))^2}  = \sum_i W_2^{(0)}(p^i,q)
\;\;\; \hbox{and} \;\;\; \forall h >1 \, , \; \sum_i W_2^{(h)}(p^i,q) =0,
\eeq
i.e.
\beq
\ovl\om_2(p,q)+\sum_{i=1}^s \om_2(p^i,q) = 0.
\eeq

The coefficient of $y^{s-1}$ is:
\beq
\sum_{i<j} Y(p^i)Y(p^j)+{1\over N^2}\om_2(p^i,p^j)  = V'(x)\sum_i \L_i + \sum_{i<j}\L_i \L_j + {1\over N} \left< \tr {V'(x)-V'(M_1)\over x-M_1}\right>.
\eeq
Notice that:
\bea
&& \sum_{i< j} \left( Y(p^i)Y(p^j)+{1\over N^2}\om_2(p^i,p^j) \right) \cr
&=& {1\over 2}\sum_{i} \left(Y(p^i)(V'(x)+\sum_j \L_j-Y(p^i)) - {1\over N^2} \ovl\om_2(p^i,p^i) \right) \cr
&=& {1\over 2}(V'(x)+\sum_j \L_j)^2 - {1\over 2}\sum_{i} \left( Y(p^i)^2 + {1\over N^2} \ovl\om_2(p^i,p^i) \right). \cr
\eea
Thus:
\beq
V'(x)^2+\sum_{i}\L_i^2 -  {2\over N} \left< \tr {V'(x)-V'(M_1)\over x-M_1}\right>
= \sum_{i} \left( Y(p^i)^2 + {1\over N^2} \ovl\om_2(p^i,p^i) \right).
\eeq

Notice that the LHS is the ratio of a polynomial in $x$ and $D(x)$: ${Q(x)\over D(x)} = V'(x)^2+\sum_{i}\L_i^2 -  {2\over N} \left< \tr {V'(x)-V'(M_1)\over x-M_1}\right>$.

The topological expansion of this equation reads for $h\geq 1$
\beq\label{Mextyexp4}
\encadremath{
\begin{array}{l}
 2 \sum_{i=0}^{d_2} y(p^i) W_{1,0}^{(h)}(p^i)dx(p) \cr
= {\sum_{i=0}^{d_2} \sum_{m=1}^{h-1} W_{1,0}^{(m)}(p^i) W_{1,0}^{(h-m)}(p^i) } + {\sum_{i=0}^{d_2} \ovl{W}_{2,0}^{(h-1)}(p^i,p^i)}  + 2 {Q^{(h)}(x(p)) dx(p)^2 \over D(x(p))}.
\end{array}
} \eeq

From now on, following the lines of \cite{CEO}, one multiplies these equations by ${1 \over 2} {dE_{p,\overline{p}}(q)
\over y(p)-y(\overline{p})} $, takes the residues when $p \to \mu_\alpha$ and sums over all the branch points
and obtains:
\beq\label{MextrecW1}
W_{1,0}^{(h)}(q) = \sum_{\alpha} \Res_{p \to \mu_\alpha} { {1 \over 2} dE_{p,\overline{p}}(q)  (W_{2,0}^{(h-1)}(p,\pbar) + \sum_{m=1}^{h-1} W_{1,0}^{(m)}(p) W_{1,0}^{(h-m)}(\pbar) ) \over(y(p)-y(\overline{p})) dx(p)}.
\eeq

Differentiating wrt the potential $V(x_i)$, one can finally write down an expression for the correlation functions:
\beq\label{MextconjecturerecW} \encadremath{
\begin{array}{rcl}
 W_{k+1,0}^{(h)}(q,p_K)
&=&   \sum_{\alpha} \Res_{p \to \mu_\alpha} {{1\over
2}dE_{p,\pbar}(q)\over (y(p)-y(\pbar))\,dx(p)}\left(
W_{k+1,0}^{(h-1)}(p,\overline{p},p_K) + \right. \cr && \;\;\; +
\left. \sum_{j,m} W_{j+1,0}^{(m)}(p,p_J) \,
W_{k+1-j,0}^{(h-m)}(\overline{p},p_{K-J}) \right) . \cr
\end{array}}\eeq

This coincides with the reccursive definition \ref{defspcorr} and ensures the equality of the correlation functions
with the former defined special "loop functions".

Keeping on following \cite{CEO}, one finds that the topological expansion of the free energy also coincides with the
special free energies defined on \ref{defspfree}, that is the $\tau$-function of the algebraic curve.

\vfill\eject

\hrule


\end{document}